%% file: DR2-SSO.tex
\documentclass[longauth]{aa_gaia} 

\usepackage{graphicx}
\usepackage{txfonts}
\usepackage{pdflscape}

\usepackage{natbib,twoopt}
\usepackage{amsmath} 
\usepackage[breaklinks=true]{hyperref} 
\bibpunct{(}{)}{;}{a}{}{,} 
\makeatletter
\newcommandtwoopt{\citeads}[3][][]{\href{http://adsabs.harvard.edu/abs/#3}%
{\def\hyper@linkstart##1##2{}%
\let\hyper@linkend\@empty\citealp[#1][#2]{#3}}}
\newcommandtwoopt{\citepads}[3][][]{\href{http://adsabs.harvard.edu/abs/#3}%
{\def\hyper@linkstart##1##2{}%
\let\hyper@linkend\@empty\citep[#1][#2]{#3}}}
\newcommandtwoopt{\citetads}[3][][]{\href{http://adsabs.harvard.edu/abs/#3}%
{\def\hyper@linkstart##1##2{}%
\let\hyper@linkend\@empty\citet[#1][#2]{#3}}}
\newcommandtwoopt{\citeyearads}[3][][]%
{\href{http://adsabs.harvard.edu/abs/#3}
{\def\hyper@linkstart##1##2{}%
\let\hyper@linkend\@empty\citeyear[#1][#2]{#3}}}
\makeatother

\newcommand{\bxi}{\boldsymbol{\xi}}
\newcommand{\bgamma}{\boldsymbol{\gamma}}

\usepackage{color}
\definecolor{mgreen}{RGB}{0,128,0}
\newcommand{\gaia}{\textit{Gaia \,}}

\newcommand{\gdrtwo}{\textit{Gaia} DR2}
\definecolor{mred}{RGB}{128,0,0}

\hypersetup{colorlinks=true,linkcolor=blue,citecolor=blue,urlcolor=blue}

\begin{document} 

   \title{\textit{Gaia} Data Release 2.}
   \subtitle{Observations of solar system objects}

\author{
{\it Gaia} Collaboration
\and F.        ~Spoto                         \inst{\ref{inst:0001},\ref{inst:0002}}
\and P.        ~Tanga                         \inst{\ref{inst:0001}}
\and F.        ~Mignard                       \inst{\ref{inst:0001}}
\and J.        ~Berthier                      \inst{\ref{inst:0002}}
\and B.        ~Carry                         \inst{\ref{inst:0001},\ref{inst:0002}}
\and A.        ~Cellino                       \inst{\ref{inst:0008}}
\and A.        ~Dell'Oro                      \inst{\ref{inst:0009}}
\and D.        ~Hestroffer                    \inst{\ref{inst:0002}}
\and K.        ~Muinonen                      \inst{\ref{inst:0011},\ref{inst:0012}}
\and T.        ~Pauwels                       \inst{\ref{inst:0013}}
\and J.-M.     ~Petit                         \inst{\ref{inst:0014}}
\and P.        ~David                         \inst{\ref{inst:0002}}
\and F.        ~De Angeli                     \inst{\ref{inst:0016}}
\and M.        ~Delbo                         \inst{\ref{inst:0001}}
\and B.        ~Fr\'{e}zouls                  \inst{\ref{inst:0018}}
\and L.        ~Galluccio                     \inst{\ref{inst:0001}}
\and M.        ~Granvik                       \inst{\ref{inst:0011},\ref{inst:0021}}
\and J.        ~Guiraud                       \inst{\ref{inst:0018}}
\and J.        ~Hern\'{a}ndez                 \inst{\ref{inst:0023}}
\and C.        ~Ord\'{e}novic                 \inst{\ref{inst:0001}}
\and J.        ~Portell                       \inst{\ref{inst:0025}}
\and E.        ~Poujoulet                     \inst{\ref{inst:0026}}
\and W.        ~Thuillot                      \inst{\ref{inst:0002}}
\and G.        ~Walmsley                      \inst{\ref{inst:0018}}
\and A.G.A.    ~Brown                         \inst{\ref{inst:0029}}
\and A.        ~Vallenari                     \inst{\ref{inst:0030}}
\and T.        ~Prusti                        \inst{\ref{inst:0031}}
\and J.H.J.    ~de Bruijne                    \inst{\ref{inst:0031}}
\and C.        ~Babusiaux                     \inst{\ref{inst:0033},\ref{inst:0034}}
\and C.A.L.    ~Bailer-Jones                  \inst{\ref{inst:0035}}
\and M.        ~Biermann                      \inst{\ref{inst:0036}}
\and D.W.      ~Evans                         \inst{\ref{inst:0016}}
\and L.        ~Eyer                          \inst{\ref{inst:0038}}
\and F.        ~Jansen                        \inst{\ref{inst:0039}}
\and C.        ~Jordi                         \inst{\ref{inst:0025}}
\and S.A.      ~Klioner                       \inst{\ref{inst:0041}}
\and U.        ~Lammers                       \inst{\ref{inst:0023}}
\and L.        ~Lindegren                     \inst{\ref{inst:0043}}
\and X.        ~Luri                          \inst{\ref{inst:0025}}
\and C.        ~Panem                         \inst{\ref{inst:0018}}
\and D.        ~Pourbaix                      \inst{\ref{inst:0046},\ref{inst:0047}}
\and S.        ~Randich                       \inst{\ref{inst:0009}}
\and P.        ~Sartoretti                    \inst{\ref{inst:0033}}
\and H.I.      ~Siddiqui                      \inst{\ref{inst:0050}}
\and C.        ~Soubiran                      \inst{\ref{inst:0051}}
\and F.        ~van Leeuwen                   \inst{\ref{inst:0016}}
\and N.A.      ~Walton                        \inst{\ref{inst:0016}}
\and F.        ~Arenou                        \inst{\ref{inst:0033}}
\and U.        ~Bastian                       \inst{\ref{inst:0036}}
\and M.        ~Cropper                       \inst{\ref{inst:0056}}
\and R.        ~Drimmel                       \inst{\ref{inst:0008}}
\and D.        ~Katz                          \inst{\ref{inst:0033}}
\and M.G.      ~Lattanzi                      \inst{\ref{inst:0008}}
\and J.        ~Bakker                        \inst{\ref{inst:0023}}
\and C.        ~Cacciari                      \inst{\ref{inst:0061}}
\and J.        ~Casta\~{n}eda                 \inst{\ref{inst:0025}}
\and L.        ~Chaoul                        \inst{\ref{inst:0018}}
\and N.        ~Cheek                         \inst{\ref{inst:0064}}
\and C.        ~Fabricius                     \inst{\ref{inst:0025}}
\and R.        ~Guerra                        \inst{\ref{inst:0023}}
\and B.        ~Holl                          \inst{\ref{inst:0038}}
\and E.        ~Masana                        \inst{\ref{inst:0025}}
\and R.        ~Messineo                      \inst{\ref{inst:0069}}
\and N.        ~Mowlavi                       \inst{\ref{inst:0038}}
\and K.        ~Nienartowicz                  \inst{\ref{inst:0071}}
\and P.        ~Panuzzo                       \inst{\ref{inst:0033}}
\and M.        ~Riello                        \inst{\ref{inst:0016}}
\and G.M.      ~Seabroke                      \inst{\ref{inst:0056}}
\and F.        ~Th\'{e}venin                  \inst{\ref{inst:0001}}
\and G.        ~Gracia-Abril                  \inst{\ref{inst:0076},\ref{inst:0036}}
\and G.        ~Comoretto                     \inst{\ref{inst:0050}}
\and M.        ~Garcia-Reinaldos              \inst{\ref{inst:0023}}
\and D.        ~Teyssier                      \inst{\ref{inst:0050}}
\and M.        ~Altmann                       \inst{\ref{inst:0036},\ref{inst:0082}}
\and R.        ~Andrae                        \inst{\ref{inst:0035}}
\and M.        ~Audard                        \inst{\ref{inst:0038}}
\and I.        ~Bellas-Velidis                \inst{\ref{inst:0085}}
\and K.        ~Benson                        \inst{\ref{inst:0056}}
\and R.        ~Blomme                        \inst{\ref{inst:0013}}
\and P.        ~Burgess                       \inst{\ref{inst:0016}}
\and G.        ~Busso                         \inst{\ref{inst:0016}}
\and G.        ~Clementini                    \inst{\ref{inst:0061}}
\and M.        ~Clotet                        \inst{\ref{inst:0025}}
\and O.        ~Creevey                       \inst{\ref{inst:0001}}
\and M.        ~Davidson                      \inst{\ref{inst:0093}}
\and J.        ~De Ridder                     \inst{\ref{inst:0094}}
\and L.        ~Delchambre                    \inst{\ref{inst:0095}}
\and C.        ~Ducourant                     \inst{\ref{inst:0051}}
\and J.        ~Fern\'{a}ndez-Hern\'{a}ndez   \inst{\ref{inst:0097}}
\and M.        ~Fouesneau                     \inst{\ref{inst:0035}}
\and Y.        ~Fr\'{e}mat                    \inst{\ref{inst:0013}}
\and M.        ~Garc\'{i}a-Torres             \inst{\ref{inst:0100}}
\and J.        ~Gonz\'{a}lez-N\'{u}\~{n}ez    \inst{\ref{inst:0064},\ref{inst:0102}}
\and J.J.      ~Gonz\'{a}lez-Vidal            \inst{\ref{inst:0025}}
\and E.        ~Gosset                        \inst{\ref{inst:0095},\ref{inst:0047}}
\and L.P.      ~Guy                           \inst{\ref{inst:0071},\ref{inst:0107}}
\and J.-L.     ~Halbwachs                     \inst{\ref{inst:0108}}
\and N.C.      ~Hambly                        \inst{\ref{inst:0093}}
\and D.L.      ~Harrison                      \inst{\ref{inst:0016},\ref{inst:0111}}
\and S.T.      ~Hodgkin                       \inst{\ref{inst:0016}}
\and A.        ~Hutton                        \inst{\ref{inst:0113}}
\and G.        ~Jasniewicz                    \inst{\ref{inst:0114}}
\and A.        ~Jean-Antoine-Piccolo          \inst{\ref{inst:0018}}
\and S.        ~Jordan                        \inst{\ref{inst:0036}}
\and A.J.      ~Korn                          \inst{\ref{inst:0117}}
\and A.        ~Krone-Martins                 \inst{\ref{inst:0118}}
\and A.C.      ~Lanzafame                     \inst{\ref{inst:0119},\ref{inst:0120}}
\and T.        ~Lebzelter                     \inst{\ref{inst:0121}}
\and W.        ~L\"{ o}ffler                  \inst{\ref{inst:0036}}
\and M.        ~Manteiga                      \inst{\ref{inst:0123},\ref{inst:0124}}
\and P.M.      ~Marrese                       \inst{\ref{inst:0125},\ref{inst:0126}}
\and J.M.      ~Mart\'{i}n-Fleitas            \inst{\ref{inst:0113}}
\and A.        ~Moitinho                      \inst{\ref{inst:0118}}
\and A.        ~Mora                          \inst{\ref{inst:0113}}
\and J.        ~Osinde                        \inst{\ref{inst:0130}}
\and E.        ~Pancino                       \inst{\ref{inst:0009},\ref{inst:0126}}
\and A.        ~Recio-Blanco                  \inst{\ref{inst:0001}}
\and P.J.      ~Richards                      \inst{\ref{inst:0134}}
\and L.        ~Rimoldini                     \inst{\ref{inst:0071}}
\and A.C.      ~Robin                         \inst{\ref{inst:0014}}
\and L.M.      ~Sarro                         \inst{\ref{inst:0137}}
\and C.        ~Siopis                        \inst{\ref{inst:0046}}
\and M.        ~Smith                         \inst{\ref{inst:0056}}
\and A.        ~Sozzetti                      \inst{\ref{inst:0008}}
\and M.        ~S\"{ u}veges                  \inst{\ref{inst:0035}}
\and J.        ~Torra                         \inst{\ref{inst:0025}}
\and W.        ~van Reeven                    \inst{\ref{inst:0113}}
\and U.        ~Abbas                         \inst{\ref{inst:0008}}
\and A.        ~Abreu Aramburu                \inst{\ref{inst:0145}}
\and S.        ~Accart                        \inst{\ref{inst:0146}}
\and C.        ~Aerts                         \inst{\ref{inst:0094},\ref{inst:0148}}
\and G.        ~Altavilla                     \inst{\ref{inst:0125},\ref{inst:0126},\ref{inst:0061}}
\and M.A.      ~\'{A}lvarez                   \inst{\ref{inst:0123}}
\and R.        ~Alvarez                       \inst{\ref{inst:0023}}
\and J.        ~Alves                         \inst{\ref{inst:0121}}
\and R.I.      ~Anderson                      \inst{\ref{inst:0155},\ref{inst:0038}}
\and A.H.      ~Andrei                        \inst{\ref{inst:0157},\ref{inst:0158},\ref{inst:0082}}
\and E.        ~Anglada Varela                \inst{\ref{inst:0097}}
\and E.        ~Antiche                       \inst{\ref{inst:0025}}
\and T.        ~Antoja                        \inst{\ref{inst:0031},\ref{inst:0025}}
\and B.        ~Arcay                         \inst{\ref{inst:0123}}
\and T.L.      ~Astraatmadja                  \inst{\ref{inst:0035},\ref{inst:0166}}
\and N.        ~Bach                          \inst{\ref{inst:0113}}
\and S.G.      ~Baker                         \inst{\ref{inst:0056}}
\and L.        ~Balaguer-N\'{u}\~{n}ez        \inst{\ref{inst:0025}}
\and P.        ~Balm                          \inst{\ref{inst:0050}}
\and C.        ~Barache                       \inst{\ref{inst:0082}}
\and C.        ~Barata                        \inst{\ref{inst:0118}}
\and D.        ~Barbato                       \inst{\ref{inst:0173},\ref{inst:0008}}
\and F.        ~Barblan                       \inst{\ref{inst:0038}}
\and P.S.      ~Barklem                       \inst{\ref{inst:0117}}
\and D.        ~Barrado                       \inst{\ref{inst:0177}}
\and M.        ~Barros                        \inst{\ref{inst:0118}}
\and M.A.      ~Barstow                       \inst{\ref{inst:0179}}
\and S.        ~Bartholom\'{e} Mu\~{n}oz      \inst{\ref{inst:0025}}
\and J.-L.     ~Bassilana                     \inst{\ref{inst:0146}}
\and U.        ~Becciani                      \inst{\ref{inst:0120}}
\and M.        ~Bellazzini                    \inst{\ref{inst:0061}}
\and A.        ~Berihuete                     \inst{\ref{inst:0184}}
\and S.        ~Bertone                       \inst{\ref{inst:0008},\ref{inst:0082},\ref{inst:0187}}
\and L.        ~Bianchi                       \inst{\ref{inst:0188}}
\and O.        ~Bienaym\'{e}                  \inst{\ref{inst:0108}}
\and S.        ~Blanco-Cuaresma               \inst{\ref{inst:0038},\ref{inst:0051},\ref{inst:0192}}
\and T.        ~Boch                          \inst{\ref{inst:0108}}
\and C.        ~Boeche                        \inst{\ref{inst:0030}}
\and A.        ~Bombrun                       \inst{\ref{inst:0195}}
\and R.        ~Borrachero                    \inst{\ref{inst:0025}}
\and D.        ~Bossini                       \inst{\ref{inst:0030}}
\and S.        ~Bouquillon                    \inst{\ref{inst:0082}}
\and G.        ~Bourda                        \inst{\ref{inst:0051}}
\and A.        ~Bragaglia                     \inst{\ref{inst:0061}}
\and L.        ~Bramante                      \inst{\ref{inst:0069}}
\and M.A.      ~Breddels                      \inst{\ref{inst:0202}}
\and A.        ~Bressan                       \inst{\ref{inst:0203}}
\and N.        ~Brouillet                     \inst{\ref{inst:0051}}
\and T.        ~Br\"{ u}semeister             \inst{\ref{inst:0036}}
\and E.        ~Brugaletta                    \inst{\ref{inst:0120}}
\and B.        ~Bucciarelli                   \inst{\ref{inst:0008}}
\and A.        ~Burlacu                       \inst{\ref{inst:0018}}
\and D.        ~Busonero                      \inst{\ref{inst:0008}}
\and A.G.      ~Butkevich                     \inst{\ref{inst:0041}}
\and R.        ~Buzzi                         \inst{\ref{inst:0008}}
\and E.        ~Caffau                        \inst{\ref{inst:0033}}
\and R.        ~Cancelliere                   \inst{\ref{inst:0213}}
\and G.        ~Cannizzaro                    \inst{\ref{inst:0214},\ref{inst:0148}}
\and T.        ~Cantat-Gaudin                 \inst{\ref{inst:0030},\ref{inst:0025}}
\and R.        ~Carballo                      \inst{\ref{inst:0218}}
\and T.        ~Carlucci                      \inst{\ref{inst:0082}}
\and J.M.      ~Carrasco                      \inst{\ref{inst:0025}}
\and L.        ~Casamiquela                   \inst{\ref{inst:0025}}
\and M.        ~Castellani                    \inst{\ref{inst:0125}}
\and A.        ~Castro-Ginard                 \inst{\ref{inst:0025}}
\and P.        ~Charlot                       \inst{\ref{inst:0051}}
\and L.        ~Chemin                        \inst{\ref{inst:0225}}
\and A.        ~Chiavassa                     \inst{\ref{inst:0001}}
\and G.        ~Cocozza                       \inst{\ref{inst:0061}}
\and G.        ~Costigan                      \inst{\ref{inst:0029}}
\and S.        ~Cowell                        \inst{\ref{inst:0016}}
\and F.        ~Crifo                         \inst{\ref{inst:0033}}
\and M.        ~Crosta                        \inst{\ref{inst:0008}}
\and C.        ~Crowley                       \inst{\ref{inst:0195}}
\and J.        ~Cuypers$^\dagger$             \inst{\ref{inst:0013}}
\and C.        ~Dafonte                       \inst{\ref{inst:0123}}
\and Y.        ~Damerdji                      \inst{\ref{inst:0095},\ref{inst:0236}}
\and A.        ~Dapergolas                    \inst{\ref{inst:0085}}
\and M.        ~David                         \inst{\ref{inst:0238}}
\and P.        ~de Laverny                    \inst{\ref{inst:0001}}
\and F.        ~De Luise                      \inst{\ref{inst:0240}}
\and R.        ~De March                      \inst{\ref{inst:0069}}
\and R.        ~de Souza                      \inst{\ref{inst:0242}}
\and A.        ~de Torres                     \inst{\ref{inst:0195}}
\and J.        ~Debosscher                    \inst{\ref{inst:0094}}
\and E.        ~del Pozo                      \inst{\ref{inst:0113}}
\and A.        ~Delgado                       \inst{\ref{inst:0016}}
\and H.E.      ~Delgado                       \inst{\ref{inst:0137}}
\and S.        ~Diakite                       \inst{\ref{inst:0014}}
\and C.        ~Diener                        \inst{\ref{inst:0016}}
\and E.        ~Distefano                     \inst{\ref{inst:0120}}
\and C.        ~Dolding                       \inst{\ref{inst:0056}}
\and P.        ~Drazinos                      \inst{\ref{inst:0252}}
\and J.        ~Dur\'{a}n                     \inst{\ref{inst:0130}}
\and B.        ~Edvardsson                    \inst{\ref{inst:0117}}
\and H.        ~Enke                          \inst{\ref{inst:0255}}
\and K.        ~Eriksson                      \inst{\ref{inst:0117}}
\and P.        ~Esquej                        \inst{\ref{inst:0257}}
\and G.        ~Eynard Bontemps               \inst{\ref{inst:0018}}
\and C.        ~Fabre                         \inst{\ref{inst:0259}}
\and M.        ~Fabrizio                      \inst{\ref{inst:0125},\ref{inst:0126}}
\and S.        ~Faigler                       \inst{\ref{inst:0262}}
\and A.J.      ~Falc\~{a}o                    \inst{\ref{inst:0263}}
\and M.        ~Farr\`{a}s Casas              \inst{\ref{inst:0025}}
\and L.        ~Federici                      \inst{\ref{inst:0061}}
\and G.        ~Fedorets                      \inst{\ref{inst:0011}}
\and P.        ~Fernique                      \inst{\ref{inst:0108}}
\and F.        ~Figueras                      \inst{\ref{inst:0025}}
\and F.        ~Filippi                       \inst{\ref{inst:0069}}
\and K.        ~Findeisen                     \inst{\ref{inst:0033}}
\and A.        ~Fonti                         \inst{\ref{inst:0069}}
\and E.        ~Fraile                        \inst{\ref{inst:0257}}
\and M.        ~Fraser                        \inst{\ref{inst:0016},\ref{inst:0274}}
\and M.        ~Gai                           \inst{\ref{inst:0008}}
\and S.        ~Galleti                       \inst{\ref{inst:0061}}
\and D.        ~Garabato                      \inst{\ref{inst:0123}}
\and F.        ~Garc\'{i}a-Sedano             \inst{\ref{inst:0137}}
\and A.        ~Garofalo                      \inst{\ref{inst:0279},\ref{inst:0061}}
\and N.        ~Garralda                      \inst{\ref{inst:0025}}
\and A.        ~Gavel                         \inst{\ref{inst:0117}}
\and P.        ~Gavras                        \inst{\ref{inst:0033},\ref{inst:0085},\ref{inst:0252}}
\and J.        ~Gerssen                       \inst{\ref{inst:0255}}
\and R.        ~Geyer                         \inst{\ref{inst:0041}}
\and P.        ~Giacobbe                      \inst{\ref{inst:0008}}
\and G.        ~Gilmore                       \inst{\ref{inst:0016}}
\and S.        ~Girona                        \inst{\ref{inst:0290}}
\and G.        ~Giuffrida                     \inst{\ref{inst:0126},\ref{inst:0125}}
\and F.        ~Glass                         \inst{\ref{inst:0038}}
\and M.        ~Gomes                         \inst{\ref{inst:0118}}
\and A.        ~Gueguen                       \inst{\ref{inst:0033},\ref{inst:0296}}
\and A.        ~Guerrier                      \inst{\ref{inst:0146}}
\and R.        ~Guti\'{e}rrez-S\'{a}nchez     \inst{\ref{inst:0050}}
\and R.        ~Haigron                       \inst{\ref{inst:0033}}
\and D.        ~Hatzidimitriou                \inst{\ref{inst:0252},\ref{inst:0085}}
\and M.        ~Hauser                        \inst{\ref{inst:0036},\ref{inst:0035}}
\and M.        ~Haywood                       \inst{\ref{inst:0033}}
\and U.        ~Heiter                        \inst{\ref{inst:0117}}
\and A.        ~Helmi                         \inst{\ref{inst:0202}}
\and J.        ~Heu                           \inst{\ref{inst:0033}}
\and T.        ~Hilger                        \inst{\ref{inst:0041}}
\and D.        ~Hobbs                         \inst{\ref{inst:0043}}
\and W.        ~Hofmann                       \inst{\ref{inst:0036}}
\and G.        ~Holland                       \inst{\ref{inst:0016}}
\and H.E.      ~Huckle                        \inst{\ref{inst:0056}}
\and A.        ~Hypki                         \inst{\ref{inst:0029},\ref{inst:0314}}
\and V.        ~Icardi                        \inst{\ref{inst:0069}}
\and K.        ~Jan{\ss}en                    \inst{\ref{inst:0255}}
\and G.        ~Jevardat de Fombelle          \inst{\ref{inst:0071}}
\and P.G.      ~Jonker                        \inst{\ref{inst:0214},\ref{inst:0148}}
\and \'{A}.L.  ~Juh\'{a}sz                    \inst{\ref{inst:0320},\ref{inst:0321}}
\and F.        ~Julbe                         \inst{\ref{inst:0025}}
\and A.        ~Karampelas                    \inst{\ref{inst:0252},\ref{inst:0324}}
\and A.        ~Kewley                        \inst{\ref{inst:0016}}
\and J.        ~Klar                          \inst{\ref{inst:0255}}
\and A.        ~Kochoska                      \inst{\ref{inst:0327},\ref{inst:0328}}
\and R.        ~Kohley                        \inst{\ref{inst:0023}}
\and K.        ~Kolenberg                     \inst{\ref{inst:0330},\ref{inst:0094},\ref{inst:0192}}
\and M.        ~Kontizas                      \inst{\ref{inst:0252}}
\and E.        ~Kontizas                      \inst{\ref{inst:0085}}
\and S.E.      ~Koposov                       \inst{\ref{inst:0016},\ref{inst:0336}}
\and G.        ~Kordopatis                    \inst{\ref{inst:0001}}
\and Z.        ~Kostrzewa-Rutkowska           \inst{\ref{inst:0214},\ref{inst:0148}}
\and P.        ~Koubsky                       \inst{\ref{inst:0340}}
\and S.        ~Lambert                       \inst{\ref{inst:0082}}
\and A.F.      ~Lanza                         \inst{\ref{inst:0120}}
\and Y.        ~Lasne                         \inst{\ref{inst:0146}}
\and J.-B.     ~Lavigne                       \inst{\ref{inst:0146}}
\and Y.        ~Le Fustec                     \inst{\ref{inst:0345}}
\and C.        ~Le Poncin-Lafitte             \inst{\ref{inst:0082}}
\and Y.        ~Lebreton                      \inst{\ref{inst:0033},\ref{inst:0348}}
\and S.        ~Leccia                        \inst{\ref{inst:0349}}
\and N.        ~Leclerc                       \inst{\ref{inst:0033}}
\and I.        ~Lecoeur-Taibi                 \inst{\ref{inst:0071}}
\and H.        ~Lenhardt                      \inst{\ref{inst:0036}}
\and F.        ~Leroux                        \inst{\ref{inst:0146}}
\and S.        ~Liao                          \inst{\ref{inst:0008},\ref{inst:0355},\ref{inst:0356}}
\and E.        ~Licata                        \inst{\ref{inst:0188}}
\and H.E.P.    ~Lindstr{\o}m                  \inst{\ref{inst:0358},\ref{inst:0359}}
\and T.A.      ~Lister                        \inst{\ref{inst:0360}}
\and E.        ~Livanou                       \inst{\ref{inst:0252}}
\and A.        ~Lobel                         \inst{\ref{inst:0013}}
\and M.        ~L\'{o}pez                     \inst{\ref{inst:0177}}
\and S.        ~Managau                       \inst{\ref{inst:0146}}
\and R.G.      ~Mann                          \inst{\ref{inst:0093}}
\and G.        ~Mantelet                      \inst{\ref{inst:0036}}
\and O.        ~Marchal                       \inst{\ref{inst:0033}}
\and J.M.      ~Marchant                      \inst{\ref{inst:0368}}
\and M.        ~Marconi                       \inst{\ref{inst:0349}}
\and S.        ~Marinoni                      \inst{\ref{inst:0125},\ref{inst:0126}}
\and G.        ~Marschalk\'{o}                \inst{\ref{inst:0320},\ref{inst:0373}}
\and D.J.      ~Marshall                      \inst{\ref{inst:0374}}
\and M.        ~Martino                       \inst{\ref{inst:0069}}
\and G.        ~Marton                        \inst{\ref{inst:0320}}
\and N.        ~Mary                          \inst{\ref{inst:0146}}
\and D.        ~Massari                       \inst{\ref{inst:0202}}
\and G.        ~Matijevi\v{c}                 \inst{\ref{inst:0255}}
\and T.        ~Mazeh                         \inst{\ref{inst:0262}}
\and P.J.      ~McMillan                      \inst{\ref{inst:0043}}
\and S.        ~Messina                       \inst{\ref{inst:0120}}
\and D.        ~Michalik                      \inst{\ref{inst:0043}}
\and N.R.      ~Millar                        \inst{\ref{inst:0016}}
\and D.        ~Molina                        \inst{\ref{inst:0025}}
\and R.        ~Molinaro                      \inst{\ref{inst:0349}}
\and L.        ~Moln\'{a}r                    \inst{\ref{inst:0320}}
\and P.        ~Montegriffo                   \inst{\ref{inst:0061}}
\and R.        ~Mor                           \inst{\ref{inst:0025}}
\and R.        ~Morbidelli                    \inst{\ref{inst:0008}}
\and T.        ~Morel                         \inst{\ref{inst:0095}}
\and D.        ~Morris                        \inst{\ref{inst:0093}}
\and A.F.      ~Mulone                        \inst{\ref{inst:0069}}
\and T.        ~Muraveva                      \inst{\ref{inst:0061}}
\and I.        ~Musella                       \inst{\ref{inst:0349}}
\and G.        ~Nelemans                      \inst{\ref{inst:0148},\ref{inst:0094}}
\and L.        ~Nicastro                      \inst{\ref{inst:0061}}
\and L.        ~Noval                         \inst{\ref{inst:0146}}
\and W.        ~O'Mullane                     \inst{\ref{inst:0023},\ref{inst:0107}}
\and D.        ~Ord\'{o}\~{n}ez-Blanco        \inst{\ref{inst:0071}}
\and P.        ~Osborne                       \inst{\ref{inst:0016}}
\and C.        ~Pagani                        \inst{\ref{inst:0179}}
\and I.        ~Pagano                        \inst{\ref{inst:0120}}
\and F.        ~Pailler                       \inst{\ref{inst:0018}}
\and H.        ~Palacin                       \inst{\ref{inst:0146}}
\and L.        ~Palaversa                     \inst{\ref{inst:0016},\ref{inst:0038}}
\and A.        ~Panahi                        \inst{\ref{inst:0262}}
\and M.        ~Pawlak                        \inst{\ref{inst:0411},\ref{inst:0412}}
\and A.M.      ~Piersimoni                    \inst{\ref{inst:0240}}
\and F.-X.     ~Pineau                        \inst{\ref{inst:0108}}
\and E.        ~Plachy                        \inst{\ref{inst:0320}}
\and G.        ~Plum                          \inst{\ref{inst:0033}}
\and E.        ~Poggio                        \inst{\ref{inst:0173},\ref{inst:0008}}
\and A.        ~Pr\v{s}a                      \inst{\ref{inst:0328}}
\and L.        ~Pulone                        \inst{\ref{inst:0125}}
\and E.        ~Racero                        \inst{\ref{inst:0064}}
\and S.        ~Ragaini                       \inst{\ref{inst:0061}}
\and N.        ~Rambaux                       \inst{\ref{inst:0002}}
\and M.        ~Ramos-Lerate                  \inst{\ref{inst:0424}}
\and S.        ~Regibo                        \inst{\ref{inst:0094}}
\and C.        ~Reyl\'{e}                     \inst{\ref{inst:0014}}
\and F.        ~Riclet                        \inst{\ref{inst:0018}}
\and V.        ~Ripepi                        \inst{\ref{inst:0349}}
\and A.        ~Riva                          \inst{\ref{inst:0008}}
\and A.        ~Rivard                        \inst{\ref{inst:0146}}
\and G.        ~Rixon                         \inst{\ref{inst:0016}}
\and T.        ~Roegiers                      \inst{\ref{inst:0432}}
\and M.        ~Roelens                       \inst{\ref{inst:0038}}
\and M.        ~Romero-G\'{o}mez              \inst{\ref{inst:0025}}
\and N.        ~Rowell                        \inst{\ref{inst:0093}}
\and F.        ~Royer                         \inst{\ref{inst:0033}}
\and L.        ~Ruiz-Dern                     \inst{\ref{inst:0033}}
\and G.        ~Sadowski                      \inst{\ref{inst:0046}}
\and T.        ~Sagrist\`{a} Sell\'{e}s       \inst{\ref{inst:0036}}
\and J.        ~Sahlmann                      \inst{\ref{inst:0023},\ref{inst:0441}}
\and J.        ~Salgado                       \inst{\ref{inst:0442}}
\and E.        ~Salguero                      \inst{\ref{inst:0097}}
\and N.        ~Sanna                         \inst{\ref{inst:0009}}
\and T.        ~Santana-Ros                   \inst{\ref{inst:0314}}
\and M.        ~Sarasso                       \inst{\ref{inst:0008}}
\and H.        ~Savietto                      \inst{\ref{inst:0447}}
\and M.        ~Schultheis                    \inst{\ref{inst:0001}}
\and E.        ~Sciacca                       \inst{\ref{inst:0120}}
\and M.        ~Segol                         \inst{\ref{inst:0450}}
\and J.C.      ~Segovia                       \inst{\ref{inst:0064}}
\and D.        ~S\'{e}gransan                 \inst{\ref{inst:0038}}
\and I-C.      ~Shih                          \inst{\ref{inst:0033}}
\and L.        ~Siltala                       \inst{\ref{inst:0011},\ref{inst:0455}}
\and A.F.      ~Silva                         \inst{\ref{inst:0118}}
\and R.L.      ~Smart                         \inst{\ref{inst:0008}}
\and K.W.      ~Smith                         \inst{\ref{inst:0035}}
\and E.        ~Solano                        \inst{\ref{inst:0177},\ref{inst:0460}}
\and F.        ~Solitro                       \inst{\ref{inst:0069}}
\and R.        ~Sordo                         \inst{\ref{inst:0030}}
\and S.        ~Soria Nieto                   \inst{\ref{inst:0025}}
\and J.        ~Souchay                       \inst{\ref{inst:0082}}
\and A.        ~Spagna                        \inst{\ref{inst:0008}}
\and U.        ~Stampa                        \inst{\ref{inst:0036}}
\and I.A.      ~Steele                        \inst{\ref{inst:0368}}
\and H.        ~Steidelm\"{ u}ller            \inst{\ref{inst:0041}}
\and C.A.      ~Stephenson                    \inst{\ref{inst:0050}}
\and H.        ~Stoev                         \inst{\ref{inst:0470}}
\and F.F.      ~Suess                         \inst{\ref{inst:0016}}
\and J.        ~Surdej                        \inst{\ref{inst:0095}}
\and L.        ~Szabados                      \inst{\ref{inst:0320}}
\and E.        ~Szegedi-Elek                  \inst{\ref{inst:0320}}
\and D.        ~Tapiador                      \inst{\ref{inst:0475},\ref{inst:0476}}
\and F.        ~Taris                         \inst{\ref{inst:0082}}
\and G.        ~Tauran                        \inst{\ref{inst:0146}}
\and M.B.      ~Taylor                        \inst{\ref{inst:0479}}
\and R.        ~Teixeira                      \inst{\ref{inst:0242}}
\and D.        ~Terrett                       \inst{\ref{inst:0134}}
\and P.        ~Teyssandier                   \inst{\ref{inst:0082}}
\and A.        ~Titarenko                     \inst{\ref{inst:0001}}
\and F.        ~Torra Clotet                  \inst{\ref{inst:0484}}
\and C.        ~Turon                         \inst{\ref{inst:0033}}
\and A.        ~Ulla                          \inst{\ref{inst:0486}}
\and E.        ~Utrilla                       \inst{\ref{inst:0113}}
\and S.        ~Uzzi                          \inst{\ref{inst:0069}}
\and M.        ~Vaillant                      \inst{\ref{inst:0146}}
\and G.        ~Valentini                     \inst{\ref{inst:0240}}
\and V.        ~Valette                       \inst{\ref{inst:0018}}
\and A.        ~van Elteren                   \inst{\ref{inst:0029}}
\and E.        ~Van Hemelryck                 \inst{\ref{inst:0013}}
\and M.        ~van Leeuwen                   \inst{\ref{inst:0016}}
\and M.        ~Vaschetto                     \inst{\ref{inst:0069}}
\and A.        ~Vecchiato                     \inst{\ref{inst:0008}}
\and J.        ~Veljanoski                    \inst{\ref{inst:0202}}
\and Y.        ~Viala                         \inst{\ref{inst:0033}}
\and D.        ~Vicente                       \inst{\ref{inst:0290}}
\and S.        ~Vogt                          \inst{\ref{inst:0432}}
\and C.        ~von Essen                     \inst{\ref{inst:0501}}
\and H.        ~Voss                          \inst{\ref{inst:0025}}
\and V.        ~Votruba                       \inst{\ref{inst:0340}}
\and S.        ~Voutsinas                     \inst{\ref{inst:0093}}
\and M.        ~Weiler                        \inst{\ref{inst:0025}}
\and O.        ~Wertz                         \inst{\ref{inst:0506}}
\and T.        ~Wevers                        \inst{\ref{inst:0016},\ref{inst:0148}}
\and \L{}.     ~Wyrzykowski                   \inst{\ref{inst:0016},\ref{inst:0411}}
\and A.        ~Yoldas                        \inst{\ref{inst:0016}}
\and M.        ~\v{Z}erjal                    \inst{\ref{inst:0327},\ref{inst:0513}}
\and H.        ~Ziaeepour                     \inst{\ref{inst:0014}}
\and J.        ~Zorec                         \inst{\ref{inst:0515}}
\and S.        ~Zschocke                      \inst{\ref{inst:0041}}
\and S.        ~Zucker                        \inst{\ref{inst:0517}}
\and C.        ~Zurbach                       \inst{\ref{inst:0114}}
\and T.        ~Zwitter                       \inst{\ref{inst:0327}}
}
\institute{
     Universit\'{e} C\^{o}te d'Azur, Observatoire de la C\^{o}te d'Azur, CNRS, Laboratoire Lagrange, Bd de l'Observatoire, CS 34229, 06304 Nice Cedex 4, France\relax                                        \label{inst:0001}
\and IMCCE, Observatoire de Paris, Universit\'{e} PSL, CNRS,  Sorbonne Universit\'{e}, Univ. Lille, 77 av. Denfert-Rochereau, 75014 Paris, France\relax                                                      \label{inst:0002}
\and INAF - Osservatorio Astrofisico di Torino, via Osservatorio 20, 10025 Pino Torinese (TO), Italy\relax                                                                                                   \label{inst:0008}
\and INAF - Osservatorio Astrofisico di Arcetri, Largo Enrico Fermi 5, 50125 Firenze, Italy\relax                                                                                                            \label{inst:0009}
\and University of Helsinki, Department of Physics, P.O. Box 64, 00014 Helsinki, Finland\relax                                                                                                               \label{inst:0011}
\and Finnish Geospatial Research Institute FGI, Geodeetinrinne 2, 02430 Masala, Finland\relax                                                                                                                \label{inst:0012}
\and Royal Observatory of Belgium, Ringlaan 3, 1180 Brussels, Belgium\relax                                                                                                                                  \label{inst:0013}
\and Institut UTINAM UMR6213, CNRS, OSU THETA Franche-Comt\'{e} Bourgogne, Universit\'{e} Bourgogne Franche-Comt\'{e}, 25000 Besan\c{c}on, France\relax                                                      \label{inst:0014}
\and Institute of Astronomy, University of Cambridge, Madingley Road, Cambridge CB3 0HA, United Kingdom\relax                                                                                                \label{inst:0016}
\and CNES Centre Spatial de Toulouse, 18 avenue Edouard Belin, 31401 Toulouse Cedex 9, France\relax                                                                                                          \label{inst:0018}
\and Department of Computer Science, Electrical and Space Engineering, Lule\aa{} University of Technology, Box 848, S-981 28 Kiruna, Sweden\relax                                                            \label{inst:0021}
\and European Space Astronomy Centre (ESA/ESAC), Camino bajo del Castillo, s/n, Urbanizacion Villafranca del Castillo, Villanueva de la Ca\~{n}ada, 28692 Madrid, Spain\relax                                \label{inst:0023}
\and Institut de Ci\`{e}ncies del Cosmos, Universitat  de  Barcelona  (IEEC-UB), Mart\'{i} i  Franqu\`{e}s  1, 08028 Barcelona, Spain\relax                                                                  \label{inst:0025}
\and AKKA for CNES Centre Spatial de Toulouse, 18 avenue Edouard Belin, 31401 Toulouse Cedex 9, France\relax                                                                                                 \label{inst:0026}
\and Leiden Observatory, Leiden University, Niels Bohrweg 2, 2333 CA Leiden, The Netherlands\relax                                                                                                           \label{inst:0029}
\and INAF - Osservatorio astronomico di Padova, Vicolo Osservatorio 5, 35122 Padova, Italy\relax                                                                                                             \label{inst:0030}
\and Science Support Office, Directorate of Science, European Space Research and Technology Centre (ESA/ESTEC), Keplerlaan 1, 2201AZ, Noordwijk, The Netherlands\relax                                       \label{inst:0031}
\and GEPI, Observatoire de Paris, Universit\'{e} PSL, CNRS, 5 Place Jules Janssen, 92190 Meudon, France\relax                                                                                                \label{inst:0033}
\and Univ. Grenoble Alpes, CNRS, IPAG, 38000 Grenoble, France\relax                                                                                                                                          \label{inst:0034}
\and Max Planck Institute for Astronomy, K\"{ o}nigstuhl 17, 69117 Heidelberg, Germany\relax                                                                                                                 \label{inst:0035}
\and Astronomisches Rechen-Institut, Zentrum f\"{ u}r Astronomie der Universit\"{ a}t Heidelberg, M\"{ o}nchhofstr. 12-14, 69120 Heidelberg, Germany\relax                                                   \label{inst:0036}
\and Department of Astronomy, University of Geneva, Chemin des Maillettes 51, 1290 Versoix, Switzerland\relax                                                                                                \label{inst:0038}
\and Mission Operations Division, Operations Department, Directorate of Science, European Space Research and Technology Centre (ESA/ESTEC), Keplerlaan 1, 2201 AZ, Noordwijk, The Netherlands\relax          \label{inst:0039}
\and Lohrmann Observatory, Technische Universit\"{ a}t Dresden, Mommsenstra{\ss}e 13, 01062 Dresden, Germany\relax                                                                                           \label{inst:0041}
\and Lund Observatory, Department of Astronomy and Theoretical Physics, Lund University, Box 43, 22100 Lund, Sweden\relax                                                                                    \label{inst:0043}
\and Institut d'Astronomie et d'Astrophysique, Universit\'{e} Libre de Bruxelles CP 226, Boulevard du Triomphe, 1050 Brussels, Belgium\relax                                                                 \label{inst:0046}
\and F.R.S.-FNRS, Rue d'Egmont 5, 1000 Brussels, Belgium\relax                                                                                                                                               \label{inst:0047}
\and Telespazio Vega UK Ltd for ESA/ESAC, Camino bajo del Castillo, s/n, Urbanizacion Villafranca del Castillo, Villanueva de la Ca\~{n}ada, 28692 Madrid, Spain\relax                                       \label{inst:0050}
\and Laboratoire d'astrophysique de Bordeaux, Univ. Bordeaux, CNRS, B18N, all{\'e}e Geoffroy Saint-Hilaire, 33615 Pessac, France\relax                                                                       \label{inst:0051}
\and Mullard Space Science Laboratory, University College London, Holmbury St Mary, Dorking, Surrey RH5 6NT, United Kingdom\relax                                                                            \label{inst:0056}
\and INAF - Osservatorio di Astrofisica e Scienza dello Spazio di Bologna, via Piero Gobetti 93/3, 40129 Bologna, Italy\relax                                                                                \label{inst:0061}
\and Serco Gesti\'{o}n de Negocios for ESA/ESAC, Camino bajo del Castillo, s/n, Urbanizacion Villafranca del Castillo, Villanueva de la Ca\~{n}ada, 28692 Madrid, Spain\relax                                \label{inst:0064}
\and ALTEC S.p.a, Corso Marche, 79,10146 Torino, Italy\relax                                                                                                                                                 \label{inst:0069}
\and Department of Astronomy, University of Geneva, Chemin d'Ecogia 16, 1290 Versoix, Switzerland\relax                                                                                                      \label{inst:0071}
\and Gaia DPAC Project Office, ESAC, Camino bajo del Castillo, s/n, Urbanizacion Villafranca del Castillo, Villanueva de la Ca\~{n}ada, 28692 Madrid, Spain\relax                                            \label{inst:0076}
\and SYRTE, Observatoire de Paris, Universit\'{e} PSL, CNRS,  Sorbonne Universit\'{e}, LNE, 61 avenue de l’Observatoire 75014 Paris, France\relax                                                          \label{inst:0082}
\and National Observatory of Athens, I. Metaxa and Vas. Pavlou, Palaia Penteli, 15236 Athens, Greece\relax                                                                                                   \label{inst:0085}
\and Institute for Astronomy, University of Edinburgh, Royal Observatory, Blackford Hill, Edinburgh EH9 3HJ, United Kingdom\relax                                                                            \label{inst:0093}
\and Instituut voor Sterrenkunde, KU Leuven, Celestijnenlaan 200D, 3001 Leuven, Belgium\relax                                                                                                                \label{inst:0094}
\and Institut d'Astrophysique et de G\'{e}ophysique, Universit\'{e} de Li\`{e}ge, 19c, All\'{e}e du 6 Ao\^{u}t, B-4000 Li\`{e}ge, Belgium\relax                                                              \label{inst:0095}
\and ATG Europe for ESA/ESAC, Camino bajo del Castillo, s/n, Urbanizacion Villafranca del Castillo, Villanueva de la Ca\~{n}ada, 28692 Madrid, Spain\relax                                                   \label{inst:0097}
\and \'{A}rea de Lenguajes y Sistemas Inform\'{a}ticos, Universidad Pablo de Olavide, Ctra. de Utrera, km 1. 41013, Sevilla, Spain\relax                                                                     \label{inst:0100}
\and ETSE Telecomunicaci\'{o}n, Universidade de Vigo, Campus Lagoas-Marcosende, 36310 Vigo, Galicia, Spain\relax                                                                                             \label{inst:0102}
\and Large Synoptic Survey Telescope, 950 N. Cherry Avenue, Tucson, AZ 85719, USA\relax                                                                                                                      \label{inst:0107}
\and Observatoire Astronomique de Strasbourg, Universit\'{e} de Strasbourg, CNRS, UMR 7550, 11 rue de l'Universit\'{e}, 67000 Strasbourg, France\relax                                                       \label{inst:0108}
\and Kavli Institute for Cosmology, University of Cambridge, Madingley Road, Cambride CB3 0HA, United Kingdom\relax                                                                                          \label{inst:0111}
\and Aurora Technology for ESA/ESAC, Camino bajo del Castillo, s/n, Urbanizacion Villafranca del Castillo, Villanueva de la Ca\~{n}ada, 28692 Madrid, Spain\relax                                            \label{inst:0113}
\and Laboratoire Univers et Particules de Montpellier, Universit\'{e} Montpellier, Place Eug\`{e}ne Bataillon, CC72, 34095 Montpellier Cedex 05, France\relax                                                \label{inst:0114}
\and Department of Physics and Astronomy, Division of Astronomy and Space Physics, Uppsala University, Box 516, 75120 Uppsala, Sweden\relax                                                                  \label{inst:0117}
\and CENTRA, Universidade de Lisboa, FCUL, Campo Grande, Edif. C8, 1749-016 Lisboa, Portugal\relax                                                                                                           \label{inst:0118}
\and Universit\`{a} di Catania, Dipartimento di Fisica e Astronomia, Sezione Astrofisica, Via S. Sofia 78, 95123 Catania, Italy\relax                                                                        \label{inst:0119}
\and INAF - Osservatorio Astrofisico di Catania, via S. Sofia 78, 95123 Catania, Italy\relax                                                                                                                 \label{inst:0120}
\and University of Vienna, Department of Astrophysics, T\"{ u}rkenschanzstra{\ss}e 17, A1180 Vienna, Austria\relax                                                                                           \label{inst:0121}
\and CITIC – Department of Computer Science, University of A Coru\~{n}a, Campus de Elvi\~{n}a S/N, 15071, A Coru\~{n}a, Spain\relax                                                                        \label{inst:0123}
\and CITIC – Astronomy and Astrophysics, University of A Coru\~{n}a, Campus de Elvi\~{n}a S/N, 15071, A Coru\~{n}a, Spain\relax                                                                            \label{inst:0124}
\and INAF - Osservatorio Astronomico di Roma, Via di Frascati 33, 00078 Monte Porzio Catone (Roma), Italy\relax                                                                                              \label{inst:0125}
\and Space Science Data Center - ASI, Via del Politecnico SNC, 00133 Roma, Italy\relax                                                                                                                       \label{inst:0126}
\and Isdefe for ESA/ESAC, Camino bajo del Castillo, s/n, Urbanizacion Villafranca del Castillo, Villanueva de la Ca\~{n}ada, 28692 Madrid, Spain\relax                                                       \label{inst:0130}
\and STFC, Rutherford Appleton Laboratory, Harwell, Didcot, OX11 0QX, United Kingdom\relax                                                                                                                   \label{inst:0134}
\and Dpto. de Inteligencia Artificial, UNED, c/ Juan del Rosal 16, 28040 Madrid, Spain\relax                                                                                                                 \label{inst:0137}
\and Elecnor Deimos Space for ESA/ESAC, Camino bajo del Castillo, s/n, Urbanizacion Villafranca del Castillo, Villanueva de la Ca\~{n}ada, 28692 Madrid, Spain\relax                                         \label{inst:0145}
\and Thales Services for CNES Centre Spatial de Toulouse, 18 avenue Edouard Belin, 31401 Toulouse Cedex 9, France\relax                                                                                      \label{inst:0146}
\and Department of Astrophysics/IMAPP, Radboud University, P.O.Box 9010, 6500 GL Nijmegen, The Netherlands\relax                                                                                             \label{inst:0148}
\and European Southern Observatory, Karl-Schwarzschild-Str. 2, 85748 Garching, Germany\relax                                                                                                                 \label{inst:0155}
\and ON/MCTI-BR, Rua Gal. Jos\'{e} Cristino 77, Rio de Janeiro, CEP 20921-400, RJ,  Brazil\relax                                                                                                             \label{inst:0157}
\and OV/UFRJ-BR, Ladeira Pedro Ant\^{o}nio 43, Rio de Janeiro, CEP 20080-090, RJ, Brazil\relax                                                                                                               \label{inst:0158}
\and Department of Terrestrial Magnetism, Carnegie Institution for Science, 5241 Broad Branch Road, NW, Washington, DC 20015-1305, USA\relax                                                                 \label{inst:0166}
\and Universit\`{a} di Torino, Dipartimento di Fisica, via Pietro Giuria 1, 10125 Torino, Italy\relax                                                                                                        \label{inst:0173}
\and Departamento de Astrof\'{i}sica, Centro de Astrobiolog\'{i}a (CSIC-INTA), ESA-ESAC. Camino Bajo del Castillo s/n. 28692 Villanueva de la Ca\~{n}ada, Madrid, Spain\relax                                \label{inst:0177}
\and Leicester Institute of Space and Earth Observation and Department of Physics and Astronomy, University of Leicester, University Road, Leicester LE1 7RH, United Kingdom\relax                           \label{inst:0179}
\and Departamento de Estad\'{i}stica, Universidad de C\'{a}diz, Calle Rep\'{u}blica \'{A}rabe Saharawi s/n. 11510, Puerto Real, C\'{a}diz, Spain\relax                                                       \label{inst:0184}
\and Astronomical Institute Bern University, Sidlerstrasse 5, 3012 Bern, Switzerland (present address)\relax                                                                                                 \label{inst:0187}
\and EURIX S.r.l., Corso Vittorio Emanuele II 61, 10128, Torino, Italy\relax                                                                                                                                 \label{inst:0188}
\and Harvard-Smithsonian Center for Astrophysics, 60 Garden Street, Cambridge MA 02138, USA\relax                                                                                                            \label{inst:0192}
\and HE Space Operations BV for ESA/ESAC, Camino bajo del Castillo, s/n, Urbanizacion Villafranca del Castillo, Villanueva de la Ca\~{n}ada, 28692 Madrid, Spain\relax                                       \label{inst:0195}
\and Kapteyn Astronomical Institute, University of Groningen, Landleven 12, 9747 AD Groningen, The Netherlands\relax                                                                                         \label{inst:0202}
\and SISSA - Scuola Internazionale Superiore di Studi Avanzati, via Bonomea 265, 34136 Trieste, Italy\relax                                                                                                  \label{inst:0203}
\and University of Turin, Department of Computer Sciences, Corso Svizzera 185, 10149 Torino, Italy\relax                                                                                                     \label{inst:0213}
\and SRON, Netherlands Institute for Space Research, Sorbonnelaan 2, 3584CA, Utrecht, The Netherlands\relax                                                                                                  \label{inst:0214}
\and Dpto. de Matem\'{a}tica Aplicada y Ciencias de la Computaci\'{o}n, Univ. de Cantabria, ETS Ingenieros de Caminos, Canales y Puertos, Avda. de los Castros s/n, 39005 Santander, Spain\relax             \label{inst:0218}
\and Unidad de Astronom\'ia, Universidad de Antofagasta, Avenida Angamos 601, Antofagasta 1270300, Chile\relax                                                                                               \label{inst:0225}
\and CRAAG - Centre de Recherche en Astronomie, Astrophysique et G\'{e}ophysique, Route de l'Observatoire Bp 63 Bouzareah 16340 Algiers, Algeria\relax                                                       \label{inst:0236}
\and University of Antwerp, Onderzoeksgroep Toegepaste Wiskunde, Middelheimlaan 1, 2020 Antwerp, Belgium\relax                                                                                               \label{inst:0238}
\and INAF - Osservatorio Astronomico d'Abruzzo, Via Mentore Maggini, 64100 Teramo, Italy\relax                                                                                                               \label{inst:0240}
\and Instituto de Astronomia, Geof\`{i}sica e Ci\^{e}ncias Atmosf\'{e}ricas, Universidade de S\~{a}o Paulo, Rua do Mat\~{a}o, 1226, Cidade Universitaria, 05508-900 S\~{a}o Paulo, SP, Brazil\relax          \label{inst:0242}
\and Department of Astrophysics, Astronomy and Mechanics, National and Kapodistrian University of Athens, Panepistimiopolis, Zografos, 15783 Athens, Greece\relax                                            \label{inst:0252}
\and Leibniz Institute for Astrophysics Potsdam (AIP), An der Sternwarte 16, 14482 Potsdam, Germany\relax                                                                                                    \label{inst:0255}
\and RHEA for ESA/ESAC, Camino bajo del Castillo, s/n, Urbanizacion Villafranca del Castillo, Villanueva de la Ca\~{n}ada, 28692 Madrid, Spain\relax                                                         \label{inst:0257}
\and ATOS for CNES Centre Spatial de Toulouse, 18 avenue Edouard Belin, 31401 Toulouse Cedex 9, France\relax                                                                                                 \label{inst:0259}
\and School of Physics and Astronomy, Tel Aviv University, Tel Aviv 6997801, Israel\relax                                                                                                                    \label{inst:0262}
\and UNINOVA - CTS, Campus FCT-UNL, Monte da Caparica, 2829-516 Caparica, Portugal\relax                                                                                                                     \label{inst:0263}
\and School of Physics, O'Brien Centre for Science North, University College Dublin, Belfield, Dublin 4, Ireland\relax                                                                                       \label{inst:0274}
\and Dipartimento di Fisica e Astronomia, Universit\`{a} di Bologna, Via Piero Gobetti 93/2, 40129 Bologna, Italy\relax                                                                                      \label{inst:0279}
\and Barcelona Supercomputing Center - Centro Nacional de Supercomputaci\'{o}n, c/ Jordi Girona 29, Ed. Nexus II, 08034 Barcelona, Spain\relax                                                               \label{inst:0290}
\and Max Planck Institute for Extraterrestrial Physics, High Energy Group, Gie{\ss}enbachstra{\ss}e, 85741 Garching, Germany\relax                                                                           \label{inst:0296}
\and Astronomical Observatory Institute, Faculty of Physics, Adam Mickiewicz University, S{\l}oneczna 36, 60-286 Pozna{\'n}, Poland\relax                                                                    \label{inst:0314}
\and Konkoly Observatory, Research Centre for Astronomy and Earth Sciences, Hungarian Academy of Sciences, Konkoly Thege Mikl\'{o}s \'{u}t 15-17, 1121 Budapest, Hungary\relax                               \label{inst:0320}
\and E\"{ o}tv\"{ o}s Lor\'and University, Egyetem t\'{e}r 1-3, H-1053 Budapest, Hungary\relax                                                                                                               \label{inst:0321}
\and American Community Schools of Athens, 129 Aghias Paraskevis Ave. \& Kazantzaki Street, Halandri, 15234 Athens, Greece\relax                                                                             \label{inst:0324}
\and Faculty of Mathematics and Physics, University of Ljubljana, Jadranska ulica 19, 1000 Ljubljana, Slovenia\relax                                                                                         \label{inst:0327}
\and Villanova University, Department of Astrophysics and Planetary Science, 800 E Lancaster Avenue, Villanova PA 19085, USA\relax                                                                           \label{inst:0328}
\and Physics Department, University of Antwerp, Groenenborgerlaan 171, 2020 Antwerp, Belgium\relax                                                                                                           \label{inst:0330}
\and McWilliams Center for Cosmology, Department of Physics, Carnegie Mellon University, 5000 Forbes Avenue, Pittsburgh, PA 15213, USA\relax                                                                 \label{inst:0336}
\and Astronomical Institute, Academy of Sciences of the Czech Republic, Fri\v{c}ova 298, 25165 Ond\v{r}ejov, Czech Republic\relax                                                                            \label{inst:0340}
\and Telespazio for CNES Centre Spatial de Toulouse, 18 avenue Edouard Belin, 31401 Toulouse Cedex 9, France\relax                                                                                           \label{inst:0345}
\and Institut de Physique de Rennes, Universit{\'e} de Rennes 1, 35042 Rennes, France\relax                                                                                                                  \label{inst:0348}
\and INAF - Osservatorio Astronomico di Capodimonte, Via Moiariello 16, 80131, Napoli, Italy\relax                                                                                                           \label{inst:0349}
\and Shanghai Astronomical Observatory, Chinese Academy of Sciences, 80 Nandan Rd, 200030 Shanghai, China\relax                                                                                              \label{inst:0355}
\and School of Astronomy and Space Science, University of Chinese Academy of Sciences, Beijing 100049, China\relax                                                                                           \label{inst:0356}
\and Niels Bohr Institute, University of Copenhagen, Juliane Maries Vej 30, 2100 Copenhagen {\O}, Denmark\relax                                                                                              \label{inst:0358}
\and DXC Technology, Retortvej 8, 2500 Valby, Denmark\relax                                                                                                                                                  \label{inst:0359}
\and Las Cumbres Observatory, 6740 Cortona Drive Suite 102, Goleta, CA 93117, USA\relax                                                                                                                      \label{inst:0360}
\and Astrophysics Research Institute, Liverpool John Moores University, 146 Brownlow Hill, Liverpool L3 5RF, United Kingdom\relax                                                                            \label{inst:0368}
\and Baja Observatory of University of Szeged, Szegedi \'{u}t III/70, 6500 Baja, Hungary\relax                                                                                                               \label{inst:0373}
\and Laboratoire AIM, IRFU/Service d'Astrophysique - CEA/DSM - CNRS - Universit\'{e} Paris Diderot, B\^{a}t 709, CEA-Saclay, 91191 Gif-sur-Yvette Cedex, France\relax                                        \label{inst:0374}
\and Warsaw University Observatory, Al. Ujazdowskie 4, 00-478 Warszawa, Poland\relax                                                                                                                         \label{inst:0411}
\and Institute of Theoretical Physics, Faculty of Mathematics and Physics, Charles University in Prague, Czech Republic\relax                                                                                \label{inst:0412}
\and Vitrociset Belgium for ESA/ESAC, Camino bajo del Castillo, s/n, Urbanizacion Villafranca del Castillo, Villanueva de la Ca\~{n}ada, 28692 Madrid, Spain\relax                                           \label{inst:0424}
\and HE Space Operations BV for ESA/ESTEC, Keplerlaan 1, 2201AZ, Noordwijk, The Netherlands\relax                                                                                                            \label{inst:0432}
\and Space Telescope Science Institute, 3700 San Martin Drive, Baltimore, MD 21218, USA\relax                                                                                                                \label{inst:0441}
\and QUASAR Science Resources for ESA/ESAC, Camino bajo del Castillo, s/n, Urbanizacion Villafranca del Castillo, Villanueva de la Ca\~{n}ada, 28692 Madrid, Spain\relax                                     \label{inst:0442}
\and Fork Research, Rua do Cruzado Osberno, Lt. 1, 9 esq., Lisboa, Portugal\relax                                                                                                                            \label{inst:0447}
\and APAVE SUDEUROPE SAS for CNES Centre Spatial de Toulouse, 18 avenue Edouard Belin, 31401 Toulouse Cedex 9, France\relax                                                                                  \label{inst:0450}
\and Nordic Optical Telescope, Rambla Jos\'{e} Ana Fern\'{a}ndez P\'{e}rez 7, 38711 Bre\~{n}a Baja, Spain\relax                                                                                              \label{inst:0455}
\and Spanish Virtual Observatory\relax                                                                                                                                                                       \label{inst:0460}
\and Fundaci\'{o}n Galileo Galilei - INAF, Rambla Jos\'{e} Ana Fern\'{a}ndez P\'{e}rez 7, 38712 Bre\~{n}a Baja, Santa Cruz de Tenerife, Spain\relax                                                          \label{inst:0470}
\and INSA for ESA/ESAC, Camino bajo del Castillo, s/n, Urbanizacion Villafranca del Castillo, Villanueva de la Ca\~{n}ada, 28692 Madrid, Spain\relax                                                         \label{inst:0475}
\and Dpto. Arquitectura de Computadores y Autom\'{a}tica, Facultad de Inform\'{a}tica, Universidad Complutense de Madrid, C/ Prof. Jos\'{e} Garc\'{i}a Santesmases s/n, 28040 Madrid, Spain\relax            \label{inst:0476}
\and H H Wills Physics Laboratory, University of Bristol, Tyndall Avenue, Bristol BS8 1TL, United Kingdom\relax                                                                                              \label{inst:0479}
\and Institut d'Estudis Espacials de Catalunya (IEEC), Gran Capita 2-4, 08034 Barcelona, Spain\relax                                                                                                         \label{inst:0484}
\and Applied Physics Department, Universidade de Vigo, 36310 Vigo, Spain\relax                                                                                                                               \label{inst:0486}
\and Stellar Astrophysics Centre, Aarhus University, Department of Physics and Astronomy, 120 Ny Munkegade, Building 1520, DK-8000 Aarhus C, Denmark\relax                                                   \label{inst:0501}
\and Argelander-Institut f\"{ ur} Astronomie, Universit\"{ a}t Bonn,  Auf dem H\"{ u}gel 71, 53121 Bonn, Germany\relax                                                                                       \label{inst:0506}
\and Research School of Astronomy and Astrophysics, Australian National University, Canberra, ACT 2611 Australia\relax                                                                                       \label{inst:0513}
\and Sorbonne Universit\'{e}s, UPMC Univ. Paris 6 et CNRS, UMR 7095, Institut d'Astrophysique de Paris, 98 bis bd. Arago, 75014 Paris, France\relax                                                          \label{inst:0515}
\and Department of Geosciences, Tel Aviv University, Tel Aviv 6997801, Israel\relax                                                                                                                          \label{inst:0517}
}
   \date{ }

 
\abstract
{The \textit{Gaia} spacecraft of the European Space Agency (ESA) has
  been securing observations of solar system objects (SSOs) since the
  beginning of its operations. Data Release 2 (DR2) contains the
  observations of a selected sample of 14,099 SSOs. These asteroids
  have been already identified and have been numbered by the Minor
  Planet Center repository. Positions are provided for each \textit{Gaia}
  observation at CCD level. As additional information, complementary
  to astrometry, the apparent brightness of SSOs in the unfiltered $G$
  band is also provided for selected observations.}
{We explain the processing of SSO data, and describe the criteria we used to
  select the sample published in \gdrtwo. We then explore the data
  set to assess its quality.}
{To exploit the main data product for the solar system in \gdrtwo,
  which is the epoch astrometry of asteroids, it is necessary to take into
  account the unusual properties of the uncertainty, as the position
  information is nearly one-dimensional. When this aspect is
handled appropriately, an orbit fit can be obtained with post-fit
  residuals that are overall consistent with the a-priori error model
  that was used to define individual values of the astrometric uncertainty. The
  role of both random and systematic errors is described. The
  distribution of residuals allowed us to identify possible
  contaminants in the data set (such as stars). Photometry in the $G$
  band was compared to computed values from reference asteroid shapes
  and to the flux registered at the corresponding epochs by the red
  and blue photometers (RP and BP).}
{The overall astrometric performance is close to the expectations,
  with an optimal range of brightness G$\sim$12-17. In this range, the
  typical transit-level accuracy is well below 1~
  mas. For fainter asteroids, the growing photon noise deteriorates
  the performance. Asteroids brighter than G$\sim$12 are affected by a
  lower performance of the processing of their signals. The dramatic
  improvement brought by \gdrtwo\ astrometry of SSOs is demonstrated by
  comparisons to the archive data and by preliminary tests on the
  detection of subtle non-gravitational effects.}
{}

   \keywords{astrometry --
                Solar System: asteroids --
                methods: data analysis --
                space vehicles: instruments
               }
   
   \titlerunning{\textit{Gaia} Data Release 2. Solar system} 
   \authorrunning{F. Spoto et al.}

   \maketitle

%

\section{Introduction}  
\label{S:intro}

The ESA \gaia mission~\citep{Prusti2016} is observing the sky since
December 2013 with a continuous and pre-determined scanning law. While
the large majority of the observations concern the stellar population
of the Milky Way, \gaia also collects data of extragalactic sources and
solar system objects (SSOs). A subset of the latter population of celestial
bodies is the topic of this work.

\gaia has been designed to map astrophysical sources of very small or
negliglible angular extension. Extended sources, like the major
planets, that do not present a narrow brightness peak are indeed
discarded by the onboard detection algorithm. This mission is
therefore a wonderful facility for the study of the population of
SSOs, including small bodies, such as asteroids, Jupiter trojans,
Centaurs, and some Transneptunian Objects (TNO) and planetary
satellites, but not the major planets.

The SSO population is currently poorly characterised, because basic
physical properties including mass, bulk density, spin
properties. shape, and albedo are not known for the vast majority of
them.

The astrometric data are continuously
updated by ground-based surveys, and they are sufficient for a general
dynamical classification. Only in rare specific situations, however,
their accuracy allows us to measure subtle effects such as
non-gravitational perturbations and/or to estimate the masses. In this
respect, \gaia represents a major step forward.

\gaia is the first global survey to provide a homogeneous data set of
positions, magnitudes, and visible spectra of SSOs,with extreme
performances on the astrometric accuracy \citep{mignard2007,
  cellino2007, tanga2008, hestroffer2010, delbo2012, tanga2012a,
  tanga2012b, spoto2017}. \gaia astrometry, for $\sim 350\,000$ SSOs by
the end of the mission, is expected to produce a real revolution. The
additional physical data (low-resolution reflection spectra, accurate
photometry) will at the same time provide a much needed physical
characterisation of SSOs.

Within this population, the \gdrtwo\ contains a sample of 14\,099 SSOs
(asteroids, Jupiter trojans, and a few TNOs) for a total of 1\,977\,702
different observations, collected during 22 months since the start of
the nominal operations in July 2014. A general description of
\gdrtwo\ is provided in \citet{brown2018}.

The main goal of releasing SSO observations in \gdrtwo\ is to
demonstrate the capabilities of \gaia in the domain of SSO astrometry
and to also allow the community to familiarise itself with \gaia SSO data and
perform initial scientific studies. For this reason, the following
fundamental properties of the release are recalled first.

\begin{itemize}
\item Only a sub-sample of well-known SSOs was selected among
  those observed by \textit{Gaia}. Moreover, this choice is not
  intended to be complete with respect to any criterion based on
  dynamics of physical categories.
\item The most relevant dynamical classes are represented, including
  near-Earth and main-belt objects, Jupiter trojans, and a few
  TNOs.
\item For each of the selected objects, all the observations obtained
  over the time frame covered by the \gdrtwo\ are included, with the
  exception of those that did not pass the quality tests described
  later in this article.
\item Photometric data are provided for only a fraction of the
  observations as a reference, but they should be considered as
  preliminary values that will be refined in future data releases.
\end{itemize}

The goals of this paper are to illustrate the main steps of the data
processing that allowed us to obtain the SSO positions from \gaia
observations and to analyse the results in order to derive the overall
accuracy of the sample, as well as to illustrate the selection criteria that
were applied to identify and eliminate the outliers.

The core of our approach is based on an accurate orbital fitting
procedure, which was applied on the \gaia data alone, for each SSO. The
data published in the DR2 contain all the quantities needed to
reproduce the same computations. The post-fit orbit residuals
generated during the preparation of this study are made available as
an auxiliary data set on the ESA
Archive\footnote{\url{https://gea.esac.esa.int/archive/}}. Its object
is to serve as a reference to evaluate the performance of independent
orbital fitting procedures that could be performed by the archive
users.
 
More technical details on the data properties and their organisation,
which are beyond the scope of this article, are illustrated in the \gdrtwo\
documentation accessible through the ESA archive.

This article is organised as follows. Section~\ref{S:data} illustrates
the main properties of the sample selected for DR2 and recalls the
features of \gaia that affect SSO observations. For a more
comprehensive description of \gaia operations, we refer
to \citet{Prusti2016}. The data reduction procedure is outlined in
Section~\ref{S:astrometric_processing}, while
Section~\ref{S:photometry} illustrates the properties of the
photometric data that complement the
astrometry. Section~\ref{S:astro_validation} is devoted to the orbital
fitting procedure, whose residuals are then used to explore the data
quality. This is described in Sections~\ref{S:astro_interpretation} and
\ref{S:orbits}.

\section{Data used} 
\label{S:data}

We recall here some basic properties of the \gaia focal plane
  that directly affect the observations. As the \gaia satellite
rotates at a constant rate, the images of the sources on the focal
plane drift continuously (in the along-scan direction, AL) across the
different CCD strips. A total of nine CCD strips exists, and the source
in the astrometric field (AF, numbered from one to nine, AF1,
  AF2... AF9) can cross up to these nine strips.

Thus each transit published in the \gdrtwo\ consists at most of nine
observations (AF instrument). Each CCD operates in time-delay
integration (TDI) mode, at a rate corresponding to the drift induced
by the satellite rotation. All observations of SSOs
published in the \gdrtwo, both for astrometry and photometry, are
based on measurements obtained by single CCDs.

The TDI rate is an instrumental constant, and the spacecraft spin rate
is calibrated on the stars. The exposure time is determined by the
crossing time of a single CCD, that is, 4.4~s. Shorter exposure times
are obtained when needed to avoid saturation, by intermediate electric
barriers (the so-called gates) that swallow all collected electrons. Their positioning on the CCD in the AL direction is chosen
in such a way that the distance travelled by the source on the CCD
itself is reduced, thus reducing the exposure time.

To drastically reduce the data volume processed on board and
transmitted to the ground, only small patches around each source
(windows) are read out from each CCDs. The window is
  assigned after the source has been detected in a first strip of CCD,
  the sky mapper (SM), and confirmed in AF1. For the vast
majority of the detected sources (G$>$16), the window has a size of
12$\times$12 pixels, but the pixels are binned in the direction
perpendicular to the scanning direction, called across-scan (AC). Only
1D information in the AL direction is thus available,
with the exception of the brightest sources (G$<$13), for which a full
2D window is transmitted. Sources of intermediate brightness are
given a slightly larger window (18$\times$12 pixels), but AC binning is
always present.

As the TDI rate corresponds to the nominal drift velocity of stars on
the focal plane, the image of an SSO that has an apparent sky motion
is slightly spread in the direction of motion. Its AL
position also moves with respect to the window centre during the
transit. The signal is thus increasingly truncated by
  the window edge. For instance, the signal of an SSO with an
apparent motion (in the AL direction) of 13.6~mas/s moves by one pixel during a single CCD crossing, with
corresponding image smearing. 

We can assume that the image is centred in the window at the
beginning of the transit, when it is detected first by the SM, and its
position is used to define the window coordinates. Then, while
drifting on the focal plane and crossing the AF CCDs, due to its
motion relative to the stars, the SSO will leave the window
center. When the AF5 strip is reached, about half of the flux will be
lost.

In practice, the uncertainty in determining the position of
the source within the window is a function of its centring and can vary
over the transit due to the image drift described above. This
contribution to the error budget is computed for each position
and published in \gdrtwo.

\subsection{Selection of the sample}
\label{S:selection}

For \gdrtwo,\ the solar system pipeline worked on a pre-determined
list of transits in the field of view (FOV) of \textit{Gaia}. To build
it, a list of accurate predictions was first created by cross-matching
the evolving position of each asteroid to the sky path of the
\textit{Gaia }FOVs. This provides a set of predictions of SSO transits that
were then matched to the observed transits. At this level, the
information on the SSO transits comes from the output of the daily
processing \citep{fabricius2016} and in particular  from the initial data
treatment (IDT).  IDT proceeds by an approximate, daily solution of
the astrometry to derive source positions with a typical uncertainty
of the order of $\sim$70-100 mas.  There was typically one SSO transit
in this list for every $100,000$ stellar transits.

\begin{figure}[h!]
\centering
\includegraphics[clip=true, trim = 0mm 0mm 0mm 0mm, width=1.0 \hsize]{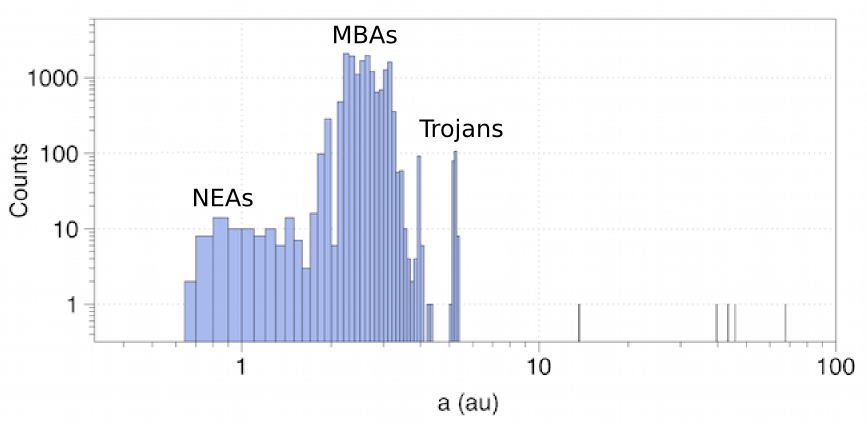}
\caption{Distribution of the semi-major axes of the 14\,125 SSOs
  contained in the final input selection. Not all the bodies shown in
  this figure are included in \gdrtwo: 26 objects were filtered
  out for different reasons (see Sect.~\ref{S:filtering}
  and~\ref{S:astro_validation}).}
\label{F:semiaxe_selection_FM}
\end{figure}

SSO targets for the \gdrtwo\ were selected following the basic
idea of assembling a satisfactory sample for the first mass processing of
sources, despite the relatively short time span covered by the
observations (22 months). The selection of the sample was based on
some simple rules:

\begin{itemize}
\item The goal was to include a significant number of SSOs,
  between 10\,000 and 15\,000.
\item The sample had to cover all classes of SSOs: near-Earth
  asteroids (NEAs), main-belt asteroids (MBAs), Jupiter trojans, and
  TNOs.
\item Each selected object was requested to have at least 12 transits
  in the 22 months covered by the \gdrtwo\ data.
\end{itemize}
The final input selection contains $14\,125$ SSOs, with a total of
$318\,290$ transits. Not all these bodies are included in \gdrtwo: 26
objects were filtered out for different reasons (see
Sect.~\ref{S:filtering} and~\ref{S:astro_validation}). The coverage in
orbital semi-major axes is represented in
Fig.~\ref{F:semiaxe_selection_FM}.

\begin{figure*}[h!]
\centering
\includegraphics[clip=true, trim = 0mm 0mm 0mm 0mm, width=1.0 \hsize]{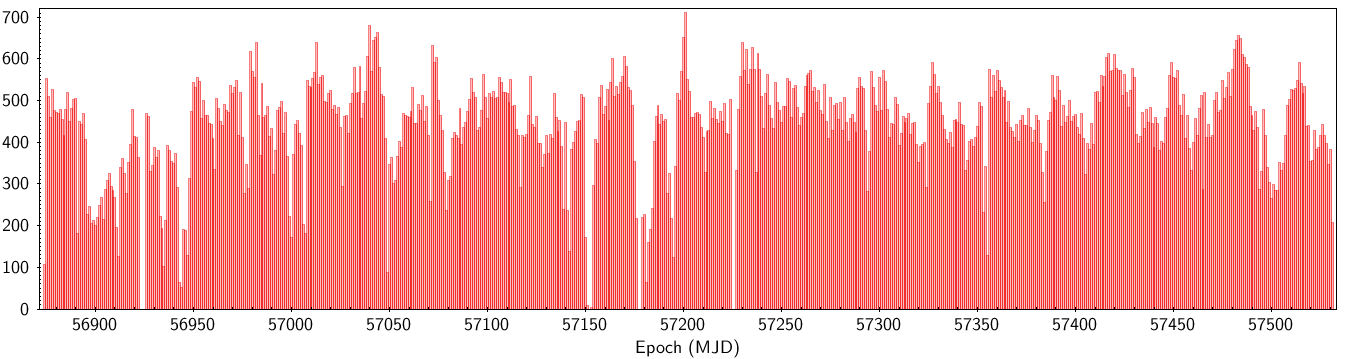}
\caption{Distribution in time of the SSO observations
  published in DR2. The bin size is one day.}
\label{F:timedistrib}
\end{figure*}
\begin{figure}[ht]
\centering
\includegraphics[clip=true, trim = 0mm 0mm 0mm 0mm, width=1.0 \hsize]{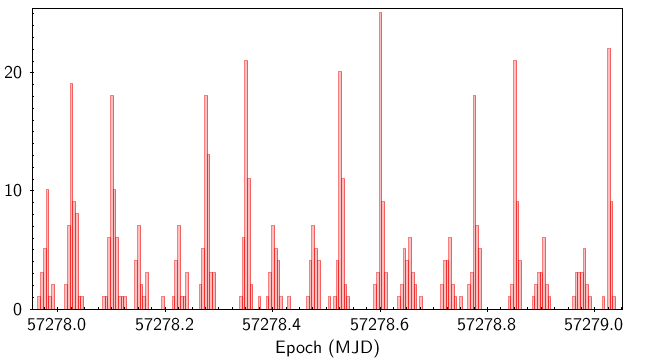}
\caption{Detail over a short time interval of the distribution shown in Fig.~\ref{F:timedistrib}}
\label{F:timedistrib_zoom}
\end{figure}

\subsection{Time coverage}
\label{S:coverage}

The \gdrtwo\ contains observations of SSOs from 5 August 2014, to 23 May, 2016\footnote{As a rule, \gdrtwo\ data start on 25 July 2014,
  but for SSOs and for technical reasons, no transits have been
  retained before August, 5.}. During the first two weeks of the
period covered by the observations, a special scanning mode was adopted
to obtain a dense coverage of the ecliptic poles \citep[the ecliptic
  pole scanning law, EPSL]{Prusti2016}. Due to the peculiar geometry
of the EPSL, the scan plane crosses the ecliptic in the perpendicular
direction with a gradual drift of the node longitude at the speed of
the Earth orbiting the Sun.

\begin{figure}[hb]
\centering
\includegraphics[clip=true, trim = 35mm 0mm 35mm 0mm, width=0.9 \hsize]{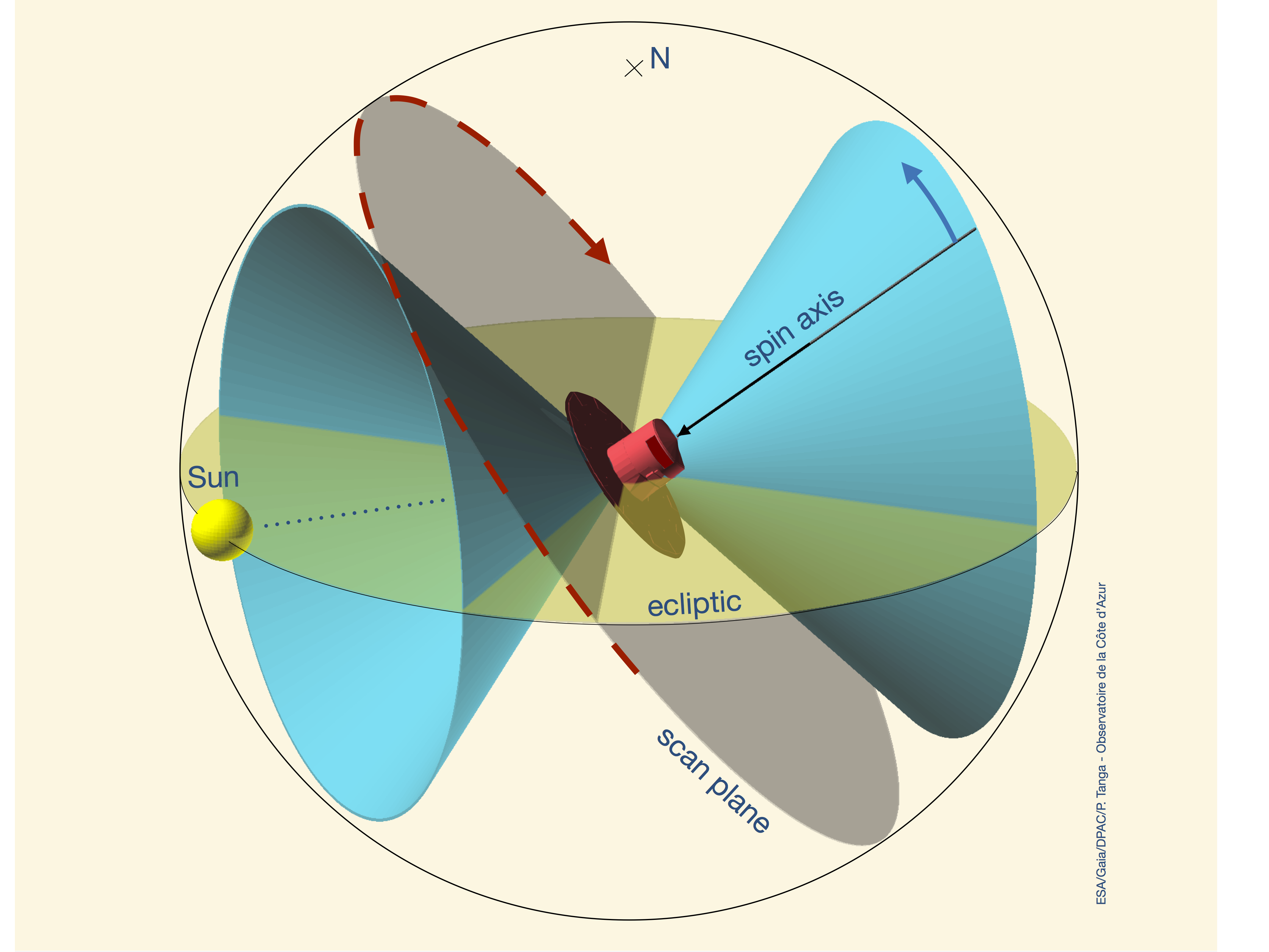}
\caption{Geometry of the \gaia NSL on the celestial sphere, with ecliptic north 
at the top. The scanning motion of \gaia is represented by the red dashed line. The precession of the 
spin axis describes the two cones, aligned on the solar--anti-solar direction, with an aperture of 90$^\circ$. 
As a consequence, the scan plane, here represented at a generic epoch, is at any time tangent to the cones. When the spin axis is
on the ecliptic plane, \gaia scans the ecliptic perpendicularly, at extreme solar elongations. The volume inside the cones is 
never explored by the scan motion.}
\label{F:observable3D}
\end{figure}

A smooth transition then occurred towards the nominal scanning
law (NSL) between 22 August and 25 September 2014 that was
maintained constant afterwards. In this configuration, the spin axis
of \gaia precesses on a cone centred in the direction of the Sun, with
a semi--aperture of 45$^\circ$ and period of 62.97 days
(Fig.~\ref{F:observable3D}). As a result, the scan plane orientation
changes continuously with respect to the ecliptic with inclinations
between 90$^\circ$ and 45$^\circ$. The nodal direction has a solar
elongation between 45$^\circ$ and 135$^\circ$.

The general distribution of the observations is rather homogeneous in
time, with very rare gaps, in general shorter than a few hours;
these are due
to maintenance operations (orbital maneuvers, telescope refocusing,
micrometeoroid hits, and other events; Fig.~\ref{F:timedistrib}).  

A more detailed view of the distribution with a resolution of several
minutes (Fig.~\ref{F:timedistrib_zoom}) reveals a general pattern
that repeats at each rotation of the satellite (6 hours) and is dominated by a
sequence of peaks that correspond to the crossing of the ecliptic
region by the two FOVs, at intervals of $\sim$106 minutes
(FOV 1 to FOV 2) and $\sim$254 minutes (FOV 2 to FOV 1). The peaks are
strongly modulated in amplitude by the evolving geometry of the scan
plane with respect to the ecliptic.

The observation dates are given in barycentric coordinate time (TCB)
\textit{Gaia}-centric\footnote{Difference between the barycentric JD time in
  TCB and 2455197.5.}, which is the primary timescale for \textit{Gaia}, and
also in coordinated universal time (UTC) \textit{Gaia}-centric. Timings
correspond to mid exposure, which is the instant of crossing of the
fiducial line on the CCD by the photocentre of the SSO image.

The accuracy of timing is granted by a time-synchronisation procedure
between the atomic master clock onboard \gaia (providing onboard
time, OBT) and OBMT, the onboard mission timeline
\citep{Prusti2016}. OBMT can then be converted into TCB at \gaia. The
absolute timing accuracy requirements for the science of \gaia is 2
$\mu s$. In practice, this requirement is achieved throughout the mission,
with a significant margin.

\subsection{Geometry of detection}
\label{S:geometry_detection}
\begin{figure}[h]
\centering
\includegraphics[clip=true, trim = 0mm 0mm 0mm 0mm, width=0.94 \hsize]{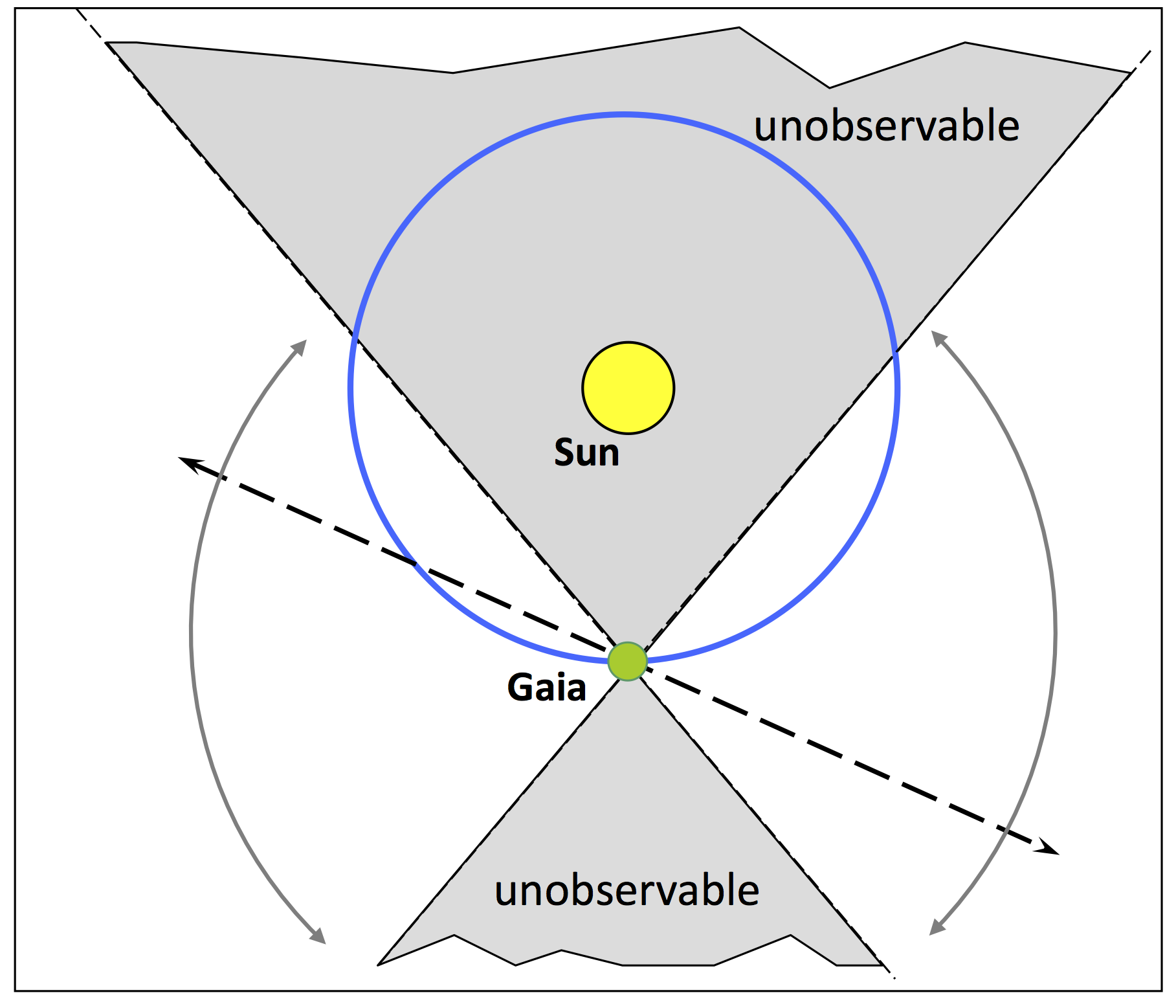}
\caption{By drawing the intersection of the possible scan plane
  orientations with the ecliptic, in the reference rotating around the
  Sun with the \gaia spacecraft, the two avoidance regions
  corresponding the the cones of Fig.~\ref{F:observable3D} emerge in
  the direction of the Sun and around opposition. The dashed line
  represents the intersection of the scanning plane and the ecliptic
  at an arbitrary epoch. During a single rotation of the satellite, the
FOVs of \gaia cross the ecliptic in two opposite
  directions. The intersection continuously scans the allowed sectors,
  as indicated by the curved arrows.}
\label{F:observable}
\end{figure}

The solar elongation is the most important geometric feature in \gaia
observations of SSOs.  By considering the intersection
of the scan plane with the ecliptic, as shown in
Fig.\ref{F:observable}, it is clear that SSOs are always observed at
solar elongations between 45$^\circ$ and 135$^\circ$.


This peculiar geometry has important consequences on solar system
observations. The SSOs are not only observed at
non-negligible phase angles (Fig.~\ref{F:phase_Gred}), in any case
never close to the opposition, but also in a variety of configurations
(high/low proper motion, smaller or larger distance, etc.), which have
some influence on many scientific applications and can affect the
detection capabilities of \gaia and the measurement accuracy.

The mean geometrical solar elongation of the scan plane on the
ecliptic is at quadrature. In this situation, the scan plane is
inclined by 45$^\circ$ with respect to the ecliptic. During the
precession cycle, the scan plane reaches the extreme inclination of
90$^\circ$ on the ecliptic. In this geometry, the SSOs with
low-inclination orbits move mainly in the AC direction when they are
observed by \textit{Gaia}. As the AC pixel size and window are $\text{approximately}$
times larger than AL, the sensitivity to the motion in terms of flux
loss, image shift, and smearing will thus be correspondingly lower.

These variations of the orientation and the distribution of the SSO orbit inclinations translate into a wide range of possible
orientations of the velocity vector on the (AL, AC) plane. Even for a
single object, a large variety of velocities and scan directions is
covered over time.

\subsection{Errors and correlations}
\label{S:error_distribution}
The SSO apparent displacement at the epoch of each observation is
clearly a major factor affecting the performance of \textit{Gaia},
even within a single transit. Other general effects acting on single
CCD observations exist, such as local CCD defects, local point
spread function (PSF)
deviations, cosmic rays, and background sources. For all these
reasons, the exploitation of the single data points must rely on a
careful analysis that takes both the geometric conditions of
the observations and appropriate error models into account.

A direct consequence of the observation strategy employed by \gaia is
the very peculiar error distribution for the single astrometric
observation.

\begin{figure}[h]
\centering
\includegraphics[clip=true, trim = 0mm 0mm 0mm 0mm, width=0.8 \hsize]{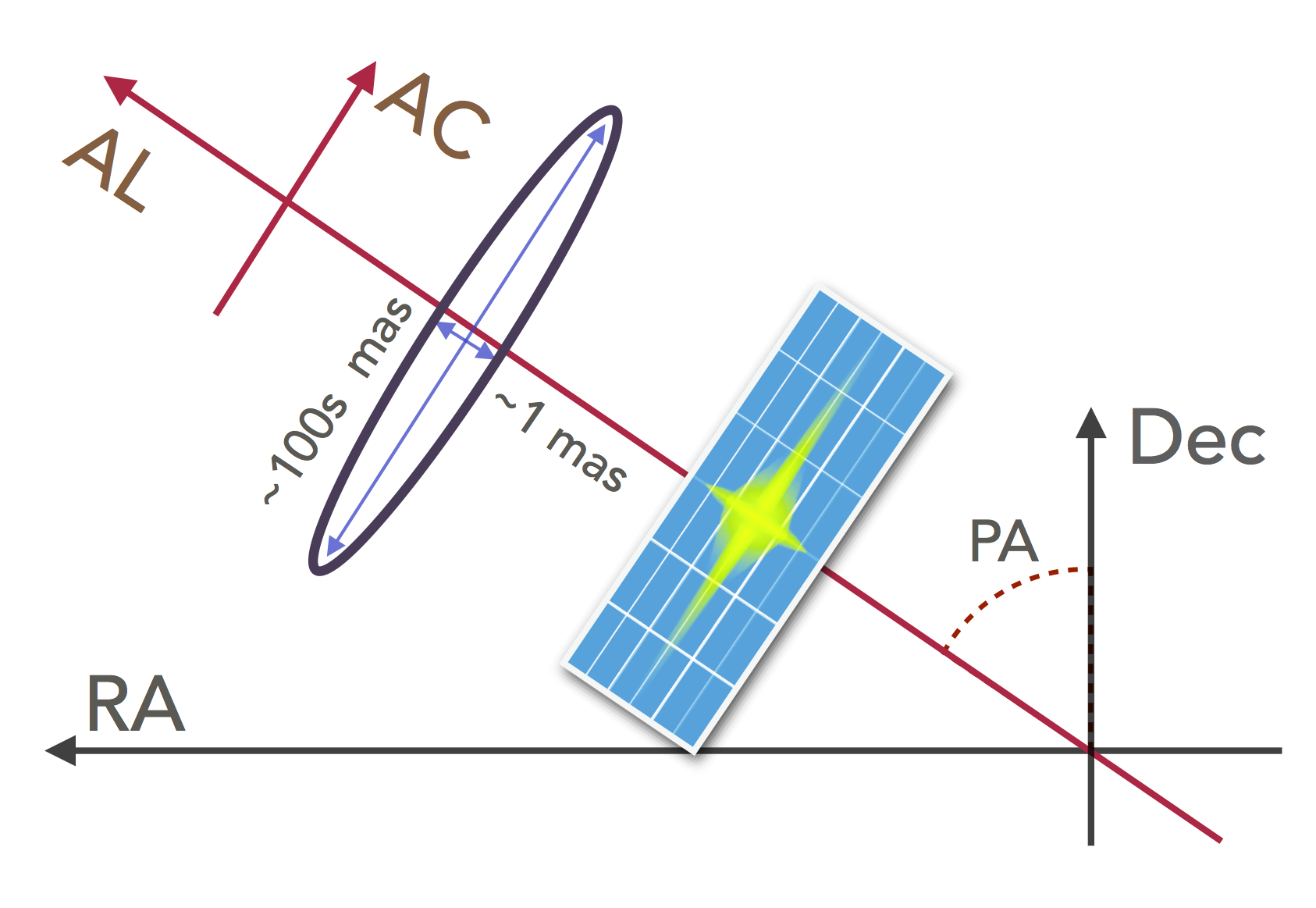}
\caption{Approximate sketch illustrating the effects of the strong
  difference between the astrometry precision in AL (reaching sub-mas
  level) and in AC (several 100s mas). The approximate uncertainty
  ellipse (not to be interpreted as a 2D Gaussian distribution) is
  extremely stretched in the AC direction. The position angle (PA) is
  the angle between the declination and the AC direction.}
\label{F:scheme_AL_AC}
\end{figure}

Because of the AC binning, most accurate astrometry in
the astrometric field for most observations is only available in the AL direction. This is a natural consequence of the design principle of \gaia, which is based
on converting an accurate measurement of time (the epoch when a source
image crosses a reference line on the focal plane) into a position. In
practical terms, the difference between AC and AL accuracy is so large
that we can say that the astrometric information is essentially
one-dimensional.

As illustrated in Fig.~\ref{F:scheme_AL_AC}, the resulting uncertainty
on the position can be represented by an ellipse that is extremely stretched
in the AC direction.  When this position is converted into another
coordinate frame (such as the equatorial reference $\alpha$, $\delta$),
a very strong correlation appears between the related uncertainties
$\sigma_{\alpha}, \sigma_{\delta}$. Therefore it is of the utmost importance
that the users take these correlations into account in their
analysis. The values are provided in the ESA Archive and must be
used to exploit the full accuracy of the \gaia astrometry and to avoid
serious misuse of the \gaia data.

\section{Outline of the data reduction process}
\label{S:astrometric_processing}

\begin{figure}[h]
\centering
\includegraphics[clip=true, trim = 0mm 0mm 0mm 0mm, width=1.0 \hsize]{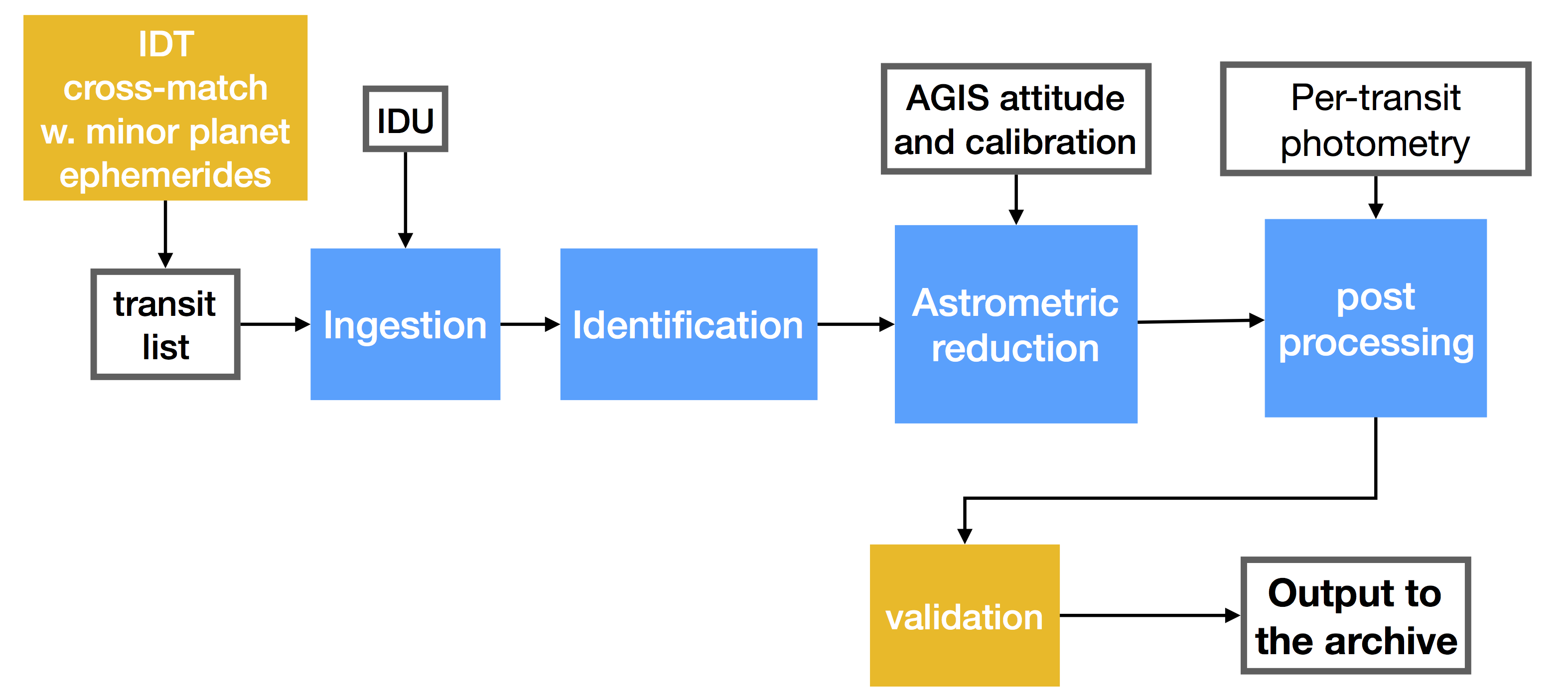}
\caption{Main step of the solar system pipeline that collects all
  the data needed to process identified transits.}
\label{F:scheme_SSO_LT}
\end{figure}

The solar system pipeline (Fig.~\ref{F:scheme_SSO_LT}) collects all
the data needed to process the identified transits (epoch of transit
on each CCD, flux, AC window coordinates, and many auxiliary pieces of
information). 

A first module, labelled "Identification" in the scheme, computes the
auxiliary data for each object, and assigns the
identifying correct identification label to each object. Focal plane coordinates are
then converted into sky coordinates by using the transformations
provided by AGIS, the astrometric global iterative solution, and the
corresponding calibrations (astrometric reduction module). This is
the procedure described below in Sect.~\ref{S:astroproc}. We
note that
this approach adopts the same principle as absolute stellar
astrometry \citep{Lindegren2018} : a local information equivalent
to the usual small--field astrometry (i.e. position relative to nearby
stars) is never used.

Many anomalous data are also rejected by the
same module. The post-processing appends the calibrated photometry
to the data of each
observation (determined by an independent
pipeline, see Sect.~\ref{S:photometry}) and groups all the
observations of a same target. Eventually, a "Validation" task rejects
anomalous data.

The origin of the anomalies are multiple: for instance, data can be
corrupted for technical reasons, or a mismatch with a nearby star on
the sky plane can enter the input list. Identifying truly anomalous
data from peculiarities of potential scientific interest is a delicate
task. Most of this article is devoted to the results obtained on the
general investigation of the overall data properties, and draws
attention to the approaches needed to exploit the accuracy of \gaia
and prepare a detailed scientific exploitation.

\subsection{Astrometric processing}
\label{S:astroproc}
We now describe the main steps of the astrometric processing. A more
comprehensive presentation is available in the \gdrtwo\ documentation
 and \citet{Lindegren2016, Lindegren2018}.  The basic
processing of the astrometric reduction for SSOs
consists of three consecutive coordinate transformations.

The first step in the processing of the astrometry is the computation
of the epoch of observations, which is the reconstructed timing of
crossing of the central line of the exposure on the CCD. The first
coordinate transformation is the conversion from the Window Reference
System (WRS) to the Scanning Reference System (SRS). The former
consist of pixel coordinates of the SSO inside the transmitted window
along with time tagging from the On Board Mission Timeline (OBMT), the
internal time scale of \textit{Gaia} \citep{Lindegren2016}. The origin
of the WRS is the reference pixel of the transmitted window. The SRS
coordinates are expressed as two angles in directions parallel and
perpendicular to the scanning direction of \textit{Gaia}, and the
origin is a conventional and fixed point near the centre of the focal
plane of \textit{Gaia}.

The second conversion is from SRS to the centre-of-mass reference
system (CoMRS), a non-rotating coordinate system with origin in the
centre of mass of \textit{Gaia}.

The CoMRS coordinates are then transformed into the barycentric
reference system (BCRS), with the origin in the barycentre of the solar
system. The latter conversion provides the instantaneous direction of
the unit vector from \textit{Gaia} to the asteroid at the epoch of the
observation after removal of the annual light aberration (i.e., as if
\textit{Gaia} were stationary relatively to the solar system barycenter). 
These positions, expressed in right ascension ($\alpha$)
and declination ($\delta$), are provided in DR2. They are similar to
astrometric positions in classical ground-based astrometry.

A caveat applies to SSO positions concerning the
relativistic bending of the light in the solar system gravity field. In \gdrtwo,\ 
this effect is over--corrected by assuming that the target is at infinite distance (i.e. a star). 
In the case of SSOs at finite distance, this assumption introduces 
a small discrepancy (always $<$2 mas) that must be corrected for to exploit
the ultimate accuracy level. 

\subsection{Filtering and internal validation}
\label{S:filtering}

An SSO transit initially includes at most nine positions, each
corresponding to one AF CCD detection (see Sec.~\ref{S:data}). However,
in many cases, fewer than nine
observations in a transit are available in the end. The actual success of the astrometric
reduction depends on the quality of the recorded data: CCD
observations of too low quality are quickly rejected; the same
holds true if an
observation occurs in the close vicinity of a star or within too
short a time from a cosmic ray event, the software fails to produce a
good position.

These problems represent only a small part of all the possible
instances encountered in the astrometric processing, which has
required an efficient filtering.  Observations have been carefully
analysed inside the pipeline to ensure that positions that probably
do not come from an SSO are rejected, as well as positions that do not
meet high quality standards. We applied the filtering both at the
level of individual positions and at the level of complete transits.
We list the main causes of rejection below.

\begin{itemize}
\item Problematic transit data. The positions were rejected when
  some transit data were too difficult to treat or if they gave rise
  to positions with uncertain precision.
\item Error-magnitude relation. Positions with reported uncertainties
that were   too large or too small for a given magnitude are presumably not real
  SSO detections, and they were discarded.
\item No linear motion. At a solar elongation of more than 45$^\circ$,
  an SSO should show a linear motion in the sky during a single
  transit, where linear means that both space coordinates are linear
  functions of time. We considered all those positions to be false detections
that did not fit
  the regression line to within the estimated uncertainties.
\item Minimum number of positions in a transit. The final check was to
  assess how many positions were left in a transit. For \gdrtwo,\ we set
  the limit to two because we relied on an a priori list of transits to
  be processed (see Sec.~\ref{S:selection}).
\end{itemize}

SSOs have also gone through a further quality check and filtering
according to internal processing requirements established to take into
account some expected peculiarities of SSO signals. Three control
levels were implemented:

\begin{itemize}
\item Standard window checking. Only centroids/fluxes from windows
  with standard characteristics were accepted and transmitted to
  the following step of the processing pipeline.
\item Checking of the quality codes in the input data, resulting from the 
signal centroiding. Only data that
  successfully passed the centroid determination were accepted.
\item A filtering depending on the magnitude and apparent motion of
  the source and the location of its centroid inside the window in
  order to reject observations with centroids close to the window
  limits, where the interplay between the distortion of the PSF due to
  motion and the signal truncation would introduce biases in
  centroid and flux measurements.
\end{itemize}

\subsection{Error model for astrometry}
\label{S:error_model_astrometry}

Between CCD positions within a transit, the errors are not entirely
independent, since in addition to the uncorrelated random noise, there
are some systematics, like the attitude error, that have a coherence
time longer than the few seconds interval between two successive
CCDs. This induces complex correlations between the errors in the
different CCDs from the same transit that are practically impossible
to account for rigorously. Hence, we adopted a simplified
approach separating the error into a systematic and a random
part. Systematic errors are the same for all positions of the same
transit, while random errors are statistically independent from one
CCD to another. One of the main error sources is the error from the
centroiding. It is propagated in the pipeline down from the signal
processing in pixels in the coordinate system (AL, AC), and it is
eventually converted into right ascension and declination. The errors in
AL and AC are usually uncorrelated, but the rotation from the system
(AL,AC) to the system ($\alpha\cos\delta$,$\delta$) makes them highly
correlated.

Along-scan uncertainties are very small (of the order of 1 mas), and
they show the extreme precision of \textit{Gaia}. The error on the centroiding
represents the main contribution to the random errors for SSOs fainter
than magnitude 16. For SSOs fainter than magnitude 13, all pixels are
binned in AC to a single window, and the only information we have is
that the object is inside the window. Therefore the position is given
as the centre of the window, and the uncertainty is given as the
dispersion of a rectangular distribution over the window. The errors
in AC are thus very large (of the order of 600 mas) and highly
non-Gaussian.  For SSOs brighter than magnitude 13, the uncertainty in
AC is smaller. In these cases, a 2D centroid fitting is
possible, but the error in AC is generally still more than three times
larger than in AL direction, essentially because of the shape of the
\textit{Gaia} pixels.

An important consequence is that uncertainties given in the
($\alpha\cos\delta$, $\delta$) coordinate system may appear to be
large as a result of the large uncertainties in AC, which contributes
to the uncertainty in both right ascension and declination after the
coordinate transformation.

Other errors also affect the total budget, such as the error
from the satellite attitude and the modelling errors that are
due to
some corrections that are not yet fully calibrated or implemented. They contribute to both the random and the systematic error and
are of the order
of a few milliarcseconds.

\section{Asteroid photometry in \gdrtwo}
\label{S:photometry}
The \textit{Gaia} Archive provides asteroid magnitudes in \gdrtwo\ in the $G$
band (measured in the AF \textsl{white band}), for 52$\%$ of the
observations. This fraction is a result of a severe selection
that is described below.

Asteroids, due to their orbital motion, move compared to stellar
sources on the focal plane of \textit{Gaia}. Hence, it is possible that they
can drift out of the window during the observations of the AFs. This
drift can be partial or total, resulting in potential loss of flux
during the AF$_1,\dots,AF_x$ with $x>1$ observations. Asteroid
photometry at this stage is processed with the same approach
as is used for stellar photometry \citep{carrasco2016, riello2018} and no
specific optimisation is currently in place to account for flux loss
in moving sources. This situation is expected to improve significantly in the
future \gaia releases.

The photometry of \gdrtwo\ is provided at transit level: the
brightness values (magnitude, flux, and flux error) repeat identically
for each entry of the \gaia archive that is associated with the same transit.
The transit flux is derived from the average of the calibrated fluxes
recorded in each CCD strip of the AF, weighted by the inverse variance
computed using the single CCD flux uncertainties. This choice
minimises effects that are related, for instance, to windows that are off-centred
with respect to the central flux peak of the signal. However, when the
de-centring becomes extreme during the transit of a moving object, or
worse, when the signal core leaves the allocated window, significant
biases propagate to the value of the transit average and increase its
associated error. This happens in particular for asteroids whose
apparent motion with respect to stars is non-negligible over the
transit duration. A main-belt asteroid with a typical motion of 5
mas/s drifts with respect to the computed window by several pixels
during the $\approx 40$s of the transit in the \gaia FOV.

\begin{figure}[ht]
\centering
\includegraphics[clip=true, trim = 0mm 0mm 0mm 0mm, width=0.8 \hsize]{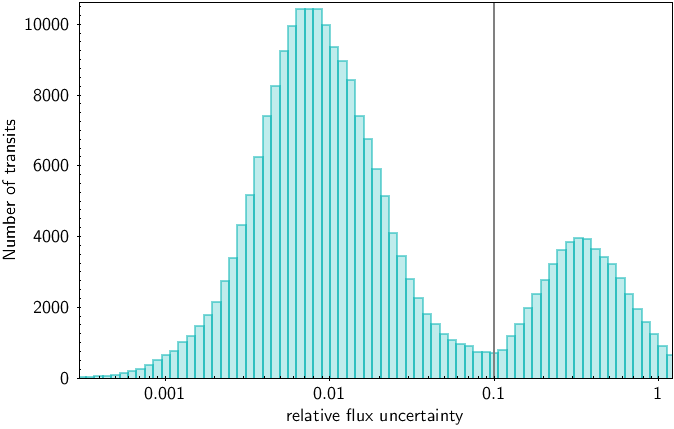}
\caption{Relative error in magnitude $\sigma_G$ for the whole sample
  of transit-level $G$ values. The vertical line at $\sigma_G\sim$0.1
  represents the cut chosen to discard the data with low reliability.}
\label{F:histo_mag_err}
\end{figure}

As provided by the photometric processing, a total of
234,123 transits of SSOs have an associated, fully calibrated
magnitude (81$\%$ of the total). Fig.\ref{F:histo_mag_err} shows the
distribution of the relative error per transit $\sigma_G$ of the
whole dataset before filtering. We found out that the sharp
bi-modality in the distribution correlates positively with transits of
fast moving objects. For this reason, we decided to discard all
transits that fell in the secondary peak of large estimated errors
$\sigma_G >$10$\%$ as they almost certainly correspond to fluxes with
a large random error and might be affected by some (unknown) bias.

A second rejection was implemented on the basis of a set of colour
indices, estimated by using the red and blue photometer (RP and BP),
the two low-resolution slitless spectrophotometers. Again due to
asteroid motion, the wavelength calibration of RP/BP can be severely
affected, and this in turn can affect the colour index that is
used to
calibrate the photometry in AF. In future processing cycles, when
the accurate information on the position of asteroids, produced by the
SSO processing system, will become available to the photometric
processing, we expect to have a significant improvement in the
calibration of the low-resolution spectra and photometric data for
these objects. After checking the distribution of the observations of
SSOs on a space defined by three colour indices (BP-RP, RP-G, and G-BP),
we decided to discard the photometric data falling outside a
reasonable range of colour indices, corresponding to the interval
(0.0, 1.0) for both RB-G and G-RP.

The two criteria above, based on the computed uncertainty and on the
colour, are not independent. Most transits that were rejected due to
poor photometry in the $G$ band also showed colour problems, which
proves
that the two issues are related.

Both filtering procedures together result in the rejection of a rather
large sample of 48$\%$ of the initial brightness measurements
available. In the end, 52$\%$ of the the transits of SSOs in
\gdrtwo\ have an associated G-band photometry.

\begin{figure}[ht]
\centering
\includegraphics[clip=true, trim = 0mm 0mm 0mm 0mm, width=0.9 \hsize]{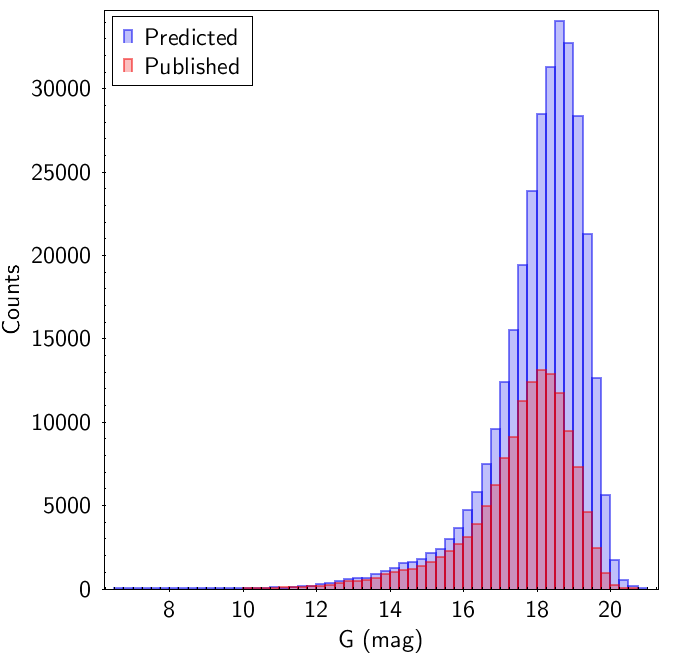}
\caption{Distribution of the apparent magnitude of the SSOs in
  \gdrtwo\, at the transit epochs. For the whole sample the brightness
  derived from ephemerides (adopting the (H,G) photometric system) is
  provided (label: ''predicted''). The sub-sample contains the
  magnitude values that are published in \gdrtwo. The shift of the
  peak towards brighter values indicates a larger fraction of ejected
  values among faint objects. }
\label{F:hist_mag}
\end{figure}

\begin{figure}[ht]
\centering
\includegraphics[clip=true, trim = 0mm 0mm 0mm 0mm, width=0.9 \hsize]{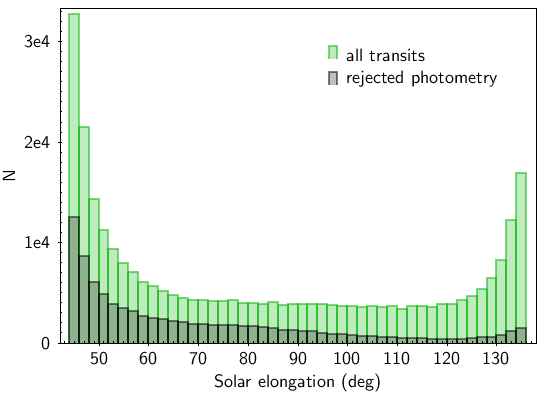}
\caption{Distribution of the asteroid sample in \gdrtwo\ as a function of
  solar elongation. The whole sample is compared to the sub-sample of
  asteroids with rejected photometric results (histogram of lower amplitude).}
\label{F:hist_elong_mag}
\end{figure}

\begin{figure}[ht]
\centering
\includegraphics[clip=true, trim = 0mm 0mm 0mm 0mm, width=0.9 \hsize]{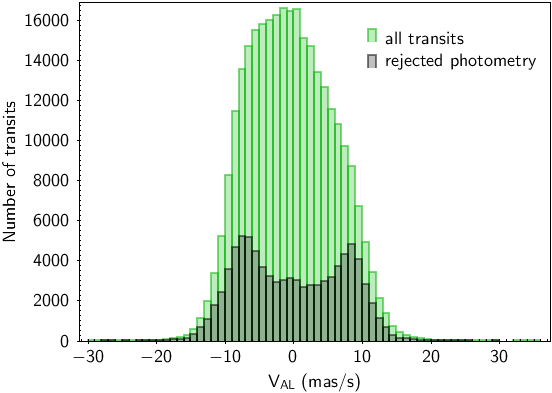}
\caption{Distribution of the asteroid sample in \gdrtwo\ as a function of
  AL velocity. The whole sample is compared to the sub-sample of
  asteroids with rejected photometric results  (histogram of lower amplitude).}
\label{F:hist_daldt_mag}
\end{figure}

Figure~\ref{F:hist_elong_mag} shows the difference in distribution of
solar elongation angles, between the entire \gdrtwo\ transit sample
and the transits for which the magnitude is
rejected. Figure~\ref{F:hist_daldt_mag} shows the same comparison on
the AL velocity distribution. The majority of rejections occurs at low
elongations, where their average apparent velocity is higher.

\begin{figure}[ht]
\centering
\includegraphics[clip=true, trim = 0mm 0mm 0mm 0mm, width=0.9 \hsize]{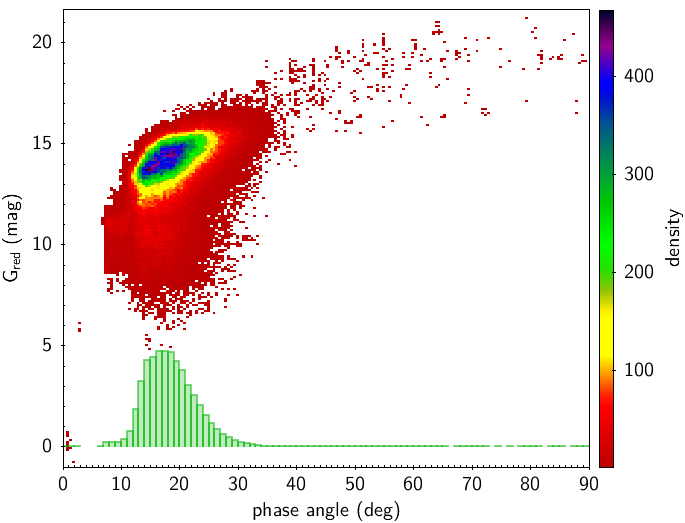}
\caption{Reduced asteroid magnitude as a function of phase angle. The
  histogram of phase angles is superposed on the bottom part
  (arbitrary vertical scale).}
\label{F:phase_Gred}
\end{figure}

The resulting distribution of phase angles and reduced magnitudes
(G$_{red}$, at 1 au distance from \gaia and the Sun) for the transits
in \gdrtwo\ is plotted in Fig.~\ref{F:phase_Gred}. In addition
to the core of the
distribution represented by MBAs, a small sample of
NEAs reaching high phase angles is visible, as well as some transits
associated with large TNOs at the smallest phase angles.

Despite the severe rejection of outliers, assessing the reliability of
the published photometry at the expected accuracy of \textit{Gaia},
specifically for solar system bodies, is not straightforward. The
intrinsic variability of the asteroids due to their changing viewing
and illumination geometry and to their complex shapes makes the
comparison of observed fluxes with theoretical ones very
challenging. Sunlight scattering effects from the asteroid surfaces
also play a role and must be modelled to reproduce the observed
brightness.

We attempted to model the observed brightness following two different
approaches, on a small sample of asteroids. First, we used a genetic inversion algorithm derived from a full inversion algorithm developed
by \citet{cellino_genetic_2009} and massively tested by
\citet{santana-ros_2015} to derive for a few selected objects the
best--fitting three--axial ellipsoid (axis ratios) from \gaia
observations alone. The procedure assumes known values of the spin
period and spin-axis direction ("asteroid pole") available in the
literature for objects that have been extensively observed from the ground, and takes
into account a linear phase-magnitude dependence. The procedure is
extensively explained in the \gdrtwo\ documentation.

Independently, we exploited the detailed shape models available for
the two asteroids \object{(21)~Lutetia} and (2867)~\v{S}teins derived by
combining ground-based data with those obtained during the ESA Rosetta
flybys to reproduce their observed \textit{Gaia} brightness. Both attempts, of
course, concern modelling the flux variations relative to a given
observation in the sample, not its absolute value.

\begin{figure*}[ht]
\centering
\includegraphics[clip=true, trim = 0mm 0mm 0mm 0mm, width=0.35 \hsize]{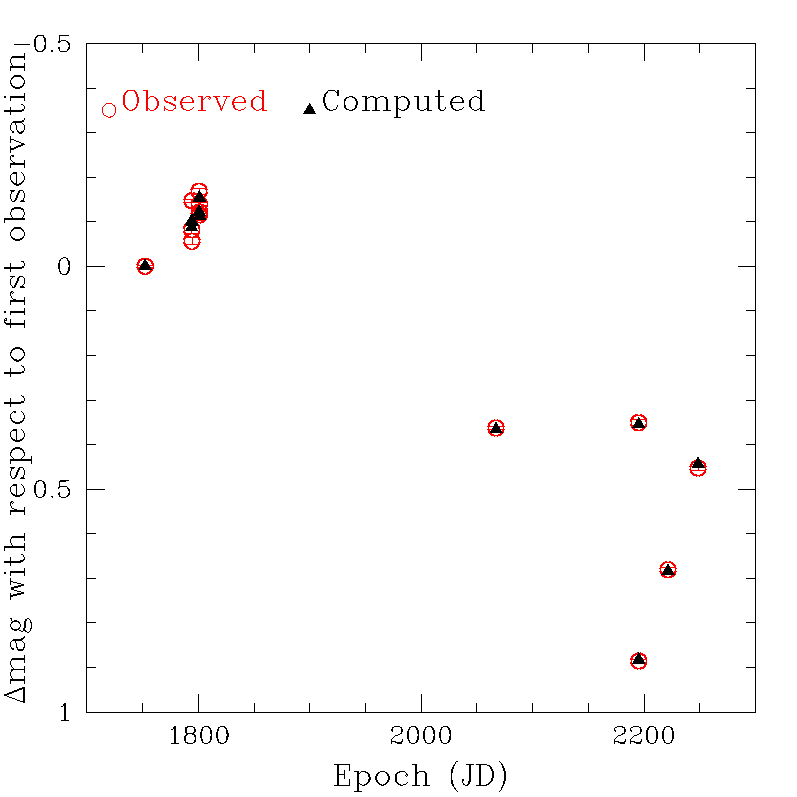}
\includegraphics[clip=true, trim = 0mm 0mm 0mm 0mm, width=0.35 \hsize]{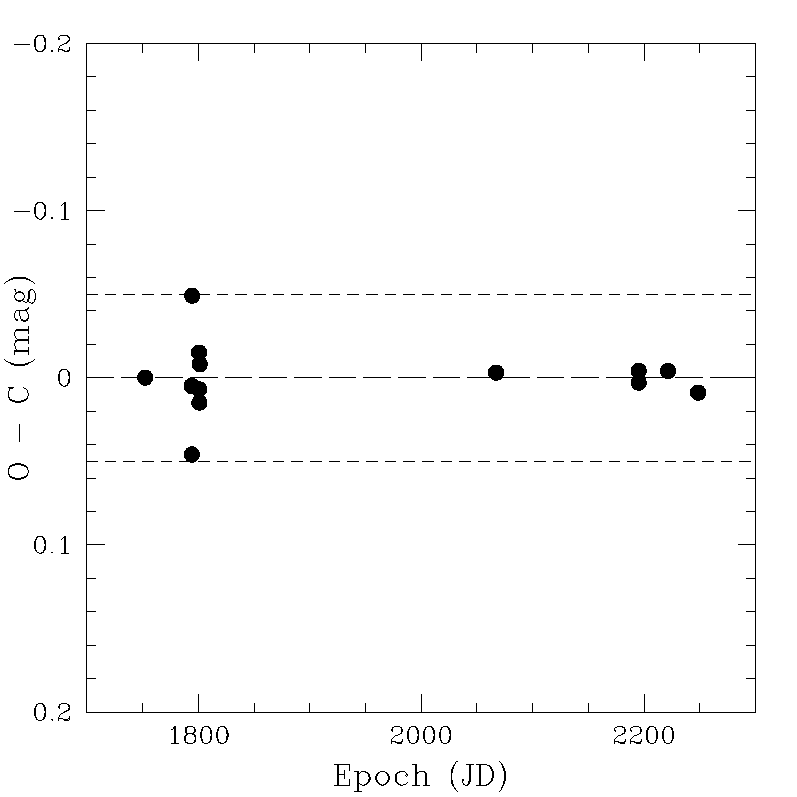}
\caption{Observed and computed magnitude from the best fit of \textit{Gaia}
  observations of an ellipsoidal model for the asteroid \object{(39)
  Laetitia}. In the right panel, we show the corresponding residuals. The
  origin of the time axis is J2010.0.}
\label{F:shapefit}
\end{figure*}

\begin{figure*}[ht]
\centering
\includegraphics[clip=true, trim = 0mm 0mm 0mm 0mm, width=0.35 \hsize]{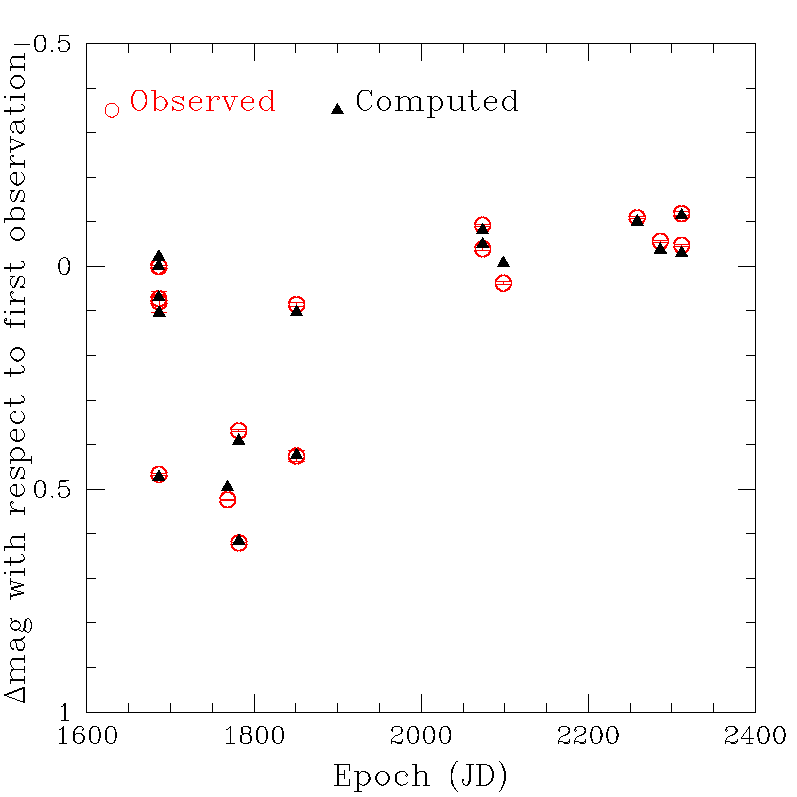}
\includegraphics[clip=true, trim = 0mm 0mm 0mm 0mm, width=0.35 \hsize]{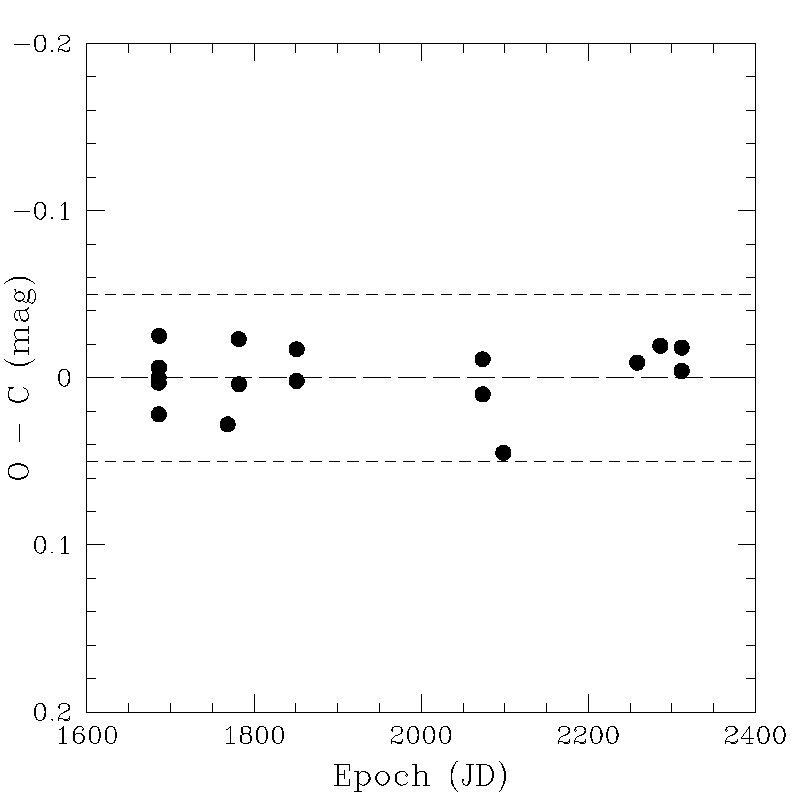}
\caption{As in Fig.~\ref{F:shapefit} for the asteroid \object{(283) Emma}.}
\label{F:shapefit2}
\end{figure*}

\begin{figure*}[ht]
\centering
\includegraphics[clip=true, trim = 0mm 0mm 0mm 0mm, width=0.35 \hsize]{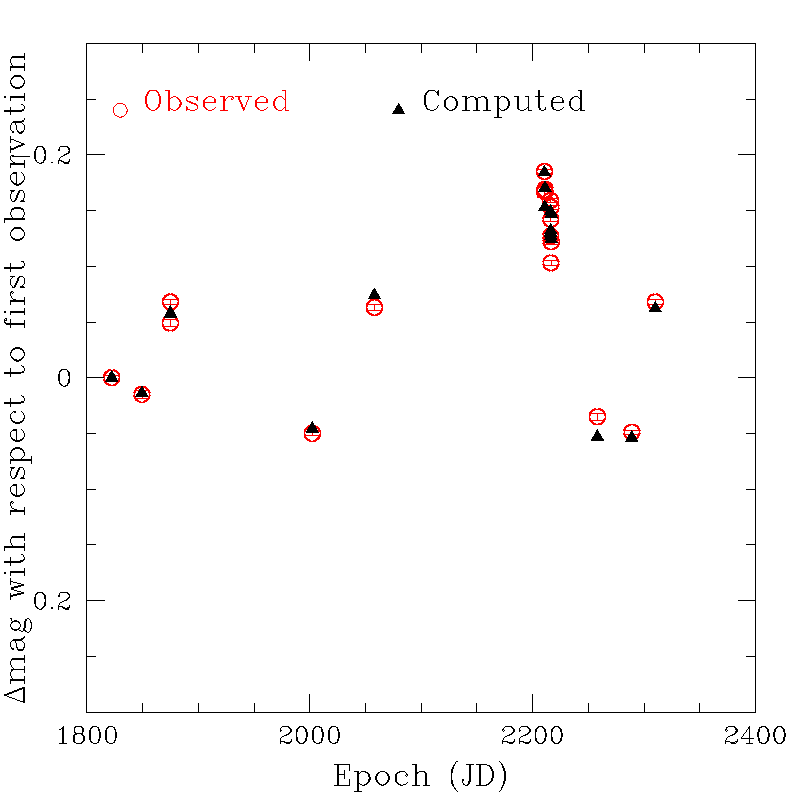}
\includegraphics[clip=true, trim = 0mm 0mm 0mm 0mm, width=0.35 \hsize]{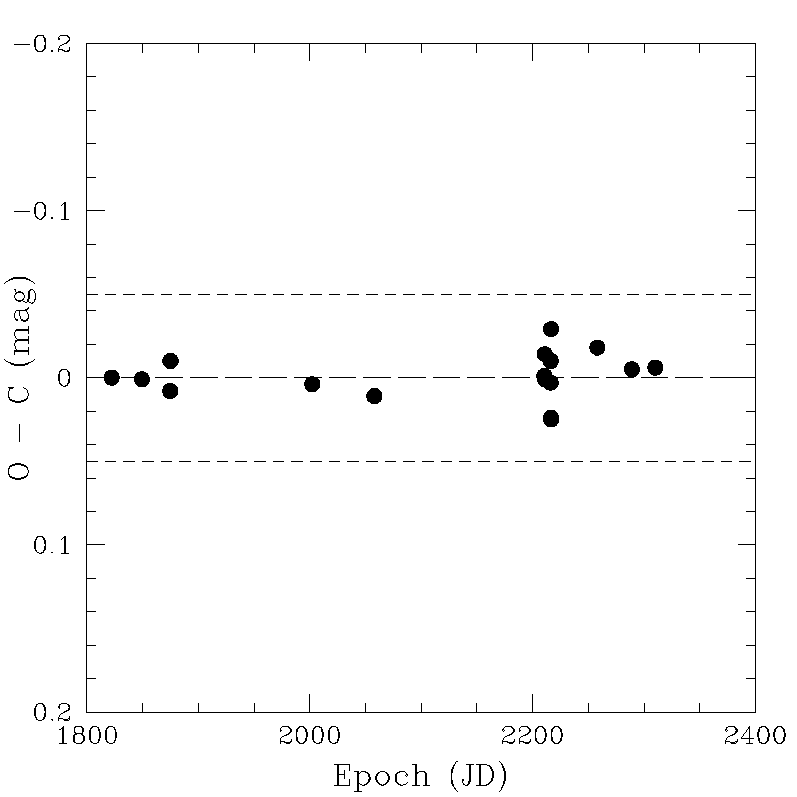}
\caption{As in Fig.~\ref{F:shapefit} for the asteroid \object{(704) Interamnia}.}
\label{F:shapefit3}
\end{figure*}

The results from the sparse photometry inversion are presented in
Fig.~\ref{F:shapefit}-\ref{F:shapefit3}. They are obtained by assuming
a Lommel-Seeliger scattering law, a realistic choice when 
a more detailed mapping of the scattering properties across the surface is not available \citep{Muinonen2015a, Muinonen2015b}. 


Despite the very simplified shape model, the residuals (observations minus computed) O-C are always 
within $\pm 0.05$ magnitudes, and the typical scatter can be estimated around 2-3$\%$. 
Using the shape models of (21)~Lutetia \citep{carry2010} and \object{(2867)~\v{S}teins} \citep{Jorda2012}, we tried to assess the photometric
accuracy limit of \textit{Gaia} on asteroids. In the case of \object{(21)~Lutetia}, it
was found that \textit{Gaia} data are in very good agreement with expectations
based upon the best available shape model of this asteroid, derived
from disk-resolved imaging by Rosetta (which only imaged one
hemisphere of the object) and a lower-resolution model based on
disk-integrated, ground-based photometry. The high-resolution shape
model reproduces the \textit{Gaia} photometry with a small RMS value of 0.025
mag, corresponding to 2.3$\%$ RMS in flux. This strongly suggests that
\textit{Gaia} photometry is probably better than 2$\%$ RMS, within the
limitations imposed by the shape model accuracy and the assumptions on the scattering model. Moreover, \textit{Gaia} data
seem to offer an opportunity to improve the currently accepted shape
solution for Lutetia, which is based partly upon ground-based data.  

The results obtained for (2867)~\v{S}teins, for which a high-resolution shape
model is also available, strongly support the conclusion that the
photometry is indeed very accurate. For (2867)~\v{S}teins two pole
solutions exist, essentially differing only by the value of the origin
of the rotational phase. By directly using the shape model to
reproduce \textit{Gaia} data, resampled at 5 degree resolution, with a
Lommel-Seeliger scattering corresponding to E-type asteroid phase
functions, the RMS value of the O-C is 1.64$\%$ and 1.51$\%$ for the
two pole solutions, a very good result. Changing the resolution to 3
degrees does not improve the fit further.
The remaining limitations in the case of (2867)~\v{S}teins are still
related to details of the shape, and to the assumptions made (and/or
scattering properties) when it was derived from Rosetta images.
 
In conclusion, our validation appears to show that \textit{Gaia} epoch
photometry, appropriately filtered to eliminate the outliers, probably has an accuracy below 1-2$\%$ up to the magnitude of
(2867)~\v{S}teins, in the range G~17-19. However, given the current
limitations on the calibration and processing, we cannot exclude that
the sample published in \gdrtwo\ still contains a non-negligible
fraction of anomalous data. For this reason, we recommend detailed
analysis and careful checks for any applications based on
\gdrtwo\ photometry of asteroids.

\section{Validation of the astrometry}
\label{S:astro_validation}
\begin{figure}[h!]
\centering
\includegraphics[clip=true, trim = 0mm 0mm 0mm 0mm, width=1.0 \hsize]{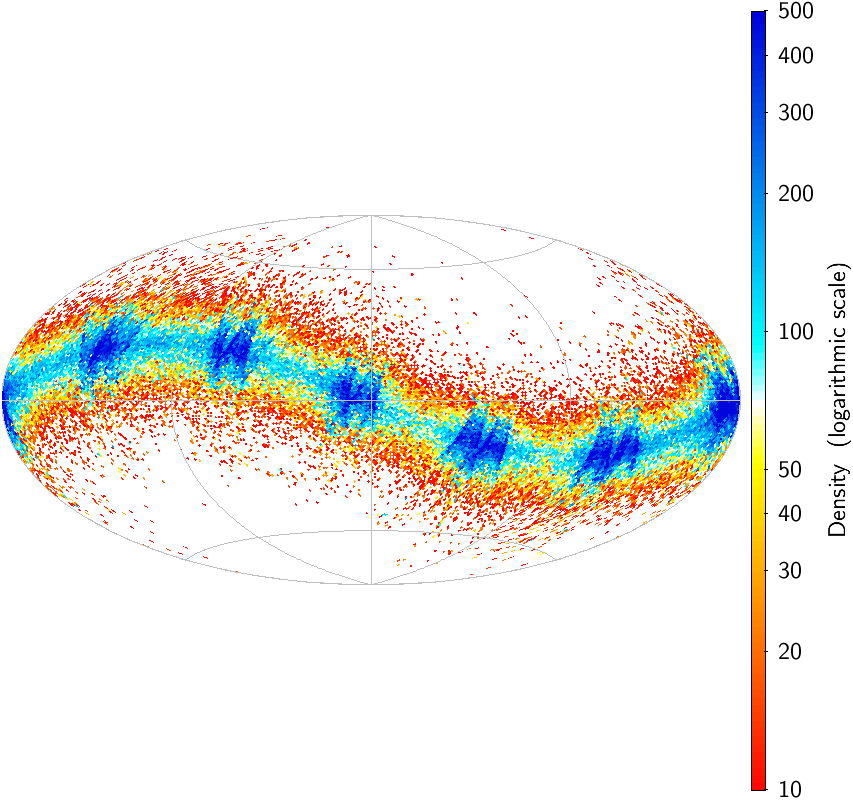}
\caption{Sky distribution (equatorial coordinates) of the 2\,005\,683
  observations for the 14\,124 asteroid in the validation sample. This
  sky map use an Aitoff projection in equatorial (ICRS) coordinates
  with $\alpha=\delta=0$ at the centre, north up, and $\alpha$
  increasing from right to left. The observation density is higher in blue
  areas. The pattern in ecliptic longitude is a consequence of the
  \gaia scanning law over a small fraction of the five-year nominal
  mission.}
\label{F:sky_distribution_validation}
\end{figure}

The processing of the solar system data described above has eventually
produced a list with 14\,124 objects (all numbered SSOs),
290\,704 transits, and 2\,005\,683 CCD observations. The sky
distribution is shown in Fig.~\ref{F:sky_distribution_validation} in a
density plot in equatorial coordinates. As expected, most SSOs are found
in a limited range of ecliptic latitudes. The distribution in
longitude is not uniform because over a relatively short duration of
22 months, the \gaia scanning returned to the same regions of the sky,
only in a limited number of areas.

Assessing the quality of the astrometry is challenging, and it needs
an ad hoc treatment. Various filters have been applied during the
activity of the astrometric reduction processing. The filtering
process ensures the rejection of a maximum number of bad detections,
while keeping the number of good positions that are rejected as small as
possible (for more details, see the \gdrtwo\ documentation). 
To prove that \gaia is already close to the
performances expected at the end of the mission, we designed an ad hoc
procedure for the external validation of the results. To this end, we
fitted an orbit (initialising the fit with the best existing orbit)
using only the available 22 months of \gaia observations, and we
examined the residuals in right ascension and declination, and also in
AL and AC (see Sec.~\ref{S:orbit_determination}). The main differences
between \gaia and ground-based observations (or any other satellite
observations) can be summarised as follows:
\begin{itemize}
\item \gaia observations are given in TCB, which is the primary timescale for \gaia.
\item Positions (right ascension and declination) are given in the
BCRS as the direction of the unit
  vector from the centre of mass of \gaia to the SSOs.
\item The observation accuracies are up to the order of few $\sim
  10^{-9}$ radians (sub-mas level) in the AL direction.
\item The error model contains the correlations in $\alpha\cos\delta$
  and $\delta$  because of the rotation from the (AL, AC) plane to the
  ($\alpha\cos\delta$, $\delta$) plane
  (Sec.~\ref{S:error_model_astrometry}).
\end{itemize}

\subsection{Orbit determination process}
\label{S:orbit_determination}
The orbit determination process usually consists of a set of
mathematical methods for computing the orbit of objects such as planets
or spacecraft, starting from their observations. For our validation
purpose, we considered only the list of numbered asteroids for which
the orbits were already well-known from ground-based (optical or
radar) /satellite observations. We used the least-squares method and
the differential correction algorithm
\citep[see][]{2010tod..book.....M} to fit orbits on 22 months of \gaia
observations, using as initial guess the known orbits of these
objects. To be consistent with the high quality of the data, we
employed a high-precision dynamical model, which includes the
Newtonian pull of the Sun, eight planets, the Moon, and Pluto based on
JPL DE431 Planetary ephemerides\footnote{We also performed the
  orbit determination process using INPOP13c~\citep{inpop13c}
  ephemerides and did not find significant differences in the
  results.}. We also added the contribution of 16 massive main-belt
asteroids (see~\ref{A:perturbing}). We used a relativistic force model
including the contribution of the Sun, the planets, and the Moon,
namely the Einstein-Infeld-Hoffman approximation~\citep{moyer2003}
or~\citep{will1993}. As a result of the orbit determination process,
we obtained for every object a corrected orbit fitted on \textit{Gaia} data
only together with the post-fit residuals.

The core of the least-squares procedure is to minimise the target
function~\citep{2010tod..book.....M},
\begin{equation}
\label{E:target_function}
Q = \frac{1}{m} \ \bxi^T  W  \bxi
,\end{equation}
where $m$ is the number of observations, $\bxi$ are the residuals
(observed positions minus computed positions), and $W$ is the weight
matrix. The solution is given by the normal equations,
\begin{equation}
\label{E:normal_equations}
 C = B^T W B; \quad D = -B^T W \bxi \quad \left(B = \frac{\delta \bxi}{\delta \boldsymbol{x}} \right)
,\end{equation}
where $\boldsymbol{x}$ is the vector of the parameters to be solved
for. The differential corrections produce the adjustments $\Delta
\boldsymbol{x}$ to be applied to the orbit:
\[
\Delta \boldsymbol{x} = C^{-1}D.
\]
It is clear from Eqs. \ref{E:target_function} and \ref{E:normal_equations}
that the weight matrix plays a fundamental role in the orbit
determination. It is usually the inverse of a diagonal matrix
($\Gamma$) that contains on the diagonal the square of the
uncertainties in right ascension and declination for each observation,
according to the existing debiasing and error models (as in
~\citet{2015Icar..245...94F}). Each \gaia observation comes with its
uncertainties on both coordinates and the correlation, which are key
quantities in the orbit determination process. Therefore the weight
matrix in our case is $W = \Gamma^{-1}$, where \small
\[
\Gamma =
\begin{bmatrix}
  \sigma_{\alpha_1}^2 &  \mathbf{cov(\alpha_1,\delta_1)} & 0 & \cdots & 0 \\
  \mathbf{cov(\alpha_1,\delta_1)}  & \sigma_{\delta_1}^2 & 0 & \cdots & 0\\
  \vdots & \vdots & \ddots\\
  0 & 0 & \cdots & \sigma_{\alpha_m}^2 & \mathbf{cov(\alpha_m,\delta_m)} \\
  0 & 0 & \cdots & \mathbf{cov(\alpha_m,\delta_m)}  & \sigma_{\delta_m}^2\\
\end{bmatrix}.
\]
\normalsize 

The uncertainties used to build the W matrix are given by the random
component of the error model, but we also take into account the
systematic contribution when this is needed, as explained in the
following section.

\subsection{Outlier rejection procedure}
\label{S:outlier_rejection}

The rejection of the outliers is a fundamental step in the orbit
determination procedure. Since we assumed that the residuals are
distributed as normal variables, the rejection was based on the post-fit
$\chi^2$ value for each observation, computed as in
\citet{2003Icar..166..248C}:
\[
\chi_i^2 = \bxi_i \bgamma^{-1}_{\bxi_i} \bxi_i^T \quad i=1,\dots,m,
\]
where $m$ is the total number of observations, $\bxi_i$ is the vector
of the residuals for each observation, and $\bgamma_{\bxi_i}$ is the
expected covariance of the residuals. Each $\chi_i^2$ has a
distribution of a $\chi^2$ variable with two degrees of freedom. We call
outlier each observation whose $\chi^2$ value is greater than
25. The choice of 25 as a threshold was driven by the fact that we
wished to keep as many good observations as possible and wished
to discard only the observations (or the transits) that are very far
from the expected \gaia performances. During this procedure,
we took random and systematic errors into account.

Firstly, we rejected all the observations whose
$\chi^2$ value was greater than $25$. Then, when the systematic part was
larger than the random part, we performed a second step in the outlier
rejection, described as follows:
\begin{itemize}
\item We computed the mean of the residuals for each transit.
\item We checked if the value of the mean is lower than the systematic
  error for the transit.
\item If the value was higher than the systematic error, we discarded the
  entire transit.
\item If the value was lower than the systematic error, we computed
  for each observation the difference between the residual and the mean value.
\item We checked whether the difference was smaller than the random error. When that was the case, we kept the observation, otherwise we discarded it.
\end{itemize}

This approach is consistent with the uncertainties produced for
\gdrtwo. Its underlying hypothesis is that the error correlations over
a single transit can be completely represented by just one quantity,
which is the value of the systematic component alone. Although this is
an approximation, we currently do not have the impression that a more
complex correlation model is required.

\subsection{Results}
\label{S:results}
We fitted the orbits of the 14\,124 asteroids contained in the validation
sample using an updated version of the OrbFit
software\footnote{\url{http://adams.dm.unipi.it/orbfit/}}, developed
to handle \gaia observations and \gaia error model
(Sections~\ref{S:orbit_determination} and~\ref{S:outlier_rejection}).

We were unable to fit the
observations to the existing orbit for only three asteroids because the time span covered by the available transits
was too short. They were removed from the final
output of \gdrtwo. We also removed $22$ bright objects (transits with
G$<10$) whose residuals were substantially larger than the
uncertainties we expected, and we considered these solutions as not
reliable.

\begin{figure}[h!]
\centering \includegraphics[clip=true, trim = 0mm 0mm 0mm 0mm,
  width=0.98\hsize]{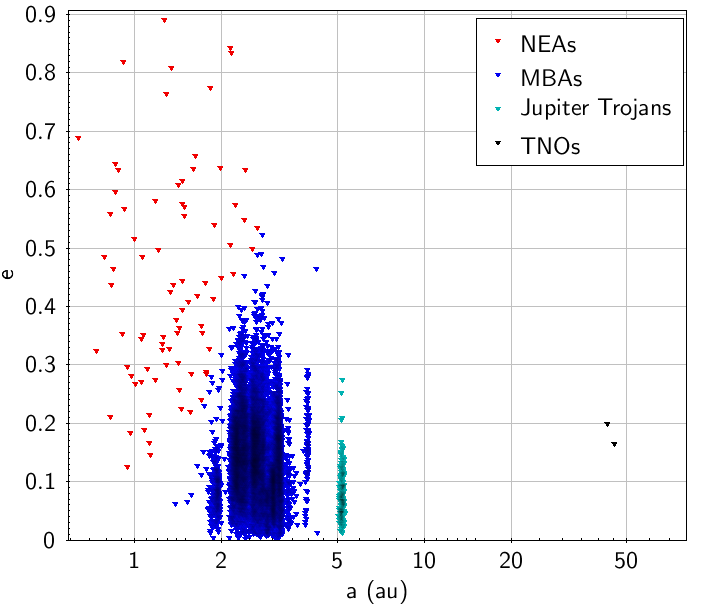}
\caption{Distribution of the 14\,099 asteroids published in
  \gdrtwo\ in semi-major axis a (au) and eccentricity e. The sample
  shows that all the broad categories of SSOs are represented (NEAs,
  MBAs, Jupiter trojans, and TNOs).}
\label{F:q_sample}
\end{figure}

Figure~\ref{F:q_sample} shows the distribution in semi-major axis and
eccentricity of the 14\,099 SSOs published in \gdrtwo. We can
distinguish the NEA population (q$<$1.3) from the MBAs (including
Jupiter trojans in this class) and the TNOs (q$>28$). 

\begin{figure}[h!]
\centering
\includegraphics[clip=true, trim = 0mm 0mm 0mm 0mm, width=1.0 \hsize]{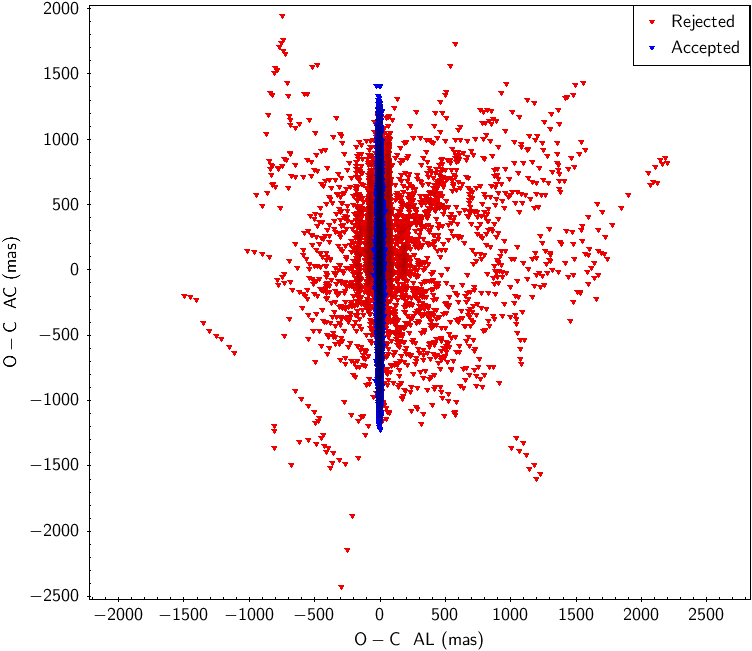}
\caption{Residuals in the (AL, AC) plane in milliarcseconds. Outliers are marked
  in red, while the blue thick line in the middle contains all the
  residuals for the accepted observations. The total number of fitted
  observations is $2\,005\,683,$ and there are $27\,981$ outliers
($\sim  1 \%$ of the total).}
\label{F:all_outliers_alac}
\end{figure}

The total number of fitted observations is 2\,005\,683, which
corresponds to 290\,704 transits. During the outlier rejection
procedure (Sec.~\ref{S:outlier_rejection}), we discarded 27\,981
observations ($\sim 1$\% of the
total). Figure~\ref{F:all_outliers_alac} shows all the residuals in
the (AL, AC) plane in mas. Outliers are the red points, while accepted
observations are all contained in the blue thick line in the centre of
the figure.

\begin{figure}[h!]
\centering
\includegraphics[clip=true, trim = 0mm 0mm 0mm 0mm, width=0.98\hsize]{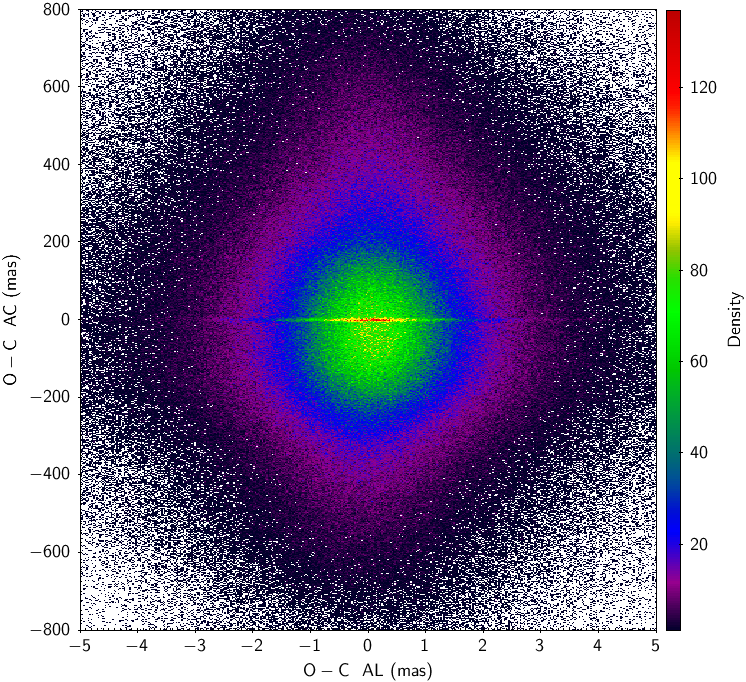}
\caption{Zoom-in of the density plot of the residuals in the (AL, AC)
  plane expressed in milliarcsecond for all the observations published in
  \gdrtwo. $96\%$ of the AL residuals fall in the interval [-5, 5], and
  $52\%$ are at sub-milliarcsecond level. Almost all the residuals in AC ($98\%$)
  fall in the interval [-800, 800].}
\label{F:good_outliers_alac}
\end{figure}

\begin{figure*}[h!]
\centering
\includegraphics[clip=true, trim = 0mm 0mm 0mm 0mm, width=0.45 \hsize]{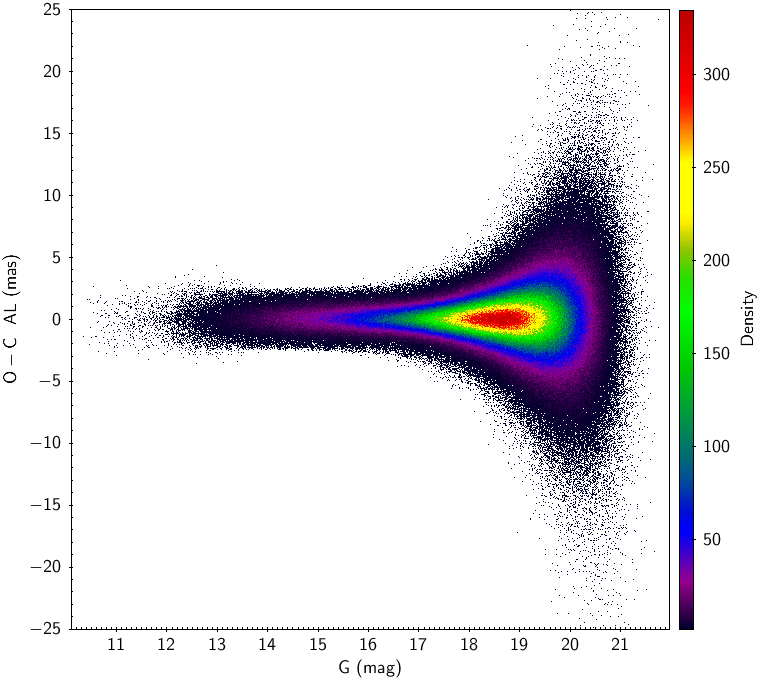}
\includegraphics[clip=true, trim = 0mm 0mm 0mm 0mm, width=0.45 \hsize]{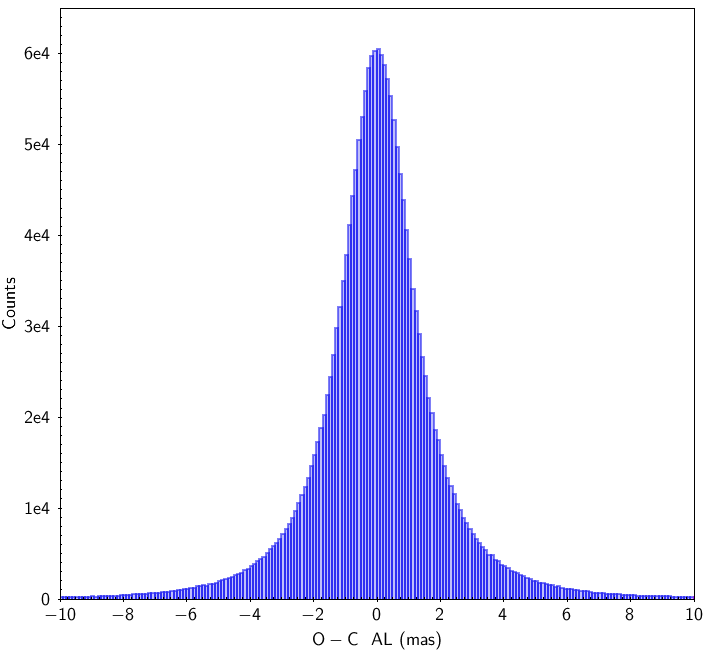}
\caption{Left: AL residuals with respect to $G$ magnitude. Right:
  Histogram of the AL residuals in the interval [-10,10] mas. The
  tails (not visible in the histogram) contain 7304 observations
  ($\sim0.4\%$ of the total number of observations published in
  \gdrtwo) for which the residuals are greater than 10 mas or smaller
  than -10 mas.  The mean of the residuals in AL is 0.05 mas and the
  standard deviation is 2.14 mas.}
\label{F:good_al}
\end{figure*}

\begin{figure*}[h!]
\centering
\includegraphics[clip=true, trim = 0mm 0mm 0mm 0mm, width=0.45 \hsize]{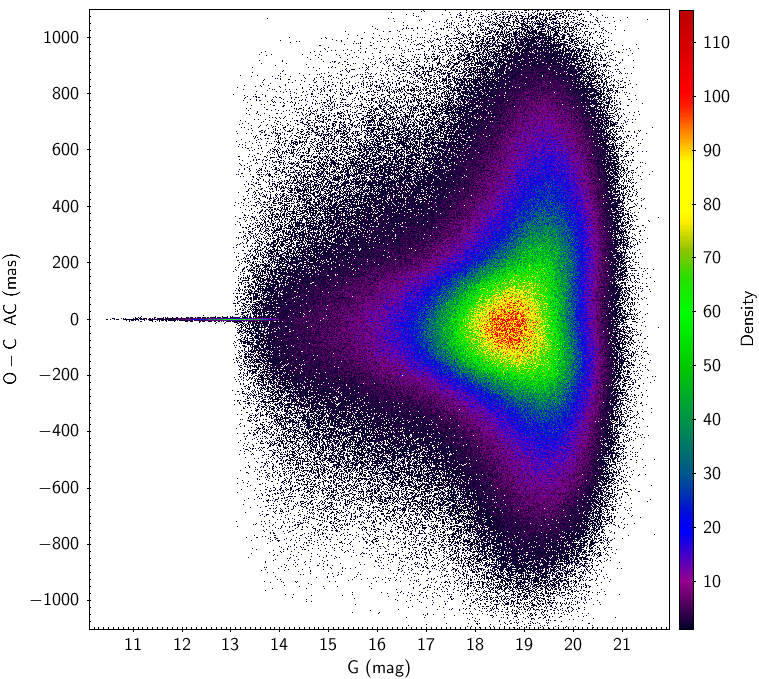}
\includegraphics[clip=true, trim = 0mm 0mm 0mm 0mm, width=0.45 \hsize]{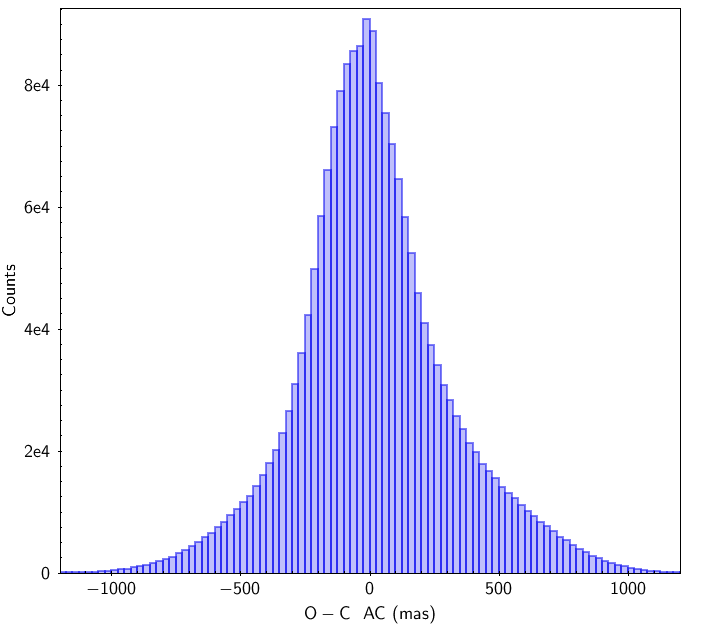}
\caption{Left: AC residuals with respect to $G$ magnitude. Right:
  Histogram of the residuals in AC. The peak around $0$ is strictly
  related to the distribution of the residuals as a function of the $G$
  magnitude. For objects brighter than G=13, the accuracy in AC and AL
  are similar, while for objects fainter than G=13, the errors in AC
  are larger.}
\label{F:good_ac}
\end{figure*}

After the filtering and the outlier rejection, \gdrtwo\ contains
1\,977\,702 observations, corresponding to 14\,099 SSOs and 287\,904 transits. Figure~\ref{F:good_outliers_alac}
represents a density plot of the residuals at CCD level in the (AL,
AC) plane, for all the observations published in \gdrtwo. This plot,
together with the plots of the residuals (Figures~\ref{F:good_al} and
\ref{F:good_ac}), shows the epoch-making change brought about by \gaia
astrometry: $96\%$ of the AL residuals fall in the interval [-5, 5]
mas and $52\%$ are at sub-milliarcsecond level. The behaviour of the residuals in
AC is markedly different as a result of the geometry of the spacecraft
observations. The AC residuals as a rule are much larger than AL,
for the reasons detailed in
Sec.~\ref{S:error_model_astrometry}. Figures~\ref{F:good_al}
and~\ref{F:good_ac} highlight the performances of
\gaia even better. They show the residuals in AL and AC with respect to the $G$
magnitude of the asteroids at the epoch of observation. The right
panels of both Fig.~\ref{F:good_al} and Fig.~\ref{F:good_ac} display the histogram of the residuals in AL and AC, respectively. The clear
peak around 0 is well visible in the histogram of the residuals in AL
(Fig.~\ref{F:good_al}), and it corresponds to the core of the density
plot in Fig.~\ref{F:good_outliers_alac}.

Even the histogram of the residuals in AC (Fig.~\ref{F:good_ac}) shows
a peak around 0, which is strictly related to the distribution of the
residuals as a function of the $G$ magnitude. For objects brighter
than $G = 13,$ the full 2D window is transmitted (see
Sec.~\ref{S:data}), thus the accuracies in AC and AL are similar
(although still slightly larger in the across direction because of the pixel
size), while for objects fainter than $G= 13$, the errors in AC are
much larger.

\section{Interpretation of asteroid residuals}
\label{S:astro_interpretation}
The residuals in AL, as in Figures~\ref{F:good_outliers_alac} and
\ref{F:good_al}, show the quality and extreme precision of
\gaia observations. We now examine the residuals in more
detail, using first as an example a main-belt asteroid, and then
showing the properties at  transit-level.
\subsection{CCD-level residuals}
\label{S:ccd_analysis}

We chose \object{(367) Amicitia} as a test case to analyse the residuals in
right ascension and declination and to explain their relationship (and
main differences) with the residuals in AL and AC. Residuals in right
ascension and declination are important because they are the direct
output of the fit and are probably more easily understood than the
residuals in AL and AC, which are closely related to the \textit{Gaia} scanning
mode. The residuals in right ascension and declination are by-products
of the orbit correction and computed as the differences between the
observed and computed positions. Afterwards, given the position angle
shown in Fig.~\ref{F:scheme_AL_AC}, we can rotate the residual vector
in $\alpha \cos\delta$ and $\delta$ to compute the residuals in AL and
AC (\ref{E:rotation}), 

\begin{equation}
  \label{E:rotation}
  \begin{pmatrix}
    \Delta AL \\
    \Delta AC 
  \end{pmatrix} =\begin{bmatrix}
   \phantom{-}\sin(PA) & \cos(PA) \\
  -\cos(PA) & \sin(PA)
  \end{bmatrix}
  \begin{pmatrix}
    \Delta \alpha\cos\delta \\ \Delta \delta
  \end{pmatrix}
.\end{equation}

For the sake of simplicity, we call ($\Delta \alpha\cos\delta$,
$\Delta\delta$) the vector of the residuals in $\alpha \cos\delta$ and
$\delta$ and ($\Delta$AL, $\Delta$AC) the vector of the residuals in
AL and AC.

\begin{figure}[h!]
\centering
\includegraphics[clip=true, trim = 0mm 0mm 0mm 0mm, width=1.0 \hsize]{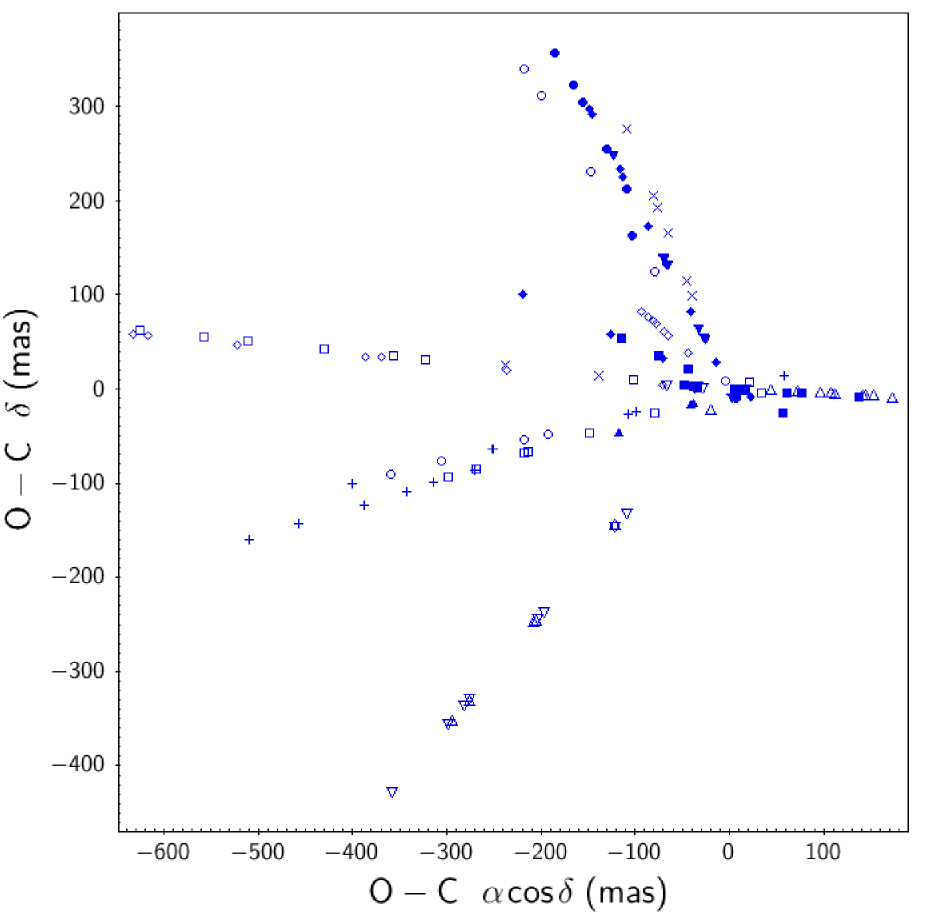}
\caption{Residuals in $\alpha\cos\delta$ and $\delta$ for the MBA
  (367) Amicitia. Different symbols correspond to the 22 different
  transits of the object.}
\label{F:367_res_ra_dec}
\end{figure}

Asteroid (367) Amicitia is a typical object among SSOs observed by
\gaia: its average $G$ magnitude is 14, which means that it is not one
of the brightest objects for which we have the 2D window. It has 22
\gaia transits, which corresponds to 132 CCD
observations. Figure~\ref{F:367_res_ra_dec} shows all the residuals in
right ascension and declination. Each transit is represented by a
different symbol.

\begin{figure}[h!]
\centering
\includegraphics[clip=true, trim = 0mm 0mm 0mm 0mm, width=1.0 \hsize]{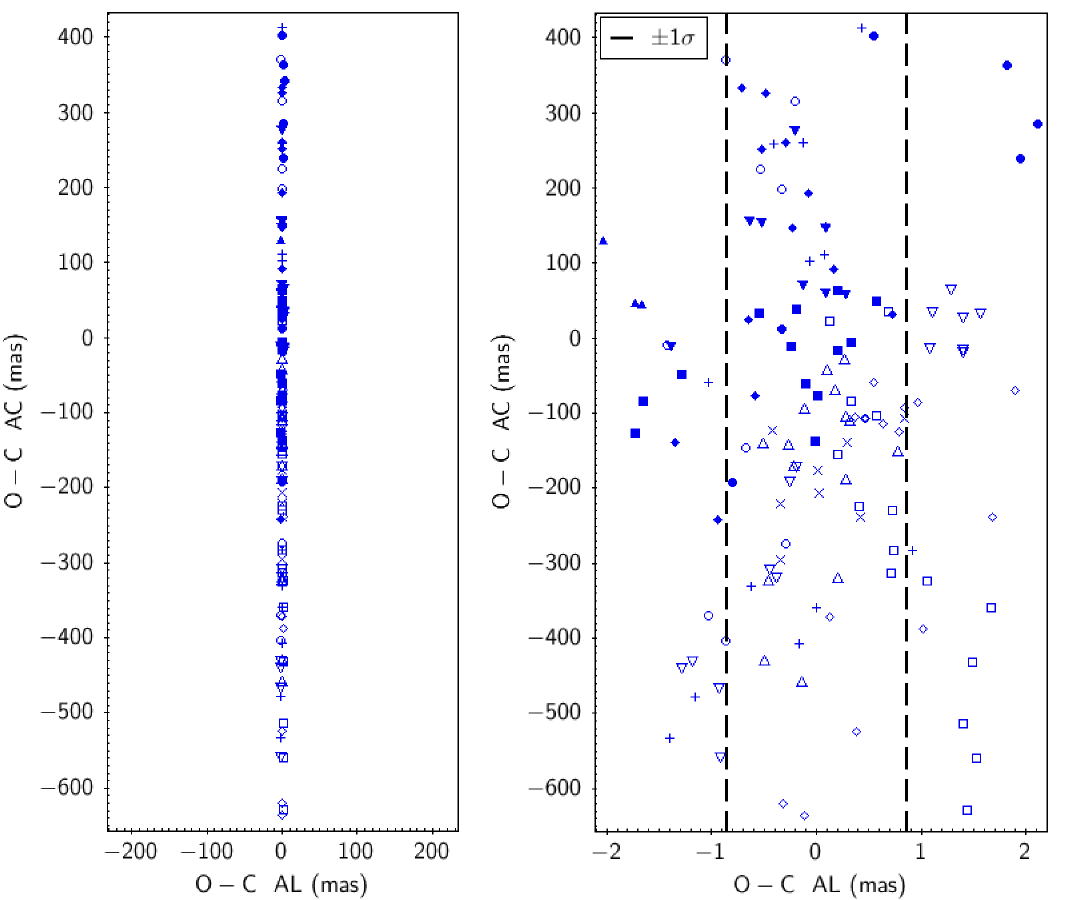}
\caption{Residuals in AL and AC for the MBA \object{(367) Amicitia}. The
  residuals are obtained as a rotation from the plane of the residuals
  ($\alpha\cos\delta$, $\delta$) to the plane of the residuals in (AL,
  AC). The right panel is a zoom-in of the residuals and shows that
  almost all the residuals are at sub-milliarcsecond level and inside 1$\sigma$
  (dashed lines).}
\label{F:367_res_al_ac}
\end{figure}

Figure~\ref{F:367_res_al_ac} shows the residuals rotated in the (AL,
AC) plane. They are clearly very small in AL and considerably larger
in AC (as expected for an asteroid of that magnitude with almost no
across-scan data). The majority of the residuals in AL are at sub-milliarcsecond
level (Fig. \ref{F:367_res_al_ac}, right panel), which means that almost
all are inside $1\sigma$ (dashed vertical lines).

We have already mentioned the particular features of the residuals in
the (AL, AC) plane (Sec.~\ref{S:error_model_astrometry} and
Sec.~\ref{S:results}). We now focus on the residuals in
($\alpha\cos\delta$, $\delta$). They are a linear combination of the
residuals in AL and AC as in Eq.~\ref{E:rotation}. Thus
they are in general larger in both coordinates (Fig.~\ref{F:367_res_ra_dec}) and
highly correlated. Even if the residuals in the (AL, AC) plane are
uncorrelated, the large difference in their standard deviations gives
rise to a very strong correlation between the residuals expressed as
$(\Delta\alpha\cos\delta, \Delta\delta)$.  This strong correlation
between $\Delta\alpha\cos\delta$ and $\Delta\delta$ is expressed visually
by the fact that the residuals for each transit lie roughly on a
straight line.

\begin{figure}[h!]
\centering
\includegraphics[clip=true, trim = 0mm 0mm 0mm 0mm, width=1.0 \hsize]{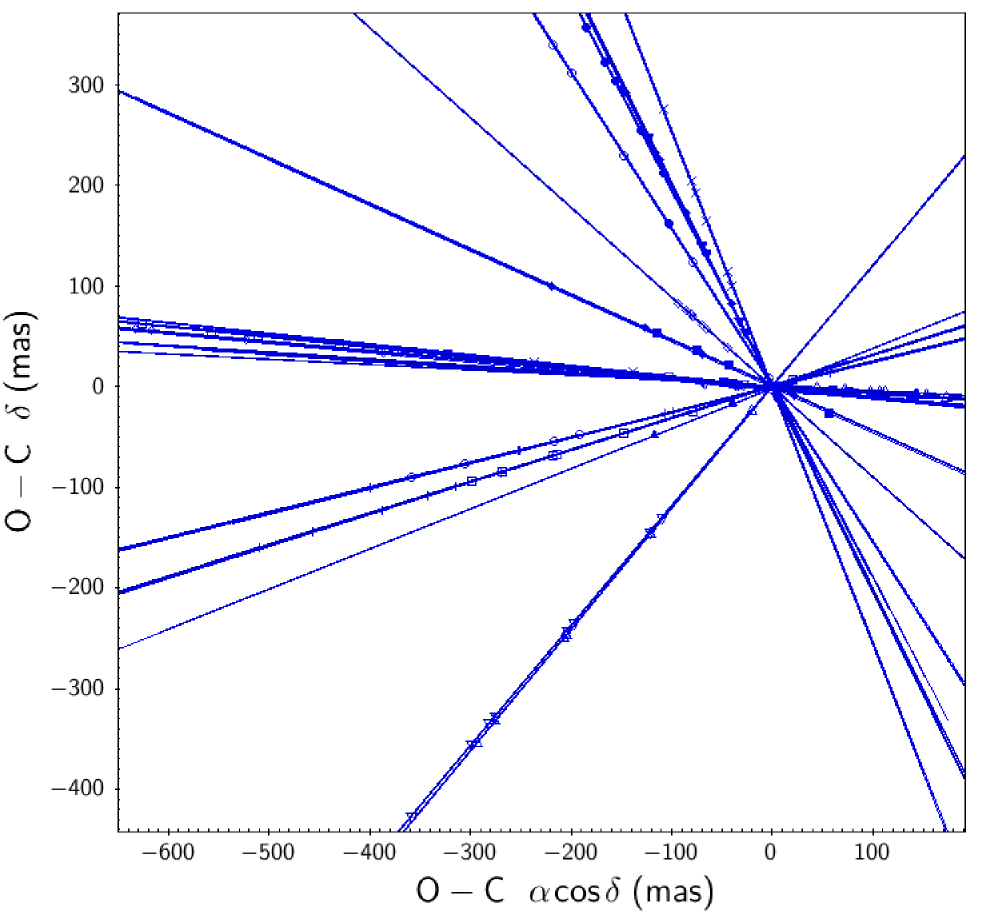}
\caption{Residuals in $\alpha\cos\delta$ and $\delta$ for the MBA
  (367) Amicitia. Different symbols correspond to different
  transits of the object. The lines represent the error ellipse for
  each observation, including the correlations, as given by the error
  model.}
\label{F:367_res_ra_dec_ellipse}
\end{figure}

The error ellipses associated with each observation, and thus with
each residual, are shown in Fig.~\ref{F:367_res_ra_dec_ellipse}. They
are very elongated as a result of the large errors in AC, and they are not
parallel to the axes because of the orientation of the \gaia scanning
law, which changes from one transit to the next.

\begin{figure}[h!]
\centering
\includegraphics[clip=true, trim = 0mm 0mm 0mm 0mm, width=1.0 \hsize]{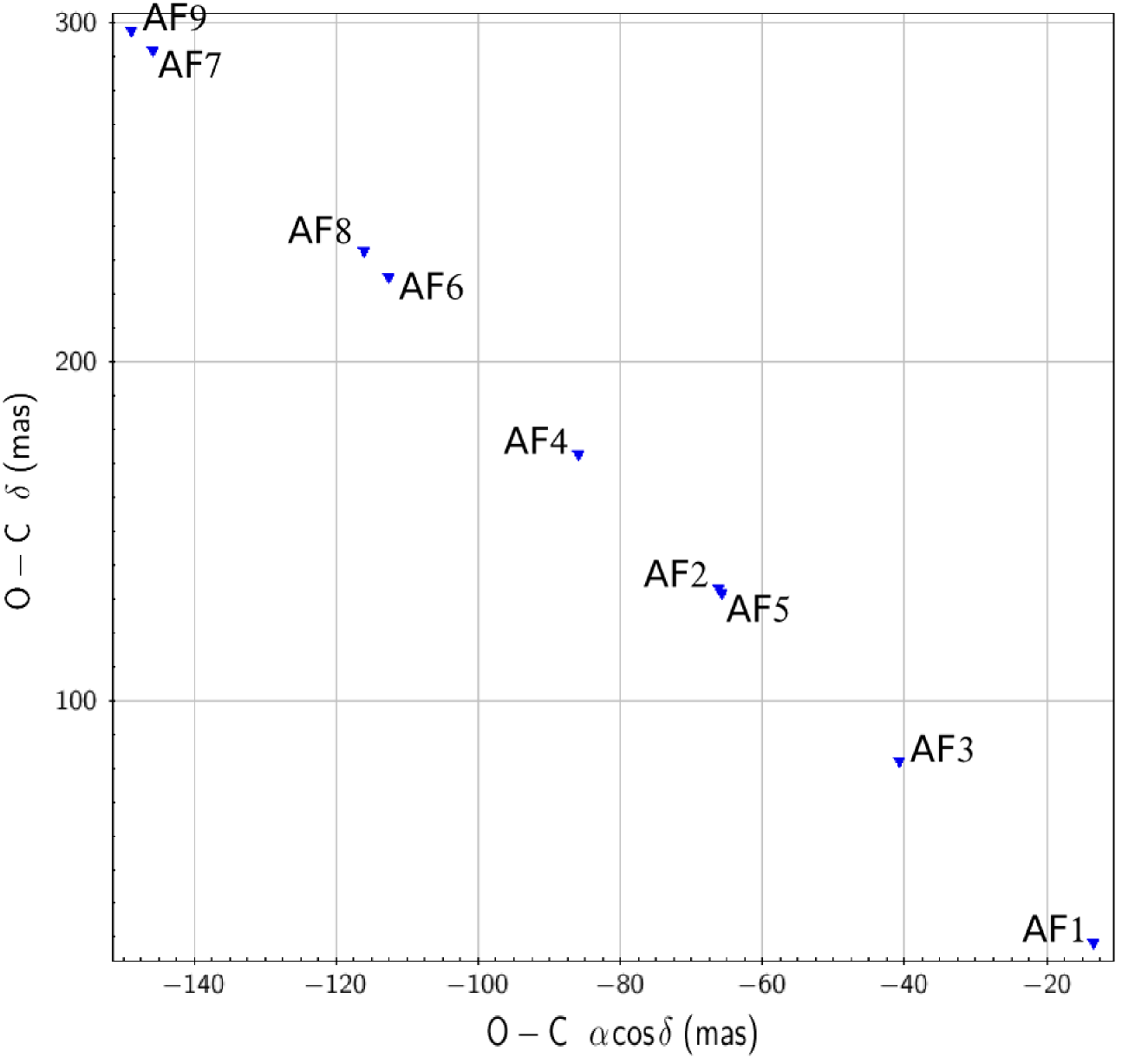}
\caption{Residuals in $\alpha\cos\delta$ and $\delta$ of one transit
  of the MBA (367) Amicitia. Each astrometric field (AF) is
  highlighted.}
\label{F:367_res_ra_dec_dispersion}
\end{figure}

To explain the distance of each point from the centre $(0,0)$ of the
residuals for each transit, we illustrate the detailed properties of a
single transit (Fig.~\ref{F:367_res_ra_dec_dispersion}).
To this end, we consider a single transit. We chose one of the
transits for which none of the nine CCD observations (AF1-AF9) was rejected
during the processing (Fig.~\ref{F:367_res_ra_dec_dispersion}). In
this case, the asteroid motion was not fast enough to move
it outside the window during the transit.

The main striking feature is the alignment of the residuals. The
direction of maximum spread corresponds to the motion across
scan, as oriented during the transit. The AC motion of the SSOs is due
both to the slow precession of the spin axis of \textit{Gaia} as imposed by the
NSL and to the proper motion of the asteroid.

Another important feature is the order of the AF CCDs, provided by
their numbering, in the sequence of residuals. While on average the
drift proceeds in the figure from AF1 at the bottom right towards AF9
at top left, the sequence is sometimes inverted in direction. This is
an effect of the quantisation of the window position AC at integer
steps of one full sample size.

The compact clustering of the residuals shows that the in the AL
direction, the determination of the position during the transit has a
very low spread, as expected thanks to the much higher accuracy.

\begin{figure}[h!]
\centering
\includegraphics[clip=true, trim = 0mm 0mm 0mm 0mm, width=1.0 \hsize]{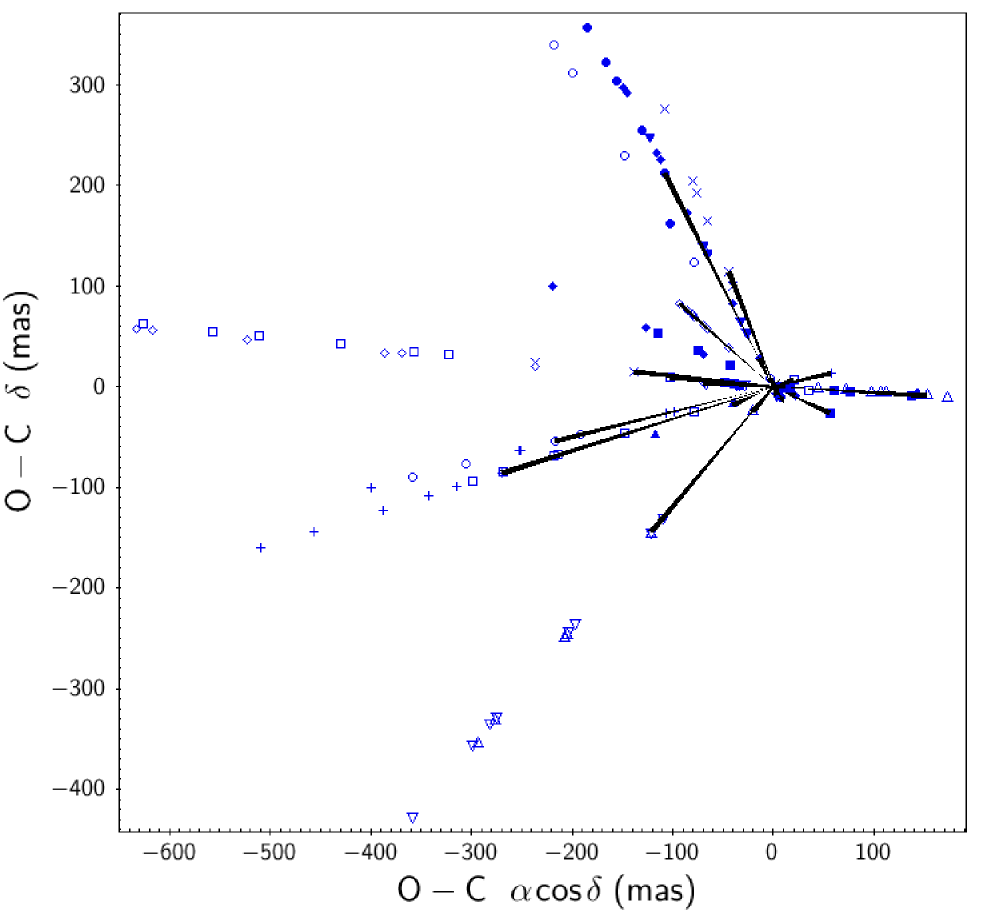}
\caption{Residuals in the ($\alpha\cos\delta$, $\delta$) plane and
  projection of the AC velocity (black lines) on the same
  plane for the MBA \object{(367) Amicitia}.}
\label{F:367_res_ra_dec_transits_velocity_vectors}
\end{figure}

The last point we need to analyse is the trend of the residuals in the
negative side of the Cartesian plane. The black vectors in
Fig.~\ref{F:367_res_ra_dec_transits_velocity_vectors} are the
projections of the AC velocity in the plane described by
$\alpha\cos\delta$ and $\delta$. The residuals follow the direction of
the velocity vectors, which explain the main dispersion of the
detections in each CCD across scan.

\subsection{Transit-level residuals}
\label{S:transit_analysis}
The observations collected during a single transit extend over a
limited period of time during which the motion of the SSOs can be
taken as linear, and over this interval its position changes by less
than 1 arcsec.  Likewise, the scan direction is $\text{approximately
}$constant at first
order. The same applies in general to the SSO orientation in space
(rotational phase, direction of the pole) and as a consequence of its
brightness.

In addition, successive transits are well separated in time, with a minimum
interval of 106 minutes (preceding the following FOV) to 6 hours (one
rotation) or much more (days, months) before the \textit{Gaia} pointing returns
to the same SSO. Since a single transit can be considered as a
coherent unit that is well separated from the others, different transits are
clearly statistically independent measurements. This is also supported
by the time resolution of the attitude solution, with nodes spaced by
5 s in AL but with a much longer coherence time. On the other hand, one
can expect that during a transit the attitude from one CCD to the next
is significantly correlated, while it is not correlated over longer
timescales.

Some systematics that can be related to the asteroid itself (such as
motion, apparent size, and photocentre position relative to centre of
mass) are also expected to be correlated during a transit, but
they are completely different in general when different transits
are compared.

For these reasons, it makes sense to group single
observations within a transit and analyse  the residuals at transit
level. We consider first the average of the residual values during a
transit and their scatter separately.

\begin{figure}[h!]
\centering
\includegraphics[clip=true, trim = 0mm 0mm 0mm 0mm, width=1.0 \hsize]{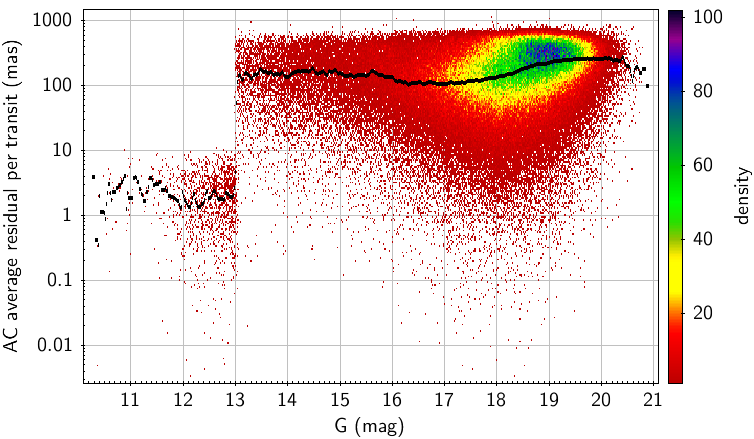}
\includegraphics[clip=true, trim = 0mm 0mm 0mm 0mm, width=1.0 \hsize]{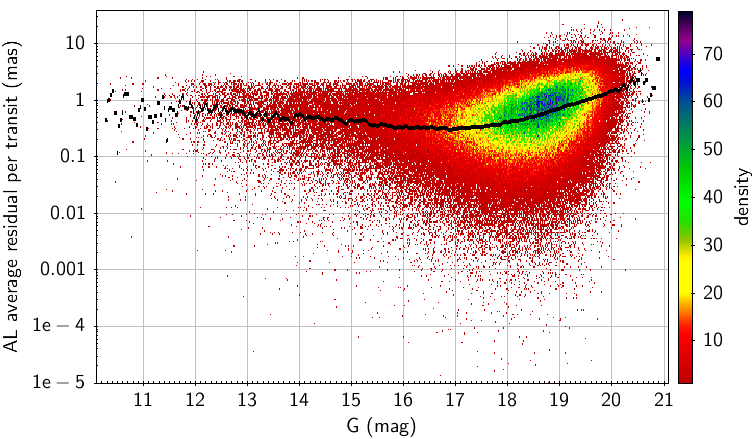}
\caption{Systematic component of transit-level residuals estimated as
  the absolute value of the average of post-fit residuals associated
  with a single position during each transit as a function of $G$
  magnitude. Transit-level residuals in AC and AL are shown in
  the top and bottom panel, respectively. The black line represents the average
  value. The transition in the AC direction at G$\sim$13 is due to the
  change of window dimension. The colour scale represents the local density of data points.}
\label{F:ave_transit_G}
\end{figure}

The average is an analogue of accuracy, as it is expected to
provide the overall discrepancy of the SSO position with respect to
the fitted orbital solution around the transit epoch.
The result of the transit-level average of residuals is shown in
Fig.~\ref{F:ave_transit_G}.  The large residuals associated with the AC
direction are due to the AF lack of resolution, with the exception of
bright (G$<$13) targets. The plot in AL is much more impressive, as it
catches the full accuracy of \gdrtwo\ data for SSOs, with an average
well below 1~mas for all sources G<19.5. The best performance appears
to be around G$\sim$17.

\begin{figure}[h!]
\centering
\includegraphics[clip=true, trim = 0mm 0mm 0mm 0mm, width=1.0 \hsize]{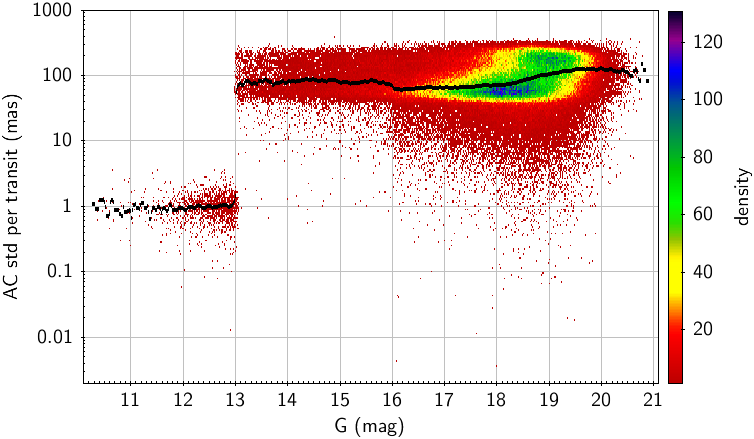}
\includegraphics[clip=true, trim = 0mm 0mm 0mm 0mm, width=1.0 \hsize]{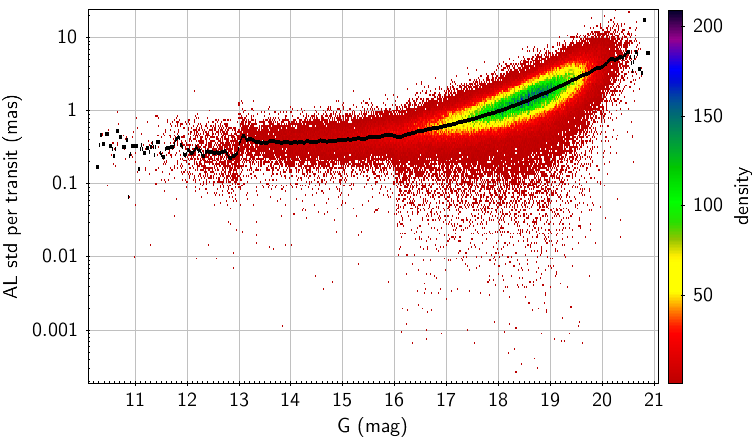}
\caption{Random component of transit-level residuals estimated as the
  standard deviation of the post-fit residuals associated with a single
  position, during each transit (top panel: AC; bottom panel: AL) as a
  function of $G$ magnitude. Only transits with more than four positions
  are considered. The black line represents the average value. The colour scale represents the local density of data points.}
\label{F:std_transit_G}
\end{figure}

The scatter of the residuals during a transit
(Fig.~\ref{F:std_transit_G}) can be considered as an indication of the
transit-level astrometric precision.  We compute this precision as a
standard deviation. Of course, given the small number of data points
(at most nine per transit), this is a rather poor statistical estimator of
the true standard deviation of the population. We show here the
results obtained with transits that had four or more observations, but
even without this cut, the general picture holds true.

The AL scatter component has a remarkable minimum at G$<$16,
around 400~$\mu$as. A transition at G$<$13 is a clear signature of the
change in window size. The scatter reaches $\sim$5~mas at G=20.

\begin{figure}[h!]
\centering
\includegraphics[clip=true, trim = 0mm 0mm 0mm 0mm, width=1.0 \hsize]{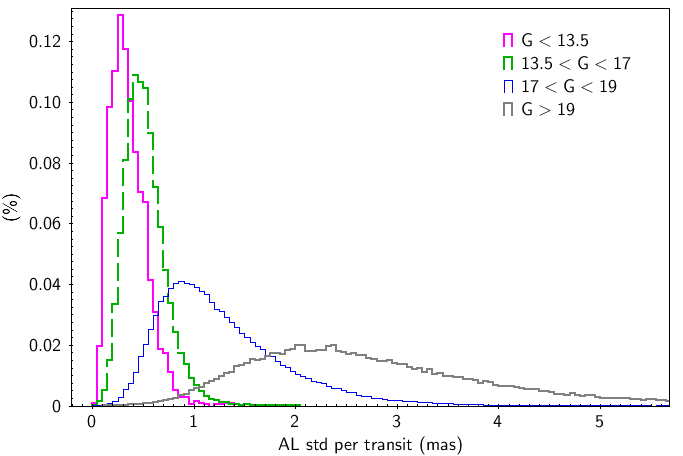}
\caption{Scattering of the transit-level residuals in
  Fig.~\ref{F:std_transit_G} for four magnitude ranges.}
\label{F:hist_AL_std}
\end{figure}

As the core of the distribution is very dense in the plots, we
collected in a set of histograms (Fig.~\ref{F:hist_AL_std}) the
distribution of the scatter for four ranges of $G$ to better illustrate
the difference in performance.

A common feature of both systematic and random residuals is the lack
of improvement for SSOs brighter than G$\sim$16. The systematic
component clearly shows a degradation. This can be due to several
overlapping effects:

\begin{itemize}
\item The apparent size of the asteroid increases with
  brightness. Although in general it remains  below the pixel size, it
  can introduce a bias in the centroid position, as no special
  treatment is applied in \gdrtwo\ to extended SSOs.
\item The difference between the centre of mass
  and the photocentre can introduce a further systematic effect.
\item The various thresholds of gating of the CCD to avoid saturation
  and the associated photon loss at G$<$13 also tend to suppress a
  gain in accuracy.
\end{itemize}

To these factors, one must also add the signal smearing that
is due to the
apparent motion of SSOs relative to the stars. While it does not
depend on brightness, it can add a noise floor to the whole
distribution of residuals.

\begin{figure}[h!]
\centering
\includegraphics[clip=true, trim = 0mm 0mm 0mm 0mm, width=1.0 \hsize]{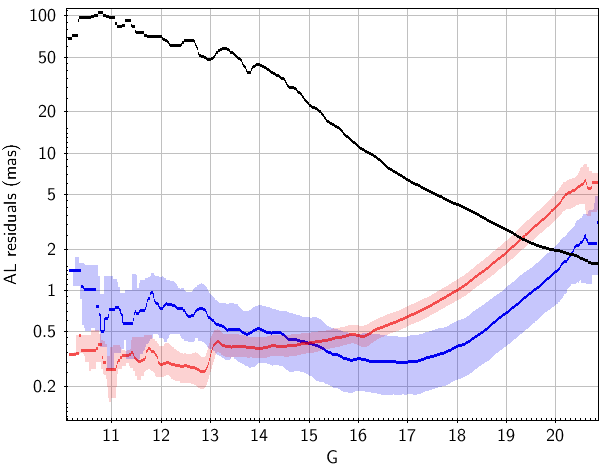}
\caption{Average and one-sigma range for the residuals in
  Fig.~\ref{F:std_transit_G} (red) and \ref{F:ave_transit_G}
  (blue). The black line is the average apparent size of the SSOs.}
\label{F:ave_res_size}
\end{figure}

Fig.~\ref{F:ave_res_size} summarises the two residual components and
compares them to the average apparent size of the observed
asteroids. This is computed by assuming a spherical shape and by
taking into account the distance of the SSO from \textit{Gaia} at the mean
epoch of the FOV transit. The physical radius used is provided
by the Wide--field Infrared Survey Explorer (WISE) telescope
\citep{Mainzer16}. In \gdrtwo, $11,984$ SSOs have a WISE size
determination. As size here is just for statistical comparison, errors
and biases on the WISE sizes are of no consequence.

The trend of the size shows that its median value is higher than any
residual for objects brighter than G$\sim$19 and reaches the AL pixel
size of (60~mas) at G$\sim$13. As the centroiding algorithm used in
\gdrtwo\ is only optimised for the stars (point sources), some
degradations from the image extension can show up, which is a likely
cause of the increase in the systematic residuals. More detailed
investigations are required and will provide indications to further
improve the astrometric quality in the next releases.

\begin{figure}[h!]
\centering
\includegraphics[clip=true, trim = 0mm 0mm 0mm 0mm, width=1.0 \hsize]{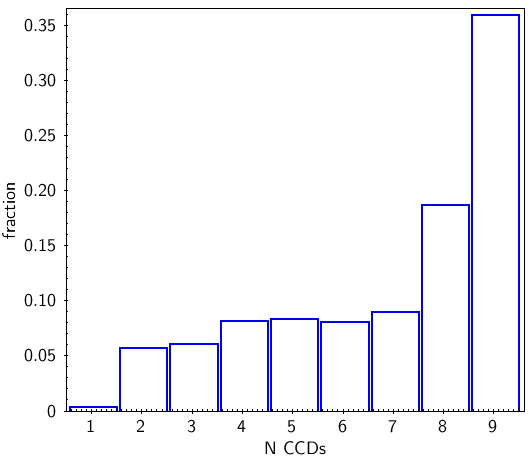}
\caption{Distribution of the number of single CCD positions.}
\label{F:hist_n_CCDs}
\end{figure}

The effect of motion can be roughly assimilated to an increased size,
as a signal elongation occurs in the direction of displacement. If the
hypothesis above on the role of size is valid, a signature should also
be found as a dependency of residuals from motion. However, the
situation is also more complex, as a displacement of the image with
respect to the window centre is expected, with an asymmetric loss of
signal from the PSF tails. For fast SSOs, the displacement can move
the signal outside the CCD window, and in this situation, the number of
valid observations per transit decreases.
Fig.~\ref{F:hist_n_CCDs} shows the distribution of the number of single CCD measurements
per transit and illustrates the reduction of the number of valid
observations due to displacement of the signal out of the allocated windows. 

We found no clear evidence of a difference between centre--of--mass and photocentre. Although the phase angle of \textit{Gaia}
observations is rather high, the typical photocentre shift can reach
at most a few $\text{percent}$ of the SSO diameter, but its direction can have any
orientation with respect to the AL position angle. When we select
asteroids with similar astrometric accuracy (i.e. similar $G$ within a
one-magnitude interval, for instance), no clear trend of the average
residuals with respect to phase angle is found. The effect can
be considered of second order with respect to the other uncertainty
sources illustrated above.

\section{Analysis of the orbits}
\label{S:orbits}

\subsection{Comparisons with existing orbital data sets.}

\begin{figure}[h!]
\centering
\includegraphics[clip=true, trim = 0mm 0mm 0mm 0mm, width=1.0 \hsize]{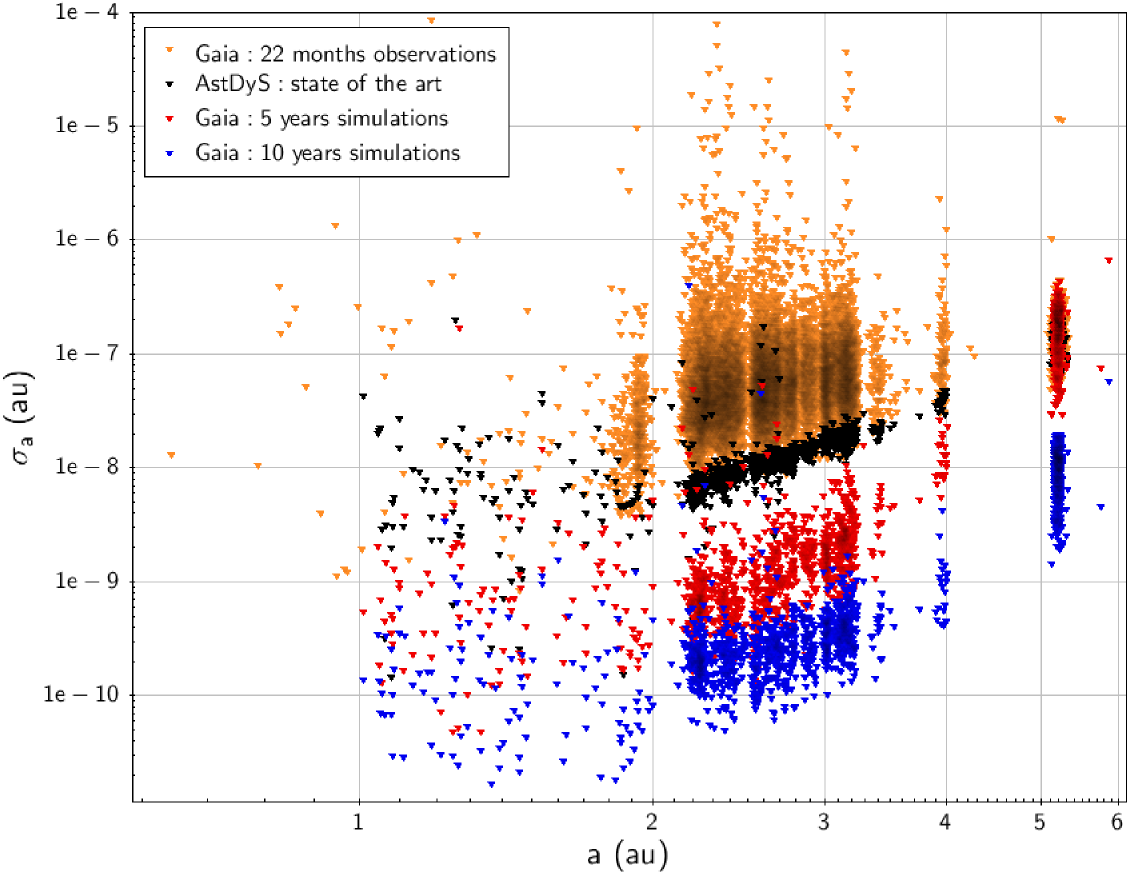}
\caption{Quality of the orbit determination measured by the post-fit
  uncertainty of the semi-major axis for the whole sample of objects
  contained in \gdrtwo\ after the nominal 5 years (red points)
  and 10 years (blue points) of mission. Black points represent the
  current situation (initial orbits taken from AstDyS). The orange
  points are the post-fit uncertainties using only 22 months of \textit{Gaia}
  observations available for the \gdrtwo.}
\label{F:a_sigmaa_all}
\end{figure}

The orbit fitting discloses interesting properties of \gaia asteroid
observations. Figure~\ref{F:a_sigmaa_all} shows the results of the
current situation (22 months of \textit{Gaia}) obtained as output of the orbit
determination process and what we may expect at the end of the
nominal mission (5 years), or of a possible extension (10 years), based
on our simulations using the same error model.

The data represent the uncertainty in the semi-major axis (a good and
easy to determine indicator of the orbit quality) as a function of
the semi-major axis $a$ (in au). Red and blue points are the results
of the simulation after the nominal 5-  or 10-year mission. Black
points are the current uncertainty from the AstDyS website. The orange
points represent the accuracy of the orbits resulting from the
processing of the 22-month \gdrtwo\ data.

The simulations show an improvement of almost a factor
10 in the orbit determination after 5 years of mission, except for
Jupiter trojans, which indeed have a period of 11 years. Thus after 10
years of mission, the improvement is clearly visible not only for main-belt asteroids (a factor 20), but also for the trojans. On the other
hand, the \gdrtwo\ short time-span of 22 months (compared to
tens/hundreds years for AstDyS, and 5 or 10 years for \textit{Gaia} itself)
represents a limitation on the expected quality of the orbits. However,
some asteroids observed by \gaia already reach a quality equivalent to
ground-based data (and in 350 cases, it is even better).

\begin{figure}[h!]
\centering
\includegraphics[clip=true, trim = 0mm 0mm 0mm 0mm, width=1.0 \hsize]{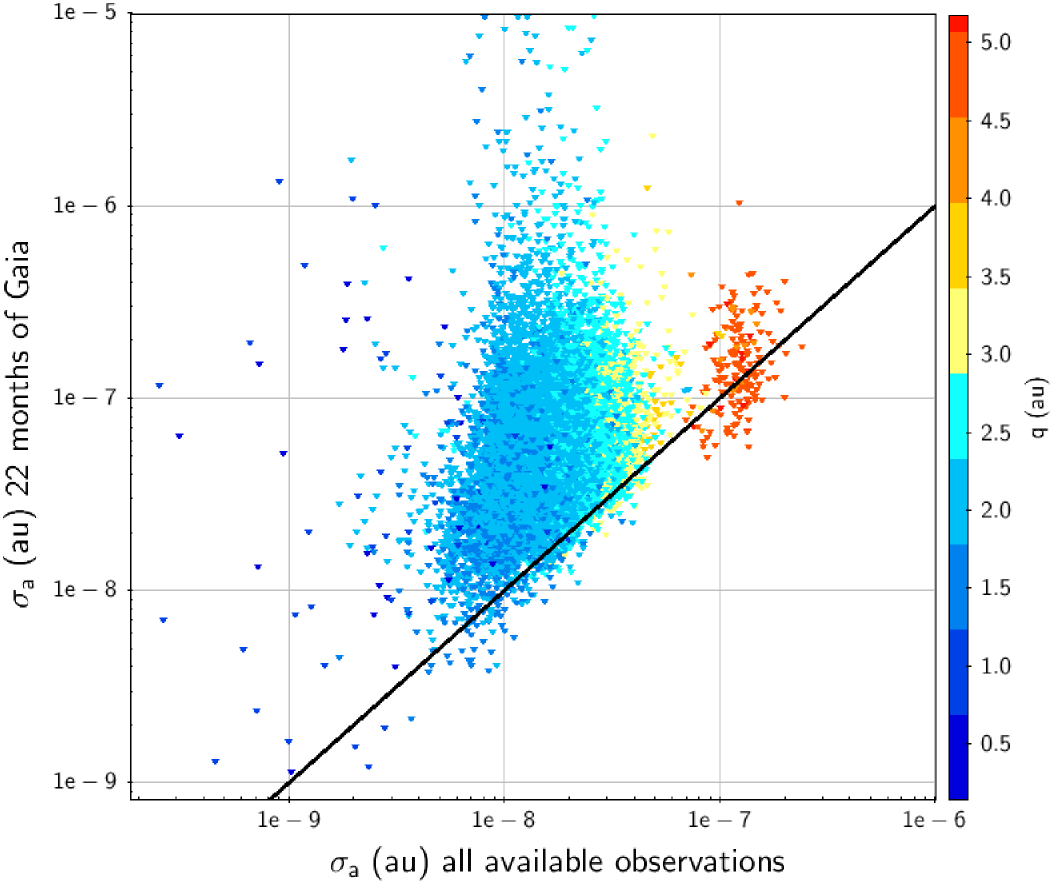}
\caption{Quality of the orbit determination measured by the post-fit
  uncertainty of the semi-major axis for the whole sample of objects
  contained in \gdrtwo\ with respect to the current measurements
  (from the AstDys website).}
\label{F:a_sigmaa_gaia_all}
\end{figure}

Figure~\ref{F:a_sigmaa_gaia_all} shows the uncertainty on the
semi-major axis obtained as a result of the orbit determination
process using only \textit{Gaia} observations, with respect to the current
uncertainty from the orbits available on the AstDyS website as
function of the perihelion distance (q). The different clouds of
points represent different classes of objects in the solar system
population: NEAs (dark blue), MBAs (different shades of light blue),
Hildas, and trojans (orange and red). The black line is the bisector of
the first quadrant: the orbits of the objects below the line have a
better estimate using only \textit{Gaia} observations. Among these objects
are also five NEAs, namely \object{(3554) Amun}, \object{(4957) BruceMurray}, \object{(10563)
Izhdubar}, \object{(12538) 1998~OH}, and \object{(161989) Cacus}.

\subsection{Perspectives: detection of non-gravitational perturbations.}

The unprecedented quality of \gaia SSO observations, demonstrated
above, opens new perspectives for study. These include the
computation of subtle non-gravitational effects.

The most important of these is the so-called Yarkovsky effect, a
small recoil force that is due to the emission of photons in the thermal
infrared from the surface of SSOs. The consequence of this emission
is a drift in the semi-major axis of asteroids, which changes their
orbits in the long term~\citep{vok2000}.

The Yarkovsky effect depends on numerous physical quantities, such as
the density and the thermal inertia of the asteroids. Usually, they
are unknown and estimated on the base of reasonable guesses. When
very accurate astrometry is available, the Yarkovsky drift can be measured directly, but this remains a challenge today. While the
Yarkovsky effect has previously been measured for about 90
NEAs~\citep{farnocchia13,delvigna18}, it has never been measured for
MBAs because their distances are larger and their temperatures and
surface properties are different. Moreover, NEA are smaller objects (the Yaskovsky
intensity decreases with size), which are close enough to enable precise
astrometry by radar ranging techniques.

\gaia is the key to change the current situation. At present,
\gdrtwo\ SSO observations are affected by a limitation through their
short time span, 22 months, which in most cases does not even cover
the whole orbit of an object. Therefore they cannot be used alone to
detect small perturbations such as the Yarkovsky effect, but need to
be combined with available ground-based observations.

The joint use of \textit{Gaia} and ground-based astrometry is possible, but because the accuracies are very different (a few milliarcseconds for \gaia astrometry
versus a few hundred milliarcsecond for most ground-based
observations), adequate techniques are required to reduce
the signature of zonal errors on SSO astrometry due to pre-\gaia
stellar catalogues. An improved error model is required as well. These
complex tasks are beyond the purpose of this article.

In some exceptional situations, however, ground-based data are so good
that the joint use of \gdrtwo\ astrometry becomes possible even
without additional optimisations. We therefore selected an NEA for which the
Yarkovsky effect has previously been measured in \citet{delvigna18},
namely \object{(2062) Aten}, and directly combined 69 \gaia astrometric
observations to the 959 pre-existing ground-based optical measurements
and 7 measurements by radar, from 1955 to 2017.

\begin{table}[h!]
\caption{Asteroid name and number, value of the $A_2$ with its
  uncertainty, S/N (obtained as
  $A_2/\sigma_{A_2}$), value of $da/dt$ with its uncertainties, and the
  corresponding references for this result.}
\label{T:yarko_compare}
\resizebox{\columnwidth}{!}{\begin{tabular}{l|c|c|c|l}
\hline
\textbf{Asteroid} & $\mathbf{A_2}$ & $\mathbf{S/N_{A2}}$ & $\mathbf{da/dt}$ & \textbf{Reference}\\
Number and name   &  $10^{-15} au/d^2$ &             & $10^{-4} au/Myr$  \\
\hline
(2062) Aten       & $-10.97 \pm 0.75$ & 14.63       & $-4.90 \pm 0.34$ & This article\\
\hline
(2062) Aten       & $-13.18 \pm 1.53$ & \phantom{0}8.64        & $-5.98 \pm 0.68$ & \citet{delvigna18}\\
\hline
\end{tabular}}
\end{table}

The Yarkovsky effect can be modelled as a transverse acceleration
depending on one parameter, the so-called $A_2$, which can be
determined in an orbital fit together with the six orbital
elements~\citep{vok2000} and then converted into $da/dt$ (the
variation of the semi-major axis due to the non-gravitational
force). Using this approach, we then compared our results with
the previously published result. Table~\ref{T:yarko_compare} shows that the
use of \gdrtwo\ observations not only changes the value significantly,
but also improves the signal-to-noise ratio (S/N) of the detection and its uncertainty.

We wish to stress that even though it is meaningful, this is only a
preliminary result that can certainly be improved further by reducing
biases (zonal errors) and introducing a better error model for the
non-\gaia observations.
The systematic exploitation of \gaia astrometry in this domain appears
to be very promising.

\section{Conclusions}
\gdrtwo\ contains SSO astrometry with an accuracy better than
2-5~mas for the faintest asteroids (around G$\sim$20.5) and reaches
sub-milliarcseond level for objects with G<17.5, as estimated from post-fit
residuals. This accuracy is essentially 1D in the direction of the
\textit{Gaia} scan, but an appropriate orbital fitting procedure shows that the
strong footprint of the orientation vanished when several transits,
and therefore different scan directions were combined. This
situation is similar to the one encountered for the astrometric solution
of stars whose parameters are optimally constrained when data
obtained during several scans in different directions are combined to
obtain precise 2D locations. For SSOs, the location is subject to great
changes at each epoch, but despite the one-dimensionality of the \textit{Gaia}
astrometric data, a unique orbital solution can be efficiently
adjusted to the trajectory on the sky, and it provides residuals close
to the expectations.

We find that the apparent size of the asteroids can indeed play a role
in the deterioration of the astrometric accuracy for the brightest
targets. The apparent size for bright SSOs (G<16) increases well
beyond the accuracy of the measurements, and it even appears to mask the effect of velocity to some extent.

Although more careful investigations are required, second-order effect
such as the displacement of the photo-centre with respect to the
centre of mass as a function of the phase angle does not seem to be
detectable in the bulk distribution of residuals. This does not
necessarily imply that their signature is entirely absent, as the
orbital fit could partially absorb and compensate it.

The SSO data in \gdrtwo\ clearly have some limitations that will be
corrected in the future releases. These include corrections to the
astrometry as a function of the object motion, size and shape, and
better instrumental calibration. The bright end is also expected
to be much
improved.

Despite these limitations, the \textit{Gaia} absolute astrometry of SSOs is clearly
the best ever obtained at optical wavelengths. Only well-observed stellar occultations can compete, but with the important
limitation that they provide only the relative positions of an SSO and
the occulted star.

\input{acknow}

\bibliographystyle{aa} 
\bibliography{refs} 

\appendix
\section{Orbit determination process: perturbing asteroids}
\label{A:perturbing}
In the orbit determination process, in addition to the Sun, the eight
planets, and the Moon, we also considered the perturbations due to
the 16 massive asteroids and Pluto (Table~\ref{T:perturbing}).
\begin{table}[h!]
  \caption{Perturbing bodies included in the dynamical model in the
    orbit determination process. The table contains the asteroid
    number and name and the corresponding mass and reference}
  \label{T:perturbing}
  \begin{tabular}{lcl}
    \hline
    \textbf{Asteroid name} & \textbf{Grav. mass} & \textbf{References}\\
                           &  (km$^3$/s$^2$) & \\
    \hline
    (1) Ceres & \phantom{0}63.20 & \citet{standish1984}\\
    (2) Palla & \phantom{0}14.30 & \citet{standish1984}\\
    (3) Juno & \phantom{00}1.98 & \citet{konopliv2011}\\
    (4) Vesta & \phantom{0}17.80 & \citet{standish1984}\\
    (6) Hebe & \phantom{00}0.93 & \citet{Carry2012}\\
    (7) Iris & \phantom{00}0.86 & \citet{Carry2012}\\
    (10) Hygiea & \phantom{00}5.78 & \citet{baer2011}\\
    (15) Eunomia & \phantom{00}2.10 & \citet{Carry2012}\\
    (16) Psyche & \phantom{00}1.81 & \citet{Carry2012}\\
    (29) Amphitrite & \phantom{00}0.86 & \citet{Carry2012}\\
    (52) Europa & \phantom{00}1.59 & \citet{Carry2012}\\
    (65) Cybele & \phantom{00}0.91 & \citet{Carry2012}\\
    (87) Sylvia & \phantom{00}0.99 & \citet{Carry2012}\\
    (88) Thisbe & \phantom{00}1.02 & \citet{Carry2012}\\
    (511) Davida & \phantom{00}2.26 & \citet{Carry2012}\\
    (704) Interamnia & \phantom{00}2.19 & \citet{Carry2012}\\
    (134340) Pluto & 977.00 & \citet{folkner2014}\\
    \hline
  \end{tabular}
\end{table}

\section{Example of queries to the ESA Archive for SSO tables.}
All the SSO data are made available through the ESA Archive
\url{https://gea.esac.esa.int/archive/}. We here provide some
examples of queries to \gdrtwo\ tables concerning asteroids.\\

\vspace{2mm}
This query calls the whole list of SSOs published in \gdrtwo, with the number and
name/provisional designation as in the Minor Planet Center website
(\url{https://www.minorplanetcenter.net/db_search}):
\begin{verbatim}
SELECT number_mp, denomination 
FROM gaiadr2.sso_source
\end{verbatim}

\vspace{2mm}
The following query selects epoch, right ascension $\alpha,$ and
declination $\delta$ for a given SSO, ordered by their observation time.
In this case, we have chosen the asteroid (8) Flora, but any asteroid can be selected from the list of objects published in \gdrtwo.
\begin{verbatim}
SELECT epoch, ra, dec  
FROM gaiadr2.sso_observation 
WHERE number_mp = 8 ORDER BY epoch
\end{verbatim}

\vspace{2mm}
The query counts the number of NEAs in \gdrtwo:
\begin{verbatim}
SELECT COUNT(number_mp)
FROM user_dr2int6.aux_sso_orbits
WHERE a*(1-eccentricity)<1.3
\end{verbatim}

\vspace{2mm} 
This query calls the observations (epoch, $\alpha$,
$delta$) and the residuals in AL and AC for a given object (in this
case, we have chosen the NEA (2062) Aten):
\begin{verbatim}
SELECT obs.epoch, obs.ra, obs.dec, 
       res.residual_al, res.residual_ac
FROM user_dr2int6.sso_observation AS obs
JOIN user_dr2int6.aux_sso_orbit_residuals AS res
USING(observation_id)
WHERE obs.number_mp = 2062
\end{verbatim}

\vspace{2mm} 
This query selects all the asteroids with the
corresponding values of $G$ magnitude, observations and residuals in AL
and AC when the $G$ magnitude is between 13 and 17.
\begin{verbatim}
SELECT obs.number_mp, obs.epoch, 
       obs.ra, obs.dec, obs.g_mag, 
       res.residual_al, res.residual_ac
FROM user_dr2int6.sso_observation AS obs
JOIN user_dr2int6.aux_sso_orbit_residuals AS res
USING(observation_id)
WHERE obs.g_mag BETWEEN 13 AND 17
ORDER BY obs.number_mp
\end{verbatim}

\vspace{2mm} 
This query selects all the observations for a given asteroid (in this
case, (8) Flora) with a non-null magnitude value.
\begin{verbatim}
SELECT obs.epoch, obs.ra, obs.dec, obs.g_mag
FROM user_dr2int6.sso_observation AS obs
WHERE obs.g_mag>0 AND obs.number_mp = 8
ORDER BY obs.epoch
\end{verbatim}

\listofobjects

\end{document}

%% file: acknow.tex
\section*{Acknowledgements}

This work presents results from the European Space Agency (ESA) space
mission \gaia. \gaia\ data are being processed by the \gaia\ Data
Processing and Analysis Consortium (DPAC). Funding for the DPAC is
provided by national institutions, in particular the institutions
participating in the \gaia\ MultiLateral Agreement (MLA). The
\gaia\ mission website is \url{https://www.cosmos.esa.int/gaia}. The
\gaia\ archive website is \url{https://archives.esac.esa.int/gaia}.

The \gaia\ mission and data processing have financially been supported by, in alphabetical order by country:
\begin{itemize}
\item the Algerian Centre de Recherche en Astronomie, Astrophysique et G\'{e}ophysique of Bouzareah Observatory;
\item the Austrian Fonds zur F\"{o}rderung der wissenschaftlichen Forschung (FWF) Hertha Firnberg Programme through grants T359, P20046, and P23737;
\item the BELgian federal Science Policy Office (BELSPO) through various PROgramme de D\'eveloppement d'Exp\'eriences scientifiques (PRODEX) grants and the Polish Academy of Sciences - Fonds Wetenschappelijk Onderzoek through grant VS.091.16N;
\item the Brazil-France exchange programmes Funda\c{c}\~{a}o de Amparo \`{a} Pesquisa do Estado de S\~{a}o Paulo (FAPESP) and Coordena\c{c}\~{a}o de Aperfeicoamento de Pessoal de N\'{\i}vel Superior (CAPES) - Comit\'{e} Fran\c{c}ais d'Evaluation de la Coop\'{e}ration Universitaire et Scientifique avec le Br\'{e}sil (COFECUB);
\item the Chilean Direcci\'{o}n de Gesti\'{o}n de la Investigaci\'{o}n (DGI) at the University of Antofagasta and the Comit\'e Mixto ESO-Chile;
\item the National Science Foundation of China (NSFC) through grants 11573054 and 11703065;  
\item the Czech-Republic Ministry of Education, Youth, and Sports through grant LG 15010, the Czech Space Office through ESA PECS contract 98058, and Charles University Prague through grant PRIMUS/SCI/17;    
\item the Danish Ministry of Science;
\item the Estonian Ministry of Education and Research through grant IUT40-1;
\item the European Commission’s Sixth Framework Programme through the European Leadership in Space Astrometry (\href{https://www.cosmos.esa.int/web/gaia/elsa-rtn-programme}{ELSA}) Marie Curie Research Training Network (MRTN-CT-2006-033481), through Marie Curie project PIOF-GA-2009-255267 (Space AsteroSeismology \& RR Lyrae stars, SAS-RRL), and through a Marie Curie Transfer-of-Knowledge (ToK) fellowship (MTKD-CT-2004-014188); the European Commission's Seventh Framework Programme through grant FP7-606740 (FP7-SPACE-2013-1) for the \gaia\ European Network for Improved data User Services (\href{http://genius-euproject.eu/}{GENIUS}) and through grant 264895 for the \gaia\ Research for European Astronomy Training (\href{https://www.cosmos.esa.int/web/gaia/great-programme}{GREAT-ITN}) network;
\item the European Research Council (ERC) through grants 320360 and 647208 and through the European Union’s Horizon 2020 research and innovation programme through grants 670519 (Mixing and Angular Momentum tranSport of massIvE stars -- MAMSIE) and 687378 (Small Bodies: Near and Far);
\item the European Science Foundation (ESF), in the framework of the \gaia\ Research for European Astronomy Training Research Network Programme (\href{https://www.cosmos.esa.int/web/gaia/great-programme}{GREAT-ESF});
\item the European Space Agency (ESA) in the framework of the \gaia\ project, through the Plan for European Cooperating States (PECS) programme through grants for Slovenia, through contracts C98090 and 4000106398/12/NL/KML for Hungary, and through contract 4000115263/15/NL/IB for Germany;
\item the European Union (EU) through a European Regional Development Fund (ERDF) for Galicia, Spain;    
\item the Academy of Finland and the Magnus Ehrnrooth Foundation;
\item the French Centre National de la Recherche Scientifique (CNRS) through action 'D\'efi MASTODONS', the Centre National d'Etudes Spatiales (CNES), the L'Agence Nationale de la Recherche (ANR) 'Investissements d'avenir' Initiatives D’EXcellence (IDEX) programme Paris Sciences et Lettres (PSL$\ast$) through grant ANR-10-IDEX-0001-02, the ANR 'D\'{e}fi de tous les savoirs' (DS10) programme through grant ANR-15-CE31-0007 for project 'Modelling the Milky Way in the Gaia era' (MOD4Gaia), the R\'egion Aquitaine, the Universit\'e de Bordeaux, and the Utinam Institute of the Universit\'e de Franche-Comt\'e, supported by the R\'egion de Franche-Comt\'e and the Institut des Sciences de l'Univers (INSU);
\item the German Aerospace Agency (Deutsches Zentrum f\"{u}r Luft- und Raumfahrt e.V., DLR) through grants 50QG0501, 50QG0601, 50QG0602, 50QG0701, 50QG0901, 50QG1001, 50QG1101, 50QG1401, 50QG1402, 50QG1403, and 50QG1404 and the Centre for Information Services and High Performance Computing (ZIH) at the Technische Universit\"{a}t (TU) Dresden for generous allocations of computer time;
\item the Hungarian Academy of Sciences through the Lend\"ulet Programme LP2014-17 and the J\'anos Bolyai Research Scholarship (L.~Moln\'ar and E.~Plachy) and the Hungarian National Research, Development, and Innovation Office through grants NKFIH K-115709, PD-116175, and PD-121203;
\item the Science Foundation Ireland (SFI) through a Royal Society - SFI University Research Fellowship (M.~Fraser);
\item the Israel Science Foundation (ISF) through grant 848/16;
\item the Agenzia Spaziale Italiana (ASI) through contracts I/037/08/0, I/058/10/0, 2014-025-R.0, and 2014-025-R.1.2015 to the Italian Istituto Nazionale di Astrofisica (INAF), contract 2014-049-R.0/1/2 to INAF dedicated to the Space Science Data Centre (SSDC, formerly known as the ASI Sciece Data Centre, ASDC), and contracts I/008/10/0, 2013/030/I.0, 2013-030-I.0.1-2015, and 2016-17-I.0 to the Aerospace Logistics Technology Engineering Company (ALTEC S.p.A.), and INAF;
\item the Netherlands Organisation for Scientific Research (NWO) through grant NWO-M-614.061.414 and through a VICI grant (A.~Helmi) and the Netherlands Research School for Astronomy (NOVA);
\item the Polish National Science Centre through HARMONIA grant 2015/18/M/ST9/00544 and ETIUDA grants 2016/20/S/ST9/00162 and 2016/20/T/ST9/00170;
\item the Portugese Funda\c{c}\~ao para a Ci\^{e}ncia e a Tecnologia (FCT) through grant SFRH/BPD/74697/2010; the Strategic Programmes UID/FIS/00099/2013 for CENTRA and UID/EEA/00066/2013 for UNINOVA;
\item the Slovenian Research Agency through grant P1-0188;
\item the Spanish Ministry of Economy (MINECO/FEDER, UE) through grants ESP2014-55996-C2-1-R, ESP2014-55996-C2-2-R, ESP2016-80079-C2-1-R, and ESP2016-80079-C2-2-R, the Spanish Ministerio de Econom\'{\i}a, Industria y Competitividad through grant AyA2014-55216, the Spanish Ministerio de Educaci\'{o}n, Cultura y Deporte (MECD) through grant FPU16/03827, the Institute of Cosmos Sciences University of Barcelona (ICCUB, Unidad de Excelencia 'Mar\'{\i}a de Maeztu') through grant MDM-2014-0369, the Xunta de Galicia and the Centros Singulares de Investigaci\'{o}n de Galicia for the period 2016-2019 through the Centro de Investigaci\'{o}n en Tecnolog\'{\i}as de la Informaci\'{o}n y las Comunicaciones (CITIC), the Red Espa\~{n}ola de Supercomputaci\'{o}n (RES) computer resources at MareNostrum, and the Barcelona Supercomputing Centre - Centro Nacional de Supercomputaci\'{o}n (BSC-CNS) through activities AECT-2016-1-0006, AECT-2016-2-0013, AECT-2016-3-0011, and AECT-2017-1-0020;
\item the Swedish National Space Board (SNSB/Rymdstyrelsen);
\item the Swiss State Secretariat for Education, Research, and Innovation through the ESA PRODEX programme, the Mesures d’Accompagnement, the Swiss Activit\'es Nationales Compl\'ementaires, and the Swiss National Science Foundation;
\item the United Kingdom Rutherford Appleton Laboratory, the United Kingdom Science and Technology Facilities Council (STFC) through grant ST/L006553/1, the United Kingdom Space Agency (UKSA) through grant ST/N000641/1 and ST/N001117/1, as well as a Particle Physics and Astronomy Research Council Grant PP/C503703/1.
\end{itemize}

Our work was eased considerably by the use of the data handling and
visualisation software
\href{http://www.starlink.ac.uk/topcat/}{TOPCAT}, and
\href{http://www.starlink.ac.uk/stilts}{STILTS}
\citepads{2005ASPC..347...29T}.

This publication makes use of data products from NEOWISE, Wide-field
Infrared Survey Explorer (WISE), a joint project of the University of
California, Los Angeles, and the Jet Propulsion Laboratory/California
Institute of Technology, funded by the National Aeronautics and Space
Administration.

In addition to the currently active DPAC (and ESA science) authors of the peer-reviewed papers accompanying \gdrtwo, there are large numbers of former DPAC members who made significant contributions to the (preparations of the) data processing. Among those are, in alphabetical order:
Christopher Agard, 
Juan Jos\'{e} Aguado, 
Alexandra Alecu, 
Peter Allan, 
France Allard, 
Walter Allasia, 
Carlos Allende Prieto, 
Antonio Amorim, 
Kader Amsif, 
Guillem Anglada-Escud\'{e}, 
Sonia Ant\'{o}n, 
Vladan Arsenijevic, 
Rajesh Kumar Bachchan, 
Angelique Barbier, 
Mickael Batailler, 
Duncan Bates, 
Mathias Beck, 
Antonio Bello Garc\'{\i}a, 
Vasily Belokurov, 
Philippe Bendjoya, 
Hans Bernstein$^\dagger$, 
Lionel Bigot, 
Albert Bijaoui, 
Fran\c{c}oise Billebaud, 
Nadejda Blagorodnova, 
Thierry Bloch, 
Klaas de Boer, 
Marco Bonfigli, 
Giuseppe Bono, 
Fran\c{c}ois Bouchy, 
Steve Boudreault, 
Guy Boutonnet, 
Pascal Branet, 
Elme Breedt Lategan, 
Scott Brown, 
Pierre-Marie Brunet, 
Peter Bunclark$^\dagger$, 
Roberto Buonanno, 
Robert Butorafuchs, 
Joan Cambras, 
Heather Campbell, 
Christophe Carret, 
Manuel Carrillo, 
C\'{e}sar Carri\'{o}n, 
Fabien Ch\'{e}reau, 
Jonathan Charnas, 
Ross Collins, 
Leonardo Corcione, 
Nick Cross, 
Jean-Charles Damery, 
Eric Darmigny, 
Peter De Cat, 
C\'{e}line Delle Luche, 
Markus Demleitner, 
S\'{e}kou Diakite, 
Carla Domingues, 
Sandra Dos Anjos, 
Laurent Douchy, 
Pierre Dubath, 
Yifat Dzigan, 
Sebastian Els, 
Wyn Evans, 
Guillaume Eynard Bontemps, 
Fernando de Felice, 
Agn\`{e}s Fienga, 
Florin Fodor, 
Aidan Fries, 
Jan Fuchs, 
Flavio Fusi Pecci, 
Diego Fustes, 
Duncan Fyfe, 
Emilien Gaudin, 
Yoann G\'{e}rard, 
Anita G\'{o}mez, 
Ana Gonz\'{a}lez-Marcos, 
Andres G\'{u}rpide, 
Eva Gallardo, 
Daniele Gardiol, 
Marwan Gebran, 
Nathalie Gerbier, 
Andreja Gomboc, 
Eva Grebel, 
Michel Grenon, 
Eric Grux, 
Pierre Guillout, 
Erik H{\o}g, 
Gordon Hopkinson$^\dagger$, 
Albert Heyrovsky, 
Andrew Holland, 
Claude Huc, 
Jason Hunt, 
Brigitte Huynh, 
Giacinto Iannicola, 
Mike Irwin, 
Yago Isasi Parache, 
Thierry Jacq, 
Laurent Jean-Rigaud, 
Isabelle J{\'e}gouzo-Giroux, 
Asif Jan, 
Anne-Marie Janotto, 
Fran\c{c}ois Jocteur-Monrozier, 
Paula Jofr\'{e}, 
Anthony Jonckheere, 
Antoine Jorissen, 
Ralf Keil, 
Dae-Won Kim, 
Peter Klagyivik, 
Jens Knude, 
Oleg Kochukhov, 
Indrek Kolka, 
Janez Kos, 
Irina Kovalenko, 
Maria Kudryashova, 
Ilya Kull, 
Alex Kutka, 
Fr\'{e}d\'{e}ric Lacoste-Seris, 
Val\'{e}ry Lainey, 
Claudia Lavalley, 
David LeBouquin, 
Vassili Lemaitre, 
Thierry Levoir, 
Chao Liu, 
Davide Loreggia, 
Denise Lorenz, 
Ian MacDonald, 
Marc Madaule, 
Tiago Magalh\~{a}es Fernandes, 
Valeri Makarov, 
Jean-Christophe Malapert, 
Herv\'{e} Manche, 
Mathieu Marseille, 
Christophe Martayan, 
Oscar Martinez-Rubi, 
Paul Marty, 
Benjamin Massart, 
Emmanuel Mercier, 
Fr\'{e}d\'{e}ric Meynadier, 
Shan Mignot, 
Bruno Miranda, 
Marco Molinaro, 
Marc Moniez, 
Alain Montmory, 
Stephan Morgenthaler, 
Ulisse Munari, 
J\'{e}r\^{o}me Narbonne, 
Anne-Th\'{e}r\`{e}se Nguyen, 
Thomas Nordlander, 
Markus Nullmeier, 
Derek O'Callaghan, 
Pierre Ocvirk, 
Joaqu\'{\i}n Ordieres-Mer\'{e}, 
Patricio Ortiz, 
Jose Osorio, 
Dagmara Oszkiewicz, 
Alex Ouzounis, 
Fabien P\'{e}turaud, 
Max Palmer, 
Peregrine Park, 
Ester Pasquato, 
Xavier Passot, 
Marco Pecoraro, 
Roselyne Pedrosa, 
Christian Peltzer, 
Hanna Pentik\"{a}inen, 
Jordi Peralta, 
Bernard Pichon, 
Tuomo Pieniluoma, 
Enrico Pigozzi, 
Bertrand Plez, 
Joel Poels$^\dagger$, 
Ennio Poretti Merate, 
Arnaud Poulain, 
Guylaine Prat, 
Thibaut Prod'homme, 
Adrien Raffy, 
Serena Rago, 
Piero Ranalli, 
Gregory Rauw, 
Andrew Read, 
Jos\'{e} Rebordao, 
Philippe Redon, 
Rita Ribeiro, 
Pascal Richard, 
Daniel Risquez, 
Brigitte Rocca-Volmerange, 
Nicolas de Roll, 
Siv Ros\'{e}n, 
Idoia Ruiz-Fuertes, 
Federico Russo, 
Jan Rybizki, 
Damien Segransan, 
Arnaud Siebert, 
Helder Silva, 
Dimitris Sinachopoulos, 
Eric Slezak, 
Riccardo Smareglia, 
Michael Soffel, 
Danuta Sosnowska, 
Maxime Spano, 
Vytautas Strai\v{z}ys, 
Dirk Terrell, 
Stephan Theil, 
Carola Tiede, 
Brandon Tingley, 
Scott Trager, 
Licia Troisi, 
Paraskevi Tsalmantza, 
David Tur, 
Mattia Vaccari, 
Fr\'{e}d\'{e}ric Vachier, 
Pau Vall\`{e}s, 
Walter Van Hamme, 
Mihaly Varadi, 
Sjoert van Velzen, 
Lionel Veltz, 
Teresa Via, 
Jenni Virtanen, 
Antonio Volpicelli, 
Jean-Marie Wallut, 
Rainer Wichmann, 
Mark Wilkinson, 
Patrick Yvard, and 
Tim de Zeeuw. 

In addition to the DPAC consortium, past and present, there are numerous people, mostly in ESA and in industry, who have made or continue to make essential contributions to \gaia, for instance those employed in science and mission operations or in the design, manufacturing, integration, and testing of the spacecraft and its modules, subsystems, and units. Many of those will remain unnamed yet spent countless hours, occasionally during nights, weekends, and public holidays, in cold offices and dark clean rooms. At the risk of being incomplete, we specifically acknowledge, in alphabetical order,
from Airbus DS (Toulouse):
Alexandre Affre,
Marie-Th\'er\`ese Aim\'e,
Audrey Albert,
Aur\'elien Albert-Aguilar,
Hania Arsalane,
Arnaud Aurousseau,
Denis Bassi,
Franck Bayle,
Pierre-Luc Bazin,
Emmanuelle Benninger,
Philippe Bertrand,
Jean-Bernard Biau,
Fran\c{c}ois Binter,
C\'edric Blanc,
Eric Blonde,
Patrick Bonzom,
Bernard Bories,
Jean-Jacques Bouisset,
Jo\"el Boyadjian, 
Isabelle Brault,
Corinne Buge,
Bertrand Calvel, 
Jean-Michel Camus,
France Canton,
Lionel Carminati, 
Michel Carrie,
Didier Castel,
Philippe Charvet, 
Fran\c{c}ois Chassat, 
Fabrice Cherouat,
Ludovic Chirouze,
Michel Choquet,
Claude Coatantiec, 
Emmanuel Collados,
Philippe Corberand,
Christelle Dauga,
Robert Davancens, 
Catherine Deblock,
Eric Decourbey,
Charles Dekhtiar,
Michel Delannoy,
Michel Delgado,
Damien Delmas,
Emilie Demange, 
Victor Depeyre,
Isabelle Desenclos,
Christian Dio,
Kevin Downes,
Marie-Ange Duro,
Eric Ecale, 
Omar Emam,
Elizabeth Estrada,
Coralie Falgayrac,
Benjamin Farcot,
Claude Faubert,
Fr\'ed\'eric Faye, 
S\'ebastien Finana,
Gr\'egory Flandin, 
Loic Floury,
Gilles Fongy,
Michel Fruit, 
Florence Fusero, 
Christophe Gabilan,
J\'er\'emie Gaboriaud,
Cyril Gallard,
Damien Galy,
Benjamin Gandon,
Patrick Gareth,
Eric Gelis,
Andr\'e Gellon,
Laurent Georges, 
Philippe-Marie Gomez,
Jos\'e Goncalves,
Fr\'ed\'eric Guedes,
Vincent Guillemier,
Thomas Guilpain,
St\'ephane Halbout,
Marie Hanne,
Gr\'egory Hazera,
Daniel Herbin,
Tommy Hercher,
Claude Hoarau le Papillon,
Matthias Holz,
Philippe Humbert, 
Sophie Jallade, 
Gr\'egory Jonniaux, 
Fr\'ed\'eric Juillard,
Philippe Jung,
Charles Koeck,
Marc Labaysse, 
R\'en\'e Laborde,
Anouk Laborie, 
J\'er\^{o}me Lacoste-Barutel,
Baptiste Laynet,
Virginie Le Gall, 
Julien L'Hermitte,
Marc Le Roy, 
Christian Lebranchu, 
Didier Lebreton,
Patrick Lelong, 
Jean-Luc Leon,
Stephan Leppke,
Franck Levallois,
Philippe Lingot,
Laurant Lobo,
C\'eline Lopez,
Jean-Michel Loupias,
Carlos Luque,
S\'ebastien Maes,
Bruno Mamdy, 
Denis Marchais,
Alexandre Marson,
Benjamin Massart, 
R\'emi Mauriac,
Philippe Mayo,
Caroline Meisse, 
Herv\'e Mercereau,
Olivier Michel,
Florent Minaire,
Xavier Moisson, 
David Monteiro ,
Denis Montperrus,
Boris Niel,
C\'edric Papot,
Jean-Fran\c{c}ois Pasquier, 
Gareth Patrick,
Pascal Paulet, 
Martin Peccia,
Sylvie Peden,
Sonia Penalva, 
Michel Pendaries,
Philippe Peres,
Gr\'egory Personne, 
Dominique Pierot,
Jean-Marc Pillot,
Lydie Pinel, 
Fabien Piquemal,
Vincent Poinsignon, 
Maxime Pomelec,
Andr\'e Porras,
Pierre Pouny, 
Severin Provost, 
S\'ebastien Ramos,
Fabienne Raux,
Florian Reuscher,
Nicolas Riguet,
Mickael Roche,
Gilles Rougier, 
Bruno Rouzier, 
Stephane Roy,
Jean-Paul Ruffie,
Fr\'ed\'eric Safa, 
Heloise Scheer, 
Claudie Serris,
Andr\'e Sobeczko, 
Jean-Fran\c{c}ois Soucaille,
Philippe Tatry, 
Th\'eo Thomas,
Pierre Thoral,
Dominique Torcheux,
Vincent Tortel,
Stephane Touzeau, 
Didier Trantoul,
Cyril V\'etel, 
Jean-Axel Vatinel,
Jean-Paul Vormus, and 
Marc Zanoni;
from Airbus DS (Friedrichshafen):
Jan Beck,
Frank Blender,
Volker Hashagen,
Armin Hauser,
Bastian Hell,
Cosmas Heller,
Matthias Holz,
Heinz-Dieter Junginger,
Klaus-Peter Koeble,
Karin Pietroboni,
Ulrich Rauscher,
Rebekka Reichle,
Florian Reuscher,
Ariane Stephan,
Christian Stierle,
Riccardo Vascotto,
Christian Hehr,
Markus Schelkle,
Rudi Kerner,
Udo Schuhmacher,
Peter Moeller,
Rene Stritter,
J\"{u}rgen Frank,
Wolfram Beckert,
Evelyn Walser,
Steffen Roetzer,
Fritz Vogel, and
Friedbert Zilly;
from Airbus DS (Stevenage):
Mohammed Ali,
David Bibby,
Leisha Carratt,
Veronica Carroll,
Clive Catley,
Patrick Chapman,
Chris Chetwood,
Tom Colegrove,
Andrew Davies,
Denis Di Filippantonio,
Andy Dyne,
Alex Elliot,
Omar Emam,
Colin Farmer,
Steve Farrington,
Nick Francis,
Albert Gilchrist,
Brian Grainger,
Yann Le Hiress,
Vicky Hodges,
Jonathan Holroyd,
Haroon Hussain,
Roger Jarvis,
Lewis Jenner,
Steve King,
Chris Lloyd,
Neil Kimbrey,
Alessandro Martis,
Bal Matharu,
Karen May,
Florent Minaire,
Katherine Mills,
James Myatt,
Chris Nicholas,
Paul Norridge,
David Perkins,
Michael Pieri,
Matthew Pigg,
Angelo Povoleri,
Robert Purvinskis,
Phil Robson,
Julien Saliege,
Satti Sangha,
Paramijt Singh,
John Standing,
Dongyao Tan,
Keith Thomas,
Rosalind Warren,
Andy Whitehouse,
Robert Wilson,
Hazel Wood,
Steven Danes,
Scott Englefield,
Juan Flores-Watson,
Chris Lord,
Allan Parry,
Juliet Morris,
Nick Gregory, and
Ian Mansell.

From ESA, in alphabetical order:
%
%
Ricard Abello, 
Ivan Aksenov, 
Matthew Allen,  
Salim Ansari,  
Philippe Armbruster,  
Alessandro Atzei,  
Liesse Ayache,  
Samy Azaz,  
Jean-Pierre Balley, 
Manuela Baroni,  
Rainer Bauske, 
Thomas Beck, 
Gabriele Bellei, 
Carlos Bielsa, 
Gerhard Billig, 
Carmen Blasco,  
Andreas Boosz, 
Bruno Bras,  
Julia Braun, 
Thierry Bru, 
Frank Budnik, 
Joe Bush, 
Marco Butkovic, 
Jacques Cande\'e, 
David Cano, 
Carlos Casas, 
Francesco Castellini, 
David Chapmann, 
Nebil Cinar, 
Mark Clements, 
Giovanni Colangelo,  
Peter Collins, 
Ana Colorado McEvoy, 
Vincente Companys, 
Federico Cordero, 
Sylvain Damiani, 
Fabienne Delhaise, 
Gianpiero Di Girolamo, 
Yannis Diamantidis, 
John Dodsworth, 
Ernesto D\"olling, 
Jane Douglas,  
Jean Doutreleau,  
Dominic Doyle,  
Mark Drapes, 
Frank Dreger, 
Peter Droll, 
Gerhard Drolshagen,  
Bret Durrett, 
Christina Eilers,  
Yannick Enginger, 
Alessandro Ercolani, 
Matthias Erdmann,  
Orcun Ergincan,  
Robert Ernst,  
Daniel Escolar,  
Maria Espina, 
Hugh Evans,  
Fabio Favata,  
Stefano Ferreri, 
Daniel Firre, 
Michael Flegel, 
Melanie Flentge, 
Alan Flowers, 
Steve Foley, 
Jens Freih\"ofer, 
Rob Furnell,  
Julio Gallegos,  
Philippe Gar\'{e},  
Wahida Gasti,  
Jos\'e Gavira,  
Frank Geerling,  
Franck Germes,  
Gottlob Gienger, 
B\'en\'edicte Girouart,  
Bernard Godard, 
Nick Godfrey, 
C\'esar G\'omez Hern\'andez,  
Roy Gouka,  
Cosimo Greco, 
Robert Guilanya, 
Kester Habermann, 
Manfred Hadwiger, 
Ian Harrison, 
Angela Head, 
Martin Hechler,  
Kjeld Hjortnaes,  
John Hoar,  
Jacolien Hoek,  
Frank Hoffmann, 
Justin Howard, 
Arjan Hulsbosch,  
Christopher Hunter,  
Premysl Janik,  
Jos\'e Jim\'enez, 
Emmanuel Joliet,  
Helma van de Kamp-Glasbergen,  
Simon Kellett, 
Andrea Kerruish, 
Kevin Kewin, 
Oliver Kiddle, 
Sabine Kielbassa, 
Volker Kirschner,  
Kees van 't Klooster,  
Jan Kolmas, 
Oliver El Korashy,  
Arek Kowalczyk, 
Holger Krag, 
Beno\^{\i}t Lain\'e,  
Markus Landgraf,  
Sven Landstroem,  
Mathias Lauer, 
Robert Launer, 
Laurence Tu-Mai Levan,  
Mark ter Linden,  
Santiago Llorente, 
Tim Lock,  
Alejandro Lopez-Lozano, 
Guillermo Lorenzo, 
Tiago Loureiro, 
James Madison, 
Juan Manuel Garcia, 
Federico di Marco, 
Jonas Marie, 
Filip Marinic, 
Pier Mario Besso, 
Arturo Mart\'{\i}n Polegre,  
Ander Mart\'{\i}nez, 
Monica Mart\'{\i}nez Fern\'{a}ndez,  
Marco Massaro, 
Paolo de Meo, 
Ana Mestre, 
Luca Michienzi, 
David Milligan, 
Ali Mohammadzadeh,  
David Monteiro,  
Richard Morgan-Owen,  
Trevor Morley, 
Prisca M\"uhlmann,  
Jana Mulacova, 
Michael M\"uller, 
Pablo Munoz, 
Petteri Nieminen,  
Alfred Nillies, 
Wilfried Nzoubou, 
Alistair O'Connell, 
Karen O'Flaherty,  
Alfonso Olias Sanz,  
Oscar Pace,  
Mohini Parameswaran, 
Ramon Pardo, 
Taniya Parikh,  
Paul Parsons,  
Panos Partheniou, 
Torgeir Paulsen,  
Dario Pellegrinetti, 
Jos\'e-Louis Pellon-Bailon, 
Joe Pereira,  
Michael Perryman,  
Christian Philippe,  
Alex Popescu,  
Fr\'{e}d\'{e}ric Raison,  
Riccardo Rampini,  
Florian Renk, 
Alfonso Rivero, 
Andrew Robson, 
Gerd R\"ossling, 
Martina Rossmann, 
Markus R\"uckert, 
Andreas Rudolph,
Fr\'ed\'eric Safa,  
Jamie Salt,  
Giovanni Santin,  
Fabio de Santis, 
Rui Santos, 
Giuseppe Sarri,  
Stefano Scaglioni, 
Melanie Schabe, 
Dominic Sch\"afer, 
Micha Schmidt, 
Rudolf Schmidt,  
Ared Schnorhk,  
Klaus-J\"urgen Schulz, 
Jean Sch\"utz, 
Julia Schwartz, 
Andreas Scior, 
J\"org Seifert, 
Christopher Semprimoschnig,  
Ed Serpell, 
I\~{n}aki Serraller Vizcaino,  
Gunther Sessler, 
Felicity Sheasby,  
Alex Short,  
Heike Sillack, 
Swamy Siram, 
Christopher Smith, 
Claudio Sollazzo, 
Steven Straw, 
Pilar de Teodoro,  
Mark Thompson, 
Giulio Tonelloto,  
Felice Torelli,  
Raffaele Tosellini,  
Cecil Tranquille,  
Irren Tsu-Silva,  
Livio Tucci, 
Aileen Urwin, 
Jean-Baptiste Valet, 
Martin Vannier,  
Enrico Vassallo, 
David Verrier, 
Sam Verstaen,  
R\"udiger Vetter, 
Jos\'e Villalvilla, 
Raffaele Vitulli,  
Mildred V\"ogele, 
Sergio Volont\'e, 
Catherine Watson, 
Karsten Weber, 
Daniel Werner, 
Gary Whitehead, 
Gavin Williams, 
Alistair Winton,  
Michael Witting,  
Peter Wright, 
Karlie Yeung, 
Marco Zambianchi, and
Igor Zayer,  
and finally Vincenzo~Innocente from CERN.

In case of errors or omissions, please contact the \href{https://www.cosmos.esa.int/web/gaia/gaia-helpdesk}{\gaia\ Helpdesk}.